\documentclass[apjl]{emulateapj}
\usepackage{comment}
\usepackage{ifthen}
\usepackage{lineno}
\usepackage{hyperref}

\newcommand{\ctbd}[1]{}

\newcommand{\lc}{light curve}
\newcommand{\lcs}{light curves}
\newcommand{\Lc}{Light curve}

\newcommand{\band}[1]{\ensuremath{#1}-band}

\newcommand{\kms}{\ensuremath{\rm km\,s^{-1}}}
\newcommand{\ms}{\ensuremath{\rm m\,s^{-1}}}

\newcommand{\gcmc}{\ensuremath{\rm g\,cm^{-3}}}
\newcommand{\ergscmsq}{\ensuremath{\rm erg\,s^{-1}\,cm^{-2}}}

\newcommand{\vsini}{\ensuremath{v \sin{i}}}
\newcommand{\feh}{\ensuremath{\rm [Fe/H]}}

\newcommand{\rsun}{\ensuremath{R_\sun}}
\newcommand{\msun}{\ensuremath{M_\sun}}
\newcommand{\lsun}{\ensuremath{L_\sun}}

\newcommand{\rstar}{\ensuremath{R_\star}}
\newcommand{\mstar}{\ensuremath{M_\star}}
\newcommand{\lstar}{\ensuremath{L_\star}}

\newcommand{\teffstar}{\ensuremath{T_{\rm eff\star}}}
\newcommand{\rhostar}{\ensuremath{\rho_\star}}
\newcommand{\loggstar}{\ensuremath{\log{g_{\star}}}}

\newcommand{\rpl}{\ensuremath{R_{p}}}
\newcommand{\mpl}{\ensuremath{M_{p}}}

\newcommand{\rhopl}{\ensuremath{\rho_{p}}}

\newcommand{\arstar}{\ensuremath{a/\rstar}}
\newcommand{\zrstar}{\ensuremath{\zeta/\rstar}}

\newcommand{\rjup}{\ensuremath{R_{\rm J}}}
\newcommand{\mjup}{\ensuremath{M_{\rm J}}}

\newcommand{\refsecl}[1]{\mbox{Section \ref{sec:#1}}}

\newcommand{\reftabl}[1]{Table~\ref{tab:#1}}

\newcommand{\hatcurhtr}{HATS552-017}                      
\newcommand{\hatcurCCra}{\ensuremath{05^{\mathrm h}52^{\mathrm m}35.22{\mathrm s}}}                     
\newcommand{\hatcurCCdec}{\ensuremath{-19{\arcdeg}01{\arcmin}54.0{\arcsec}}}                    
\newcommand{\hatcurCCtwomass}{2MASS~05523523-1901539}     
\newcommand{\hatcurCCapassmV}{\ensuremath{15.160\pm0.024}} 
\newcommand{\hatcurCCapassmVshort}{\ensuremath{15.2}}      
\newcommand{\hatcurCCapassmB}{\ensuremath{16.664\pm0.043}} 
\newcommand{\hatcurCCtassmishort}{\ensuremath{13.7}}      
\newcommand{\hatcurCCtwomassJmag}{\ensuremath{12.046\pm0.024}} 
\newcommand{\hatcurCCtwomassHmag}{\ensuremath{11.397\pm0.023}} 
\newcommand{\hatcurCCtwomassKmag}{\ensuremath{11.224\pm0.023}} 
\newcommand{\hatcurLCPprec}{\ensuremath{3.3252725}}       
\newcommand{\hatcurLCPshort}{\ensuremath{3.3253}}         
\newcommand{\hatcurISOm}{\ensuremath{0.574_{-0.027}^{+0.020}}} 
\newcommand{\hatcurISOmshort}{\ensuremath{0.57}}          
\newcommand{\hatcurISOrshort}{\ensuremath{0.57}}          
\newcommand{\hatcurPPm}{\ensuremath{0.319\pm0.070}}       
\newcommand{\hatcurPPmshort}{\ensuremath{0.32}}           
\newcommand{\hatcurPPr}{\ensuremath{0.998\pm0.019}}       
\newcommand{\hatcurPPrshort}{\ensuremath{1.00}}           
\newcommand{\hatcurPPteff}{\ensuremath{712.8\pm5.1}}      
\newcommand{\hatcurPPfluxavgdim}{\ensuremath{7}}          
\newcommand{\hatcurLCrprstarcircdartmouth}{\ensuremath{0.17974\pm0.00069}} 
\newcommand{\hatcurLCimpcircdartmouth}{\ensuremath{0.428_{-0.022}^{+0.016}}} 
\newcommand{\hatcurLCzetacircdartmouth}{\ensuremath{28.608\pm0.066}}   
\newcommand{\hatcurLCdurcircdartmouth}{\ensuremath{0.08497\pm0.00036}} 
\newcommand{\hatcurLCingdurcircdartmouth}{\ensuremath{0.01542\pm0.00035}} 
\newcommand{\hatcurLCPcircdartmouth}{\ensuremath{3.3252721\pm0.0000020}} 
\newcommand{\hatcurLCTcircdartmouth}{\ensuremath{2456643.740560\pm0.000089}} 
\newcommand{\hatcurSMEiteffcircdartmouth}{\ensuremath{3770\pm100}}     
\newcommand{\hatcurSMEizfehcircdartmouth}{\ensuremath{0.200\pm0.091}}  
\newcommand{\hatcurLBizcircdartmouth}{\ensuremath{0.2843}}             
\newcommand{\hatcurLBiizcircdartmouth}{\ensuremath{0.3315}}            
\newcommand{\hatcurLBiicircdartmouth}{\ensuremath{0.4092}}             
\newcommand{\hatcurLBiiicircdartmouth}{\ensuremath{0.2831}}            
\newcommand{\hatcurLBigcircdartmouth}{\ensuremath{0.4748}}             
\newcommand{\hatcurLBiigcircdartmouth}{\ensuremath{0.3277}}            
\newcommand{\hatcurLBircircdartmouth}{\ensuremath{0.4403}}             
\newcommand{\hatcurLBiircircdartmouth}{\ensuremath{0.3333}}            
\newcommand{\hatcurLBiRcircdartmouth}{\ensuremath{0.4091}}             
\newcommand{\hatcurLBiiRcircdartmouth}{\ensuremath{0.3382}}            
\newcommand{\hatcurISOmcircdartmouth}{\ensuremath{0.595\pm0.017}}      
\newcommand{\hatcurISOmlongcircdartmouth}{\ensuremath{0.595\pm0.017}}  
\newcommand{\hatcurISOrcircdartmouth}{\ensuremath{0.5759\pm0.0092}}    
\newcommand{\hatcurISOrlongcircdartmouth}{\ensuremath{0.5759\pm0.0092}} 
\newcommand{\hatcurISOrholongcircdartmouth}{\ensuremath{4.40\pm0.13}}  
\newcommand{\hatcurISOloggcircdartmouth}{\ensuremath{4.6920\pm0.0077}} 
\newcommand{\hatcurISOlumcircdartmouth}{\ensuremath{0.0670\pm0.0054}}  
\newcommand{\hatcurISOmvcircdartmouth}{\ensuremath{9.43\pm0.19}}       
\newcommand{\hatcurISOagecircdartmouth}{\ensuremath{8.1\pm4.3}}        
\newcommand{\hatcurISOteffcircdartmouth}{\ensuremath{3872\pm53}}       
\newcommand{\hatcurISOMKcircdartmouth}{\ensuremath{5.079\pm0.065}}     
\newcommand{\hatcurRVKcircdartmouth}{\ensuremath{68\pm15}}             
\newcommand{\hatcurRVjitterAcircdartmouth}{\ensuremath{26\pm14}}       
\newcommand{\hatcurRVjitterBcircdartmouth}{\ensuremath{51\pm56}}       
\newcommand{\hatcurRVjitterCcircdartmouth}{\ensuremath{0\pm15}}        
\newcommand{\hatcurPPicircdartmouth}{\ensuremath{88.210\pm0.093}}      
\newcommand{\hatcurPPloggcircdartmouth}{\ensuremath{2.940_{-0.130}^{+0.064}}} 
\newcommand{\hatcurPParcircdartmouth}{\ensuremath{13.70\pm0.13}}       
\newcommand{\hatcurPParelcircdartmouth}{\ensuremath{0.03667\pm0.00035}} 
\newcommand{\hatcurPPrhocircdartmouth}{\ensuremath{0.431_{-0.111}^{+0.076}}} 
\newcommand{\hatcurPPmlongcircdartmouth}{\ensuremath{0.357_{-0.089}^{+0.059}}} 
\newcommand{\hatcurPPrlongcircdartmouth}{\ensuremath{1.008\pm0.019}}   
\newcommand{\hatcurPPmrcorrcircdartmouth}{\ensuremath{-0.00}}          
\newcommand{\hatcurPPteffcircdartmouth}{\ensuremath{740\pm12}}         
\newcommand{\hatcurPPthetacircdartmouth}{\ensuremath{0.0436_{-0.0112}^{+0.0069}}} 
\newcommand{\hatcurPPfluxavgcircdartmouth}{\ensuremath{6.80\pm0.44}}   
\newcommand{\hatcurXdistcircdartmouth}{\ensuremath{172.6\pm5.5}}       

\newcommand{\hatcurLCrprstarcircempirical}{\ensuremath{0.17978\pm0.00077}} 
\newcommand{\hatcurLCimpcircempirical}{\ensuremath{0.427_{-0.021}^{+0.020}}} 
\newcommand{\hatcurLCzetacircempirical}{\ensuremath{28.612\pm0.077}}   
\newcommand{\hatcurLCdurcircempirical}{\ensuremath{0.08504\pm0.00045}} 
\newcommand{\hatcurLCingdurcircempirical}{\ensuremath{0.01545\pm0.00039}} 
\newcommand{\hatcurLCPcircempirical}{\ensuremath{3.3252725\pm0.0000021}} 
\newcommand{\hatcurLCTcircempirical}{\ensuremath{2456643.740580\pm0.000084}} 
\newcommand{\hatcurISOmcircempirical}{\ensuremath{0.574_{-0.027}^{+0.020}}} 
\newcommand{\hatcurISOmlongcircempirical}{\ensuremath{0.574_{-0.027}^{+0.020}}} 
\newcommand{\hatcurISOrcircempirical}{\ensuremath{0.570\pm0.011}}      
\newcommand{\hatcurISOrlongcircempirical}{\ensuremath{0.570\pm0.011}}  
\newcommand{\hatcurISOrholongcircempirical}{\ensuremath{4.36\pm0.15}}  
\newcommand{\hatcurISOloggcircempirical}{\ensuremath{4.683\pm0.010}}   
\newcommand{\hatcurISOlumcircempirical}{\ensuremath{0.0558\pm0.0024}}  
\newcommand{\hatcurISOmvcircempirical}{\ensuremath{9.209\pm0.075}}     
\newcommand{\hatcurISOteffcircempirical}{\ensuremath{3724\pm18}}       
\newcommand{\hatcurISOMKcircempirical}{\ensuremath{5.387\pm0.043}}     
\newcommand{\hatcurRVKcircempirical}{\ensuremath{63\pm14}}             
\newcommand{\hatcurRVjitterAcircempirical}{\ensuremath{26\pm11}}       
\newcommand{\hatcurRVjitterBcircempirical}{\ensuremath{41\pm45}}       
\newcommand{\hatcurRVjitterCcircempirical}{\ensuremath{0\pm20}}        
\newcommand{\hatcurPPicircempirical}{\ensuremath{88.210\pm0.084}}      
\newcommand{\hatcurPPloggcircempirical}{\ensuremath{2.90\pm0.12}}      
\newcommand{\hatcurPParcircempirical}{\ensuremath{13.65\pm0.15}}       
\newcommand{\hatcurPParelcircempirical}{\ensuremath{0.03623_{-0.00057}^{+0.00042}}} 
\newcommand{\hatcurPPrhocircempirical}{\ensuremath{0.399\pm0.089}}     
\newcommand{\hatcurPPmlongcircempirical}{\ensuremath{0.319\pm0.070}}   
\newcommand{\hatcurPPrlongcircempirical}{\ensuremath{0.998\pm0.019}}   
\newcommand{\hatcurPPmrcorrcircempirical}{\ensuremath{0.08}}           
\newcommand{\hatcurPPteffcircempirical}{\ensuremath{712.8\pm5.1}}      
\newcommand{\hatcurPPthetacircempirical}{\ensuremath{0.0404\pm0.0089}} 
\newcommand{\hatcurPPfluxavgcircempirical}{\ensuremath{5.84\pm0.17}}   
\newcommand{\hatcurXdistcircempirical}{\ensuremath{148.4\pm3.3}}       

\newcommand{\hatcurLCrprstareccenempirical}{\ensuremath{0.17986\pm0.00070}} 
\newcommand{\hatcurLCimpeccenempirical}{\ensuremath{0.429_{-0.020}^{+0.017}}} 
\newcommand{\hatcurLCzetaeccenempirical}{\ensuremath{28.605\pm0.066}}  
\newcommand{\hatcurLCdureccenempirical}{\ensuremath{0.08508\pm0.00037}} 
\newcommand{\hatcurLCingdureccenempirical}{\ensuremath{0.01549\pm0.00034}} 
\newcommand{\hatcurLCPeccenempirical}{\ensuremath{3.3252726\pm0.0000021}} 
\newcommand{\hatcurLCTeccenempirical}{\ensuremath{2456643.740600\pm0.000094}} 
\newcommand{\hatcurISOmeccenempirical}{\ensuremath{0.573\pm0.031}}     
\newcommand{\hatcurISOmlongeccenempirical}{\ensuremath{0.573\pm0.031}} 
\newcommand{\hatcurISOreccenempirical}{\ensuremath{0.567\pm0.037}}     
\newcommand{\hatcurISOrlongeccenempirical}{\ensuremath{0.567\pm0.037}} 
\newcommand{\hatcurISOrholongeccenempirical}{\ensuremath{4.42_{-0.47}^{+0.75}}} 
\newcommand{\hatcurISOloggeccenempirical}{\ensuremath{4.689\pm0.038}}  
\newcommand{\hatcurISOlumeccenempirical}{\ensuremath{0.0555\pm0.0073}} 
\newcommand{\hatcurISOmveccenempirical}{\ensuremath{9.22\pm0.15}}      
\newcommand{\hatcurISOteffeccenempirical}{\ensuremath{3722\pm19}}      
\newcommand{\hatcurISOMKeccenempirical}{\ensuremath{5.40\pm0.14}}      
\newcommand{\hatcurRVKeccenempirical}{\ensuremath{64\pm15}}            
\newcommand{\hatcurRVrkeccenempirical}{\ensuremath{0.09\pm0.19}}       
\newcommand{\hatcurRVrheccenempirical}{\ensuremath{-0.03\pm0.16}}      
\newcommand{\hatcurRVkeccenempirical}{\ensuremath{0.015_{-0.028}^{+0.085}}} 
\newcommand{\hatcurRVheccenempirical}{\ensuremath{-0.004_{-0.058}^{+0.031}}} 
\newcommand{\hatcurRVjitterAeccenempirical}{\ensuremath{26.2\pm9.9}}   
\newcommand{\hatcurRVjitterBeccenempirical}{\ensuremath{39\pm41}}      
\newcommand{\hatcurRVjitterCeccenempirical}{\ensuremath{0\pm50}}       
\newcommand{\hatcurRVecceneccenempirical}{\ensuremath{0.053\pm0.060}}  
\newcommand{\hatcurRVomegaeccenempirical}{\ensuremath{230\pm120}}      
\newcommand{\hatcurPPieccenempirical}{\ensuremath{88.23\pm0.16}}       
\newcommand{\hatcurPPloggeccenempirical}{\ensuremath{2.91\pm0.12}}     
\newcommand{\hatcurPPareccenempirical}{\ensuremath{13.71_{-0.50}^{+0.73}}} 
\newcommand{\hatcurPPareleccenempirical}{\ensuremath{0.03621\pm0.00065}} 
\newcommand{\hatcurPPrhoeccenempirical}{\ensuremath{0.406_{-0.098}^{+0.133}}} 
\newcommand{\hatcurPPmlongeccenempirical}{\ensuremath{0.321\pm0.076}}  
\newcommand{\hatcurPPrlongeccenempirical}{\ensuremath{0.994\pm0.065}}  
\newcommand{\hatcurPPmrcorreccenempirical}{\ensuremath{0.16}}          
\newcommand{\hatcurPPteffeccenempirical}{\ensuremath{711\pm18}}        
\newcommand{\hatcurPPthetaeccenempirical}{\ensuremath{0.0409\pm0.0100}} 
\newcommand{\hatcurPPfluxavgeccenempirical}{\ensuremath{5.78\pm0.60}}  
\newcommand{\hatcurXdisteccenempirical}{\ensuremath{147.8_{-9.4}^{+7.0}}} 
\newcommand{\hatcurISOrhoeccendartmouth}{\ensuremath{14.8\pm2.9}}      
\newcommand{\hatcurRVjitterAeccendartmouth}{\ensuremath{0.1\pm5.6}}    
\newcommand{\hatcurRVecceneccendartmouth}{\ensuremath{0.404\pm0.058}}  
\newcommand{\hatcurPPtcirceccendartmouth}{\ensuremath{342\pm35}}       

\newcommand{\hatcur}{HATS-6}
\newcommand{\hatcurb}{HATS-6b}

\newcommand{\hatcurjhkfilset}{ESO}

\newcommand{\rprstar}{\ensuremath{\rpl/\rstar}}

\newboolean{emulateapj}
\setboolean{emulateapj}{true}

\newboolean{rvtablelong}
\setboolean{rvtablelong}{true}

\newboolean{astroph}
\setboolean{astroph}{true}

\shortauthors{Hartman et al.}
\shorttitle{\hatcur\lowercase{b}}
\ifthenelse{\boolean{emulateapj}}{
    
}{
    
}

\begin{document}


\title{\hatcur\lowercase{b}: A Warm Saturn Transiting an Early M Dwarf Star, and a Set of Empirical Relations for Characterizing K and M Dwarf Planet Hosts \altaffilmark{$\dagger$}}

\author{
J.~D.~Hartman\altaffilmark{1},
D.~Bayliss\altaffilmark{2},
R.~Brahm\altaffilmark{3,4},
G.~\'A.~Bakos\altaffilmark{1,$\star$,$\star\star$},
L.~Mancini\altaffilmark{5},
A.~Jord\'an\altaffilmark{3,4},
K.~Penev\altaffilmark{1},   
M.~Rabus\altaffilmark{3,5},
G.~Zhou\altaffilmark{2},
R.~P.~Butler\altaffilmark{6},
N.~Espinoza\altaffilmark{3,4},
M.~de~Val-Borro\altaffilmark{1},
W.~Bhatti\altaffilmark{1}, 
Z.~Csubry\altaffilmark{1},
S.~Ciceri\altaffilmark{5},
T.~Henning\altaffilmark{5},
B.~Schmidt\altaffilmark{2},
P.~Arriagada\altaffilmark{6},
S.~Shectman\altaffilmark{7},
J.~Crane\altaffilmark{7},
I.~Thompson\altaffilmark{7},
V.~Suc\altaffilmark{3},
B.~Cs\'ak\altaffilmark{5},
T.~G.~Tan\altaffilmark{8},
R.~W.~Noyes\altaffilmark{9},
J.~L\'az\'ar\altaffilmark{10}, 
I.~Papp\altaffilmark{10}, 
P.~S\'ari\altaffilmark{10}
}
\altaffiltext{1}{Department of Astrophysical Sciences,
    Princeton University, NJ 08544, USA;
    email: jhartman@astro.princeton.edu}

\altaffiltext{2}{The Australian National University, Canberra,
    Australia}

\altaffiltext{3}{Instituto de Astrof\'isica, Facultad de F\'isica,
    Pontificia Universidad Cat\'olica de Chile, Av.\ Vicu\~na Mackenna
    4860, 7820436 Macul, Santiago, Chile}

\altaffiltext{4}{Millennium Institute of Astrophysics, Av.\ Vicu\~na Mackenna
    4860, 7820436 Macul, Santiago, Chile}

\altaffiltext{5}{Max Planck Institute for Astronomy, Heidelberg,
    Germany}

\altaffiltext{6}{Department of Terrestrial Magnetism, Carnegie Institution of Washington, 5241 Broad Branch Road NW, Washington, DC 20015-1305, USA}

\altaffiltext{7}{The Observatories of the Carnegie Institution of Washington, 813 Santa Barbara Street, Pasadena, CA 91101, USA}

\altaffiltext{8}{Perth Exoplanet Survey Telescope, Perth, Australia}

\altaffiltext{9}{Harvard-Smithsonian Center for Astrophysics,
    Cambridge, MA, USA}

\altaffiltext{10}{Hungarian Astronomical Association, Budapest, Hungary}

\altaffiltext{$\star$}{Alfred P. Sloan Research Fellow}

\altaffiltext{$\star\star$}{Packard Fellow}

\altaffiltext{$\dagger$}{
The HATSouth network is operated by a collaboration consisting of
Princeton University (PU), the Max Planck Institute f\"ur Astronomie
(MPIA), the Australian National University (ANU), and the Pontificia
Universidad Cat\'olica de Chile (PUC).  The station at Las Campanas
Observatory (LCO) of the Carnegie Institute is operated by PU in
conjunction with PUC, the station at the High Energy Spectroscopic
Survey (H.E.S.S.) site is operated in conjunction with MPIA, and the
station at Siding Spring Observatory (SSO) is operated jointly with
ANU.
This paper includes data gathered with the 6.5\,m Magellan Telescopes
located as Las Campanas Observatory, Chile. Based in part on
observations made with the MPG~2.2\,m Telescope and the ESO~3.6\,m
Telescope at the ESO Observatory in La Silla.  This
paper uses observations obtained with facilities of the Las Cumbres
Observatory Global Telescope.
}


\begin{abstract}

We report the discovery by the HATSouth survey of \hatcurb{}, an
extrasolar planet transiting a V=\hatcurCCapassmVshort\,mag,
$i=\hatcurCCtassmishort$\,mag M1V star with a mass of
\hatcurISOmshort\,\msun\ and a radius of \hatcurISOrshort\,\rsun.
\hatcurb{} has a period of $P\approx\hatcurLCPshort$\,d, mass of $\mpl
\approx \hatcurPPmshort$\,\mjup, radius of $\rpl \approx
\hatcurPPrshort$\,\rjup, and zero-albedo equilibrium temperature of
$T_{\rm eq} = \hatcurPPteff$\,K. \hatcur{} is one of the lowest mass
stars known to host a close-in gas giant planet, and its transits are
among the deepest of any known transiting planet system. We discuss
the follow-up opportunities afforded by this system, noting that
despite the faintness of the host star, it is expected to have the
highest $K$-band S/N transmission spectrum among known gas giant
planets with $T_{\rm eq} < 750$\,K. In order to characterize the star
we present a new set of empirical relations between the density,
radius, mass, bolometric magnitude, and $V$, $J$, $H$ and $K$-band
bolometric corrections for main sequence stars with $M < 0.80$\,\msun,
or spectral types later than K5. These relations are calibrated using
eclipsing binary components as well as members of resolved binary
systems. We account for intrinsic scatter in the relations in a
self-consistent manner. We show that from the transit-based stellar
density alone it is possible to measure the mass and radius of a $\sim
0.6$\,\msun\ star to $\sim 7$\% and $\sim 2$\% precision,
respectively. Incorporating additional information, such as the $V-K$
color, or an absolute magnitude, allows the precision to be improved
by up to a factor of two.


\setcounter{footnote}{0}
\end{abstract}



\section{Introduction}
\label{sec:introduction}

One of the goals in the study of exoplanetary systems is to determine
how the properties of planets depend on the properties of their host
stars. An important parameter in this respect is the host star
mass. Results from both radial velocity and transit surveys indicate
that the occurrence rate of gas giant planets is a steep function of
stellar mass scaling approximately as $N \propto M_{\star}$
\citep{johnson:2010} for main sequence stars of type M through F
(smaller planets, on the other hand, appear to be {\em more} frequent
around M dwarfs than around hotter stars, \citealp{howard:2012}). The
low occurrence rate of these planets around M dwarfs, coupled with the
fact that most surveys primarily target FGK dwarf stars, means that
only one hot Jupiter has so far been discovered around an M0 dwarf
\citep[Kepler-45;][]{johnson:2012}, and only a handful of others have
been found around very late K dwarf stars (WASP-80,
\citealp[][]{triaud:2013}; WASP-43, \citealp{hellier:2011}; and
HAT-P-54, \citealp{bakos:2014:hat54}, being the only three known
transiting-hot-Jupiter-hosting K dwarfs with $M < 0.65\,\msun$).

In addition to enabling the study of planetary properties as a
function of stellar mass, finding planets around late-type stars has
at least two other advantages. The small sizes of these stars coupled
with their low luminosities means that a planet with a given radius
and orbital period around a late-type star will produce deeper
transits and have a cooler equilibrium temperature than if it were
around a larger star. This makes planets around late-type stars
attractive targets for carrying out detailed follow-up observations,
such as atmospheric characterizations.  A second, perhaps
under appreciated, advantage of these stars is that they are
remarkably simple in their bulk properties. Whereas, due to stellar
evolution, the radius of a solar-metallicity, solar-mass star varies
by $\sim 40\%$ over the 10\,Gyr age of the Galactic disk, the radius
of a $0.6$\,\msun\ star varies by less than $\sim 5$\% over the same
time-span. The parameters of low-mass stars follow tight main-sequence
relations, enabling high-precision measurements of their properties
from only a single (or a few) observable(s). The precision of the
stellar parameters feeds directly into the precision of the planetary
parameters, so that in principle planets around low-mass stars may be
characterized with higher precision than those around higher mass
stars.

Here we present the discovery of a transiting, short-period, gas-giant planet
around an M1 dwarf star. This planet, \hatcurb{}, was discovered by the
HATSouth survey, a global network of fully-automated wide-field
photometric instruments searching for transiting planets
\citep{bakos:2013:hatsouth}. HATSouth uses larger-diameter optics than
other similar projects allowing an enhanced sensitivity to faint K and M
dwarfs. 

We also present a new set of empirical relations to use in
characterizing the properties of transiting planet host stars with $M
< 0.80$\,\msun. While there has been much discussion in the literature
of the apparent discrepancy between observations and various
theoretical models for these stars \citep[e.g.,][and references
  therein]{torres:2002,ribas:2003,torres:2013,zhou:2014:mebs}, the
growing set of well-characterized low-mass stars has revealed that, as
expected, these stars do follow tight main sequence relations. We show
that from the bulk density of the star alone, which is determined from
the transit light curve and RV observations, it is possible to measure
the mass and radius of a $\sim 0.6$\,\msun\ star to $\sim 7\%$ and $\sim
2\%$ precision, respectively. Incorporating additional information,
such as the $V-K$ color, or an absolute magnitude, allows the
precision to be improved by a factor of two.

The layout of the paper is as follows. In \refsecl{obs} we report the
detection of the photometric signal and the follow-up spectroscopic
and photometric observations of \hatcur{}.  In \refsecl{analysis} we
describe the analysis of the data, beginning with ruling out false
positive scenarios, continuing with our global modelling of the
photometry and radial velocities, and finishing with the determination
of the stellar parameters, and planetary parameters which depend on
them, using both theoretical stellar models as well as the empirical
relations which we derive here.  Our findings are discussed in
\refsecl{discussion}.

\section{Observations}
\label{sec:obs}

\begin{figure}[!ht]
\plotone{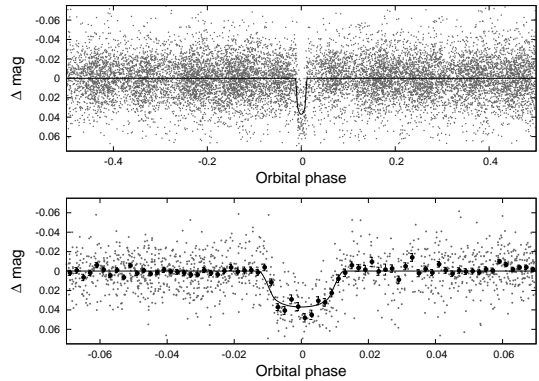}
\caption{
        Unbinned instrumental \band{r} \lc{} of \hatcur{} folded with
        the period $P = \hatcurLCPprec$\,days resulting from the global
        fit described in \refsecl{analysis}.  The solid line shows the
        best-fit transit model (see \refsecl{analysis}).  In the lower
        panel we zoom-in on the transit; the dark filled points here
        show the light curve binned in phase using a bin-size of 0.002.\\
\label{fig:hatsouth}}
\end{figure}

\begin{figure*}[!ht]
\plotone{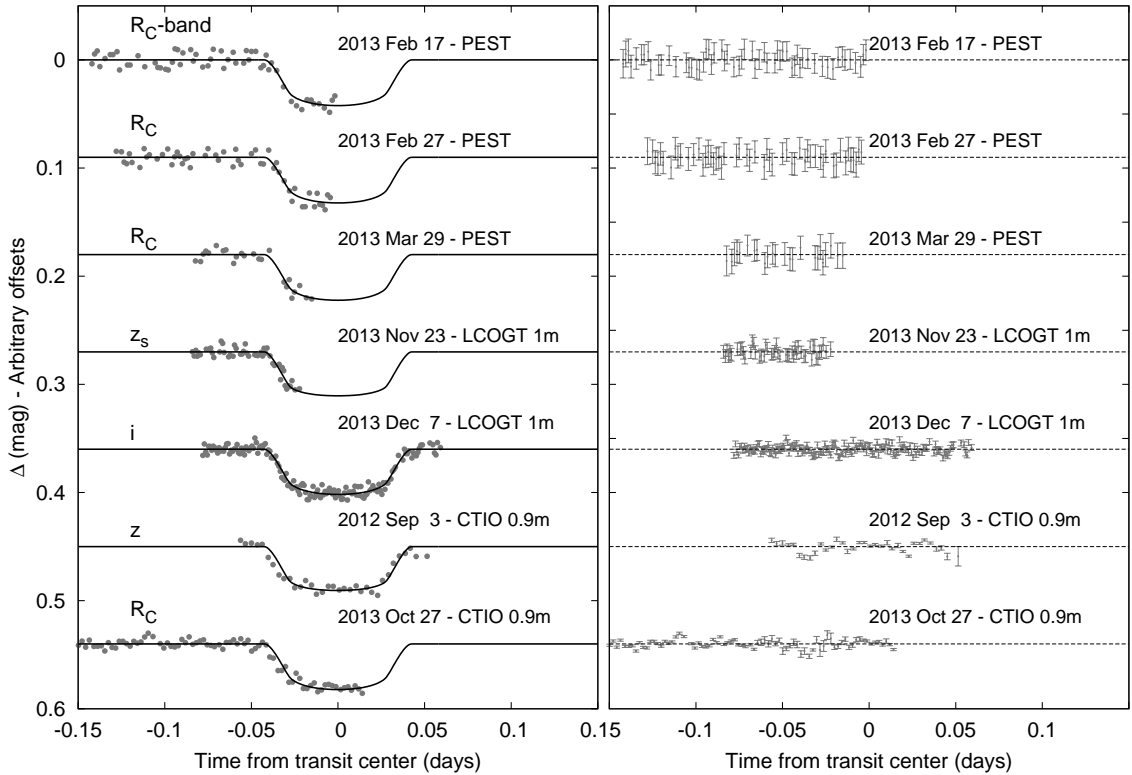}
\caption{
        Left: Unbinned follow-up transit \lcs{} of \hatcur{}.  The
        dates, filters and instruments used for each event are
        indicated.  The light curves have been detrended using the EPD
        process.  Curves after the first are shifted for clarity.  Our
        best fit is shown by the solid lines.  Right: Residuals from
        the fits in the same order as the curves at left. Additional
        follow-up light curves from GROND are shown in
        Figure~\ref{fig:lcgrond}.\\
\label{fig:lc}} \end{figure*}

\begin{figure*}[!ht]
\plotone{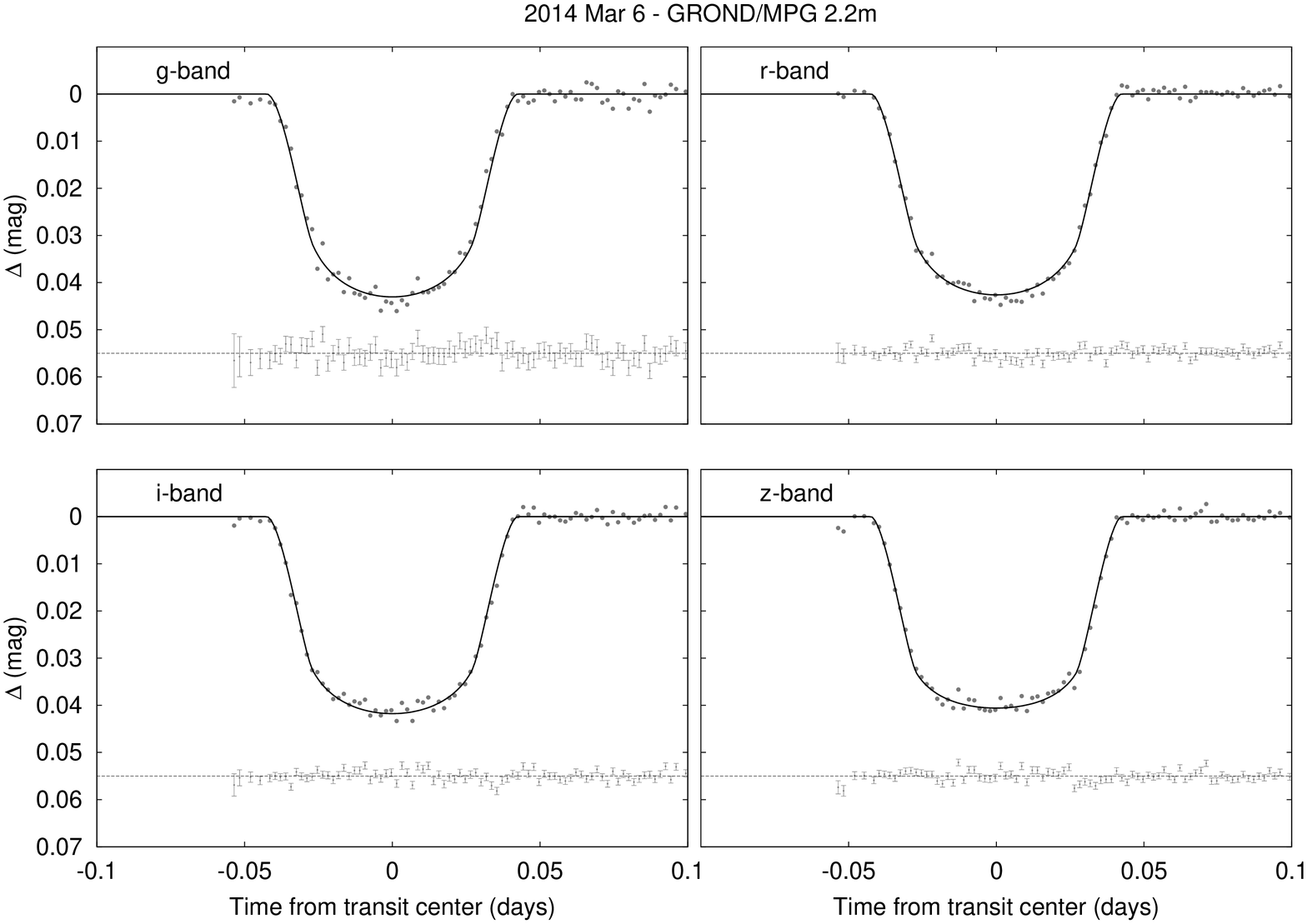}
\caption{
        Unbinned instrumental Sloan $g$-, $r$-, $i$- and \band{z}
        transit \lcs{} of \hatcur{} obtained with GROND on the
        MPG~2.2\,m on the night of UT 2014 March 6. The light curves
        have been detrended using the EPD process, and the best-fit is
        shown by the solid lines. Below each light curve we plot the
        residuals.\\
\label{fig:lcgrond}} \end{figure*}

\subsection{Photometric detection}
\label{sec:detection}

Observations of a field containing \hatcur{} (see \reftabl{stellar}
for identifying information) were carried out with the HS-2, HS-4 and
HS-6 units of the HATSouth network (located at Las Campanas
Observatory in Chile, the H.E.S.S. gamma-ray telescope site in
Namibia, and Siding Spring Observatory in Australia, respectively; see
\citealp{bakos:2013:hatsouth} for a detailed description of the
HATSouth network) between UT 2009-09-17 and UT 2010-09-10. A total
5695, 5544 and 88 images included in our final trend and
outlier-filtered light curves were obtained with HS-2, HS-4 and HS-6,
respectively. Observations were made through a Sloan $r$ filter, using
an exposure time of $240$\,s and a median cadence of $293$\,s (see
also Table~\ref{tab:photobs}).

The data were reduced to trend-filtered light curves following
\cite{bakos:2013:hatsouth}. We apply two empirical trend filtering
techniques to the data. The first is to decorrelate the light curves
against a set of measured parameters which vary from image to image,
including the $X$ and $Y$ sub-pixel coordinates of the star, three
parameters describing the shape of the image point spread function,
the hour angle of the observations, the zenith distance, and the sky
background near the target. We refer to this filtering as External
Parameter Decorrelation (EPD; \citealp{bakos:2010:hat11}). The second
filtering technique is to use the Trend Filtering Algorithm (TFA) due
to \cite{kovacs:2005:TFA}. In this method we select a list of $\sim 800$
template light curves uniformly distributed in position across the
field, and coming from stars with a broad range of magnitudes. We then
fit the EPD-filtered light curve for \hatcur{} as a linear combination
of these $\sim 800$ templates, and subtract the fit from the
observations. The filtered light curve for \hatcur{} has a
point-to-point RMS of 0.02\,mag and is dominated by noise from the
background sky. Transits were identified in the resulting HATSouth
light curve of \hatcur{} using the Box Least Squares
\citep[BLS;][]{kovacs:2002:bls} algorithm. Figure~\ref{fig:hatsouth}
shows the phase-folded HATSouth light curve of \hatcur{} together with
our best-fit transit model, while the photometric measurements are
provided in Table~\ref{tab:phfu}.

We searched the residual HATSouth light curve for additional transit
signals using BLS but found no significant detection. We also searched
for continuous quasi-periodic variability using both the Discrete
Fourier Transform \citep{kurtz:1985} and the Discrete Autocorrelation
Function \citep{edelson:1988}. We find a possible signal in the
pre-TFA light curve with a period of $P = 35.1$\,d and an S/N,
measured in the power spectrum, of 12.3. This signal is not seen in
the Autocorrelation Function, nor is it seen in the light curve after
processing with TFA. While potentially due to stellar rotation, this
may also be time-correlated noise. We therefore do not claim a
measurement of the photometric rotation period for this star.

\subsection{Photometric follow-up}
\label{sec:phfu}

In order to confirm the transit signal, and determine the parameters
of the system with better accuracy, we carried out follow-up
photometric observations of \hatcur{} using the 0.3\,m Perth Exoplanet
Survey Telescope (PEST), the CTIO 0.9\,m, telescopes in the LCOGT 1\,m
network \citep{brown:2013:lcogt}, and GROND on the MPG~2.2\,m
\citep{greiner:2008}. Key aspects of these observations, including the
dates of the observations, the number of images obtained, the cadence,
filter used, and the precision of the resulting \hatcur{} light
curves, as measured from the RMS of the residuals from our best-fit
model, are summarized in Table~\ref{tab:photobs}. The light curves are
plotted in Figures~\ref{fig:lc} and~\ref{fig:lcgrond}, while the
measurements are provided in Table~\ref{tab:phfu}. Details regarding
the PEST instrument, our observational procedure, as well as our
reduction and photometry methods can be found in
\citet{zhou:2014:mebs}; see \citet{penev:2013:hats1} and
\citet{mohlerfischer:2013:hats2} for similar information regarding the
GROND observations. Because this is the first time we have used the
CTIO~0.9\,m and the LCOGT 1\,m network, we describe our use of these
facilities in more detail below.

On the nights of 2012 Sep.~3 and 2013 Oct.~27, we performed
photometric observations of \hatcur{} using the CTIO~0.9\,m telescope,
which has a CCD with a $13\farcm6 \times 13\farcm6$ field of view. On
the first night we used a Gunn $z^{\prime}$ filter and on the latter
one a Kron-Cousins $R_{C}$ filter. We defocused the telescope in order
to broaden the point spread function.  Images from both nights were
calibrated (bias subtracted and flat fielded) with custom Python
routines. A fringing effect was seen in the images taken during our
full moon night in $z$-band. However, images taken with the $R_{C}$ filter
did not show fringing. Therefore, we also took 53 empty field regions
with the Gunn $z^{\prime}$ filter during our 2012 Sep.~CTIO run. We
combined these images and subtracted the sky background, and then
scaled the combined image to match, and remove, the additive fringing
effect seen in the science images. To reduce the calibrated images to
light curves, we chose a reference image and calculated the shift of
all images with respect to the reference image. From the reference
image we extracted the position of the stars. Following
\citet{deeg:2001}, the time series photometry was generated from these
observations using optimized aperture photometry that maximizes the
signal-to-noise ratio (S/N) for each star. For all images in one night
we used three fixed apertures and choose these to be much larger than
the typical point spread function in order to minimize the impact of
the time-variable point spread functions. The resulting light curves
of \hatcur{} have precisions of $4.7$\,mmag and $3.7$\,mmag on the
first and second nights, respectively.

Observations of \hatcur{} were carried out using the LCOGT 1\,m
network on the UT nights of 2013-11-23 and 2013-12-07. We used one of
the three telescopes installed at the South African Astronomical
Observatory (SAAO) on the night of 2013-11-23, and one of the
telescopes at CTIO on the night of 2013-12-07. In both cases we used
the SBIG STX-16803 4K$\times$4K imagers with which the telescopes
were initially deployed. These imagers provided a $16\arcmin \times
16\arcmin$ field of view with a pixel scale of $0\farcs23$. We used a
Pan-STARSS $z_{S}$ filter for the first night, and a Sloan
$i^{\prime}$ filter for the second night. Other details regarding the
instrumentation can be found in \citet{brown:2013:lcogt}. Calibrated
science images were provided by the LCOGT 1\,m pipeline. We performed
astrometry and aperture photometry using tools from the {\sc FITSH}
package \citep{pal:2012} on these images following methods that
we have previously applied to observations from the Faulkes
Telescopes, North and South, as well as to observations from Keplercam
on the Fred Lawrence Whipple Observatory 1.2\,m telescope. This
procedure is described in \citet{bakos:2010:hat11}.

\ifthenelse{\boolean{emulateapj}}{
    \begin{deluxetable*}{llrrrr}
}{
    \begin{deluxetable}{llrrrr}
}
\tablewidth{0pc}
\tabletypesize{\scriptsize}
\tablecaption{
    Summary of photometric observations
    \label{tab:photobs}
}
\tablehead{
    \multicolumn{1}{c}{Facility}          &
    \multicolumn{1}{c}{Date(s)}             &
    \multicolumn{1}{c}{Number of Images}\tablenotemark{a}      &
    \multicolumn{1}{c}{Cadence (s)}\tablenotemark{b}         &
    \multicolumn{1}{c}{Filter}               &
    \multicolumn{1}{c}{Precision (mmag)}
}
\startdata
~~~~HS-2 & 2009 Sep--2010 Sep & 5695 & 295 & $r$ & 21.3 \\
~~~~HS-4 & 2009 Sep--2010 Sep & 5544 & 293 & $r$ & 21.0 \\
~~~~HS-6 & 2010 Aug--2010 Sep & 88 & 296 & $r$ & 18.6 \\
~~~~CTIO~0.9\,m & 2012 Sep 03 & 34 & 297 & $z$ & 4.7 \\
~~~~PEST~0.3\,m & 2013 Feb 17 & 58 & 205 & $R$ & 6.1 \\
~~~~PEST~0.3\,m & 2013 Feb 27 & 53 & 131 & $R$ & 5.6 \\
~~~~PEST~0.3\,m & 2013 Mar 29 & 24 & 261 & $R$ & 4.8 \\
~~~~CTIO~0.9\,m & 2013 Oct 27 & 99 & 181 & $R_{C}$ & 3.7 \\
~~~~LCOGT~1\,m & 2013 Nov 23 & 48 & 76 & $z$ & 4.2 \\
~~~~LCOGT~1\,m & 2013 Dec 07 & 149 & 74 & $i$ & 3.8 \\
~~~~GROND/MPG~2.2\,m & 2014 Mar 06 & 95 & 155 & $g$ & 1.8 \\
~~~~GROND/MPG~2.2\,m & 2014 Mar 06 & 95 & 155 & $r$ & 1.1 \\
~~~~GROND/MPG~2.2\,m & 2014 Mar 06 & 95 & 155 & $i$ & 1.1 \\
~~~~GROND/MPG~2.2\,m & 2014 Mar 06 & 95 & 155 & $z$ & 1.1 \\
[-1.5ex]
\enddata 
\tablenotetext{a}{
  Excludes images which were rejected as significant outliers in the
  fitting procedure.
}
\tablenotetext{b}{
  The mode time difference rounded to the nearest second between
  consecutive points in each light curve.  Due to visibility, weather,
  pauses for focusing, etc., none of the light curves have perfectly
  uniform time sampling.\\
}
\ifthenelse{\boolean{emulateapj}}{
    \end{deluxetable*}
}{
    \end{deluxetable}
}

\ifthenelse{\boolean{emulateapj}}{
        \begin{deluxetable*}{lrrrrr} }{
        \begin{deluxetable}{lrrrrr} 
    }
        \tablewidth{0pc}
        \tablecaption{Differential photometry of
        \hatcur\label{tab:phfu}} \tablehead{ \colhead{BJD} &
        \colhead{Mag\tablenotemark{a}} &
        \colhead{\ensuremath{\sigma_{\rm Mag}}} &
        \colhead{Mag(orig)\tablenotemark{b}} & \colhead{Filter} &
        \colhead{Instrument} \\ \colhead{\hbox{~~~~(2\,400\,000$+$)~~~~}}
        & \colhead{} & \colhead{} & \colhead{} & \colhead{} &
        \colhead{} } \startdata $ 55185.60957 $ & $  -0.00387 $ & $   0.01084 $ & $ \cdots $ & $ r$ &         HS\\
$ 55145.70640 $ & $  -0.07475 $ & $   0.03251 $ & $ \cdots $ & $ r$ &         HS\\
$ 55095.82750 $ & $  -0.01258 $ & $   0.01255 $ & $ \cdots $ & $ r$ &         HS\\
$ 55275.39406 $ & $   0.02245 $ & $   0.01229 $ & $ \cdots $ & $ r$ &         HS\\
$ 55195.58763 $ & $  -0.00021 $ & $   0.01587 $ & $ \cdots $ & $ r$ &         HS\\
$ 55182.28728 $ & $   0.00324 $ & $   0.01174 $ & $ \cdots $ & $ r$ &         HS\\
$ 55145.70968 $ & $  -0.00768 $ & $   0.02964 $ & $ \cdots $ & $ r$ &         HS\\
$ 55185.61296 $ & $   0.02073 $ & $   0.01078 $ & $ \cdots $ & $ r$ &         HS\\
$ 55095.83089 $ & $  -0.02280 $ & $   0.01379 $ & $ \cdots $ & $ r$ &         HS\\
$ 55105.80688 $ & $  -0.02863 $ & $   0.02168 $ & $ \cdots $ & $ r$ &         HS\\

        [-1.5ex]
\enddata \tablenotetext{a}{
     The out-of-transit level has been subtracted. For the HATSouth
     light curve (rows with ``HS'' in the Instrument column), these
     magnitudes have been detrended using the EPD and TFA procedures
     prior to fitting a transit model to the light curve. Primarily as
     a result of this detrending, but also due to blending from
     neighbors, the apparent HATSouth transit depth is somewhat
     shallower than that of the true depth in the Sloan~$r$ filter
     (the apparent depth is 90\% that of the true depth). For the
     follow-up light curves (rows with an Instrument other than
     ``HS'') these magnitudes have been detrended with the EPD
     procedure, carried out simultaneously with the transit fit (the
     transit shape is preserved in this process).
}
\tablenotetext{b}{
        Raw magnitude values without application of the EPD
        procedure.  This is only reported for the follow-up light
        curves.
}
\tablecomments{
        This table is available in a machine-readable form in the
        online journal.  A portion is shown here for guidance
        regarding its form and content. The data are also available on
        the HATSouth website at \url{http://www.hatsouth.org}.
} \ifthenelse{\boolean{emulateapj}}{ \end{deluxetable*} }{ \end{deluxetable} }

\subsection{Spectroscopy}
\label{sec:spec}

\reftabl{specobssummary} summarizes the follow-up spectroscopic
observations which we obtained for \hatcur{}.

\subsubsection{Reconnaissance Spectroscopy}
\label{sec:reconspec}

Initial reconnaissance spectroscopic observations of \hatcur{} were
carried out using the Wide Field Spectrograph
\citep[WiFeS;][]{dopita:2007} on the ANU~2.3\,m telescope at SSO
together with the Echelle spectrograph on the du~Pont\,2.5\,m
telescope at LCO. The ANU~2.3\,m data were reduced and analyzed
following \cite{bayliss:2013:hats3}, while for the du~Pont data we
used the pipeline we have previously developed
\citep{jordan:2014:hats4} to analyze data from the Coralie and FEROS
spectrographs, adapted for the different spectral format of the
instrument. A single WiFeS spectrum was obtained with a resolution of
$R \equiv \lambda/\Delta\lambda = 3000$ to use in measuring the
effective temperature, surface gravity, and metallicity of the star,
while four observations were obtained at a resolution of $R = 7000$ to
check for RV variations with amplitude $\ga 5$\,\kms\ that would
indicate that the transiting companion is of stellar mass.  The two
du~Pont spectra each had a resolution of $R = 40000$ covering a
wavelength range of $3700$--$7000$\,\AA, and were used to measure the
effective temperature, gravity, metallicity, projected rotation
velocity and radial velocity of the star. Like the WiFeS spectra, the
RV precision of the du~Pont observations ($\sim 500$\,\ms) is not high
enough to detect velocity variations due to a planet, but is
sufficient to rule out stellar-mass companions.

Our analysis of the $R = 3000$ WiFeS spectrum indicated an effective
temperature of $\teffstar = 3600 \pm 300$\,K, while the du~Pont
spectra yielded $\teffstar = 3700 \pm 100$\,K. This effective
temperature corresponds to a spectral type of M1
\citep{rajpurohit:2013}. The spectrum shows clear TiO absorption
bands, and is consistent with an M1V spectral classification. This
spectrum also shows that \hatcur{} is a quiet M-dwarf, with no
evidence of emission in the H$\alpha$ or Ca~II H and K line
cores. Additional indications that \hatcur{} is a quiet star are the
lack of any obvious star-spot crossing events in the photometric
follow-up light curves (Figures~\ref{fig:lc} and~\ref{fig:lcgrond}),
and the lack of large-amplitude photometric variability in the
HATSouth light curve. The WiFeS spectrum also indicated a dwarf-like
surface gravity ($\loggstar = 3.9 \pm 0.3$; c.g.s.~units) and a
possibly sub-solar metallicity (\feh$ = -1.0 \pm 0.5$), while our
analysis of the du~Pont spectra yielded a somewhat lower surface
gravity ($\loggstar = 3.2 \pm 0.5$) and metallicity (\feh$ = -1.5 \pm
0.5$) and a moderately high rotation velocity ($\vsini = 7.5 \pm
2.0$\,\kms). The analysis, however, relies on synthetic templates
(MARCS models in the case of WiFeS and models from
\citealp{coelho:2005} in the case of du~Pont) which are known to be
unreliable for M type stars. This means that systematic errors in the
determined parameters are most likely greater than the estimated
uncertainties (especially when $\teffstar$, $\loggstar$ and \feh,
which are strongly correlated with each other, are all allowed to vary
in fitting the spectra). As an example of this, note that based on the
Dartmouth single stellar evolution models \citep{dotter:2008} the
minimum surface gravity realized for a $\teffstar = 3700$\,K dwarf
star within 13.8\,Gyr is $\loggstar = 4.73$ which is significantly
higher than the values determined from the spectroscopic modelling
(the only evolved single stars that reach $\teffstar = 3700$\,K have
$\loggstar \la 2$). We therefore do not consider the $\loggstar$\,
\feh, or $\vsini$ measurements from this analysis to be reliable. More
reliable estimate of \teffstar\ and \feh\ are presented in the next
section.

\subsubsection{Confirmation Spectroscopy}
\label{sec:confspec}

\begin{figure} [ht]
\plotone{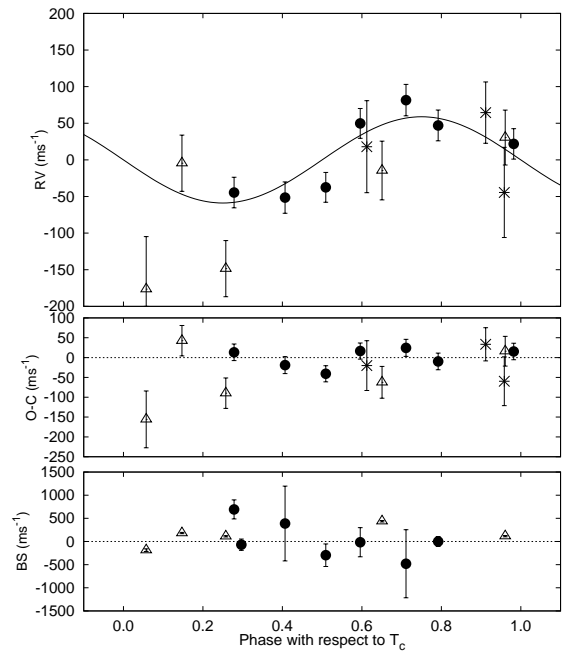}
\caption{
    {\em Top panel:} High-precision RV measurements for
    \hbox{\hatcur{}} from Magellan/PFS (dark filled circles),
    MPG~2.2\,m/FEROS (open triangles), and ESO~3.6\,m/HARPS (stars),
    together with our best-fit circular orbit model.  Zero phase
    corresponds to the time of mid-transit.  The center-of-mass
    velocity has been subtracted.  The orbital fit is primarily
    constrained by the PFS observations. {\em Second panel:} Velocity
    $O\!-\!C$ residuals from the best-fit model.  The error bars for
    each instrument include the jitter which is varied in the fit.
    {\em Third panel:} Bisector spans (BS), with the mean value
    subtracted.  Note the different vertical scales of the panels.\\
\label{fig:rvbis}}
\end{figure}

In order to confirm \hatcur{} as a transiting planet system through a
detection of the RV orbital variation we obtained high-resolution
spectra with three facilities capable of achieving $\sim 10$\,\ms\ or
better RV precision. These are the FEROS spectrograph
\citep{kaufer:1998} on the MPG~2.2\,m telescope at La Silla
Observatory (LSO), the Planet Finder Spectrograph
\citep[PFS;][]{crane:2010} on the Magellan Clay~6.5\,m telescope at
LCO, and the High Accuracy Radial Velocity Planet Searcher
\citep[HARPS;][]{mayor:2003} spectrograph on the ESO~3.6\,m telescope
at LSO. Figure~\ref{fig:rvbis} shows the phased RV measurements and
bisector spans from these observations, together with our best-fit
orbit, while Table~\ref{tab:rvs} lists the individual measurements.

A total of 8 $R=48000$ spectra were obtained with FEROS between 2013
Mar 24 and 2013 May 13. Details on the FEROS spectra as used by
HATSouth have been provided by \citet{mohlerfischer:2013:hats2}. For
the observations reported we did not follow the reduction procedure
described previously, and instead reduced the data using an adapted
version of the pipeline described by \citet{jordan:2014:hats4}. This
pipeline, which utilizes cross-correlation against binary
templates, was originally developed for the CORALIE spectrograph on
the Euler~1.2\,m telescope at LSO. We found that applying this
pipeline to FEROS data yields a precision of $\sim$8\,\ms\ for RV
standard stars, which is significantly better than the $\ga
20$\,\ms\ precision previously achieved for this instrument. This will
be described in additional detail in a separate paper (Brahm et al.~in
preparation).

The PFS observations consisted of an I$_{2}$-free template spectrum
obtained on the night of UT~2013 Nov 8, and 7 observations taken
through an I$_{2}$ cell obtained between 2013 Nov 7 and
16. Observations were carried out with a $0\farcs5\times2\farcs5$
slit, using $2\times2$ binning and slow read-out mode. The spectra
were reduced to RVs in the bary-centric frame of the solar system
following the method of \citet{butler:1996}. We computed bisector
spans in a manner similar to that used by \citet{torres:2007:hat3}
to calculate bisector spans for Keck/HIRES data. The presence of the
I$_{2}$ cell restricts the spectral range over which the bisector
spans may be computed. Whereas in \citet{torres:2007:hat3}, and
previous studies using Keck/HIRES data, the bisector span analysis was
done on the bluest spectral orders which are free of I$_{2}$
absorption, the faintness and extreme red color of \hatcur{} makes the
signal blueward of $5000$\,\AA\ too low to be used for a bisector
analysis. We therefore use five orders covering the spectral range
$6200$\,\AA\ to $6540$\,\AA. The limited spectral range reduces the
precision of the bisector spans. In fact we find that the bisector
spans calculated from the MPG~2.2\,m/FEROS data have lower scatter
than those from Magellan/PFS. Nonetheless there is no significant
trend in the bisector spans, nor is there any evidence for more than
one stellar component in the Magellan/PFS cross-correlation functions.

We used our I$_{2}$-free template spectrum from PFS to measure $T_{\rm
  eff}$ and [Fe/H] for \hatcur{} following the method of
\citet{neves:2014}. We made use of the Python routines referenced in
that paper to perform this analysis, after modifying it for the
spectral range of PFS (we verified that the code reproduces the
temperature and metallicity of the two HARPS spectra supplied with the
routines, when applied to the restricted spectral range). We find
$\teffstar = \hatcurSMEiteffcircdartmouth{}$\,K and [Fe/H]$=\hatcurSMEizfehcircdartmouth{}$, where
the errors are determined from Table~6 of \citet{neves:2014} scaling
by $\sqrt{N_{\rm all}/N_{\rm used}}$ where $N_{\rm all}$ is the number
of spectral lines used by \citet{neves:2014}, and $N_{\rm used}$ is
the number of lines within the PFS spectral range.

The HARPS observations consisted of three exposures taken on the
nights of UT~2013 Dec 7--9. We used an exposure time of 1200\,s on the
first night and 1800\,s on the following two nights. Observations were
carried out in the ``object+sky'' mode (due to the faintness of the
target, contamination from scattered moonlight is substantial), and
reduced to RVs using the facility Data Reduction Software (DRS)
together with a K star spectral mask (at present this is the coolest
facility mask available). The spectra have a resolution of $R =
115000$ covering a range of 378\,nm--691\,nm.

In practice, due to the faintness of \hatcur{} and the low mass of the
planet \hatcurb{}, the orbital variation is detected with significance
only from the PFS data. We include the data from HARPS and FEROS in
our RV model (Section~\ref{sec:globmod}) for completeness, and to
ensure that the fit accounts for all data, including non-detections.


\ifthenelse{\boolean{emulateapj}}{
    \begin{deluxetable*}{llrrrrr}
}{
    \begin{deluxetable}{llrrr}
}
\tablewidth{0pc}
\tabletypesize{\scriptsize}
\tablecaption{
    Summary of spectroscopic observations\label{tab:specobssummary}
}
\tablehead{
    \multicolumn{1}{c}{Telescope/Instrument} &
    \multicolumn{1}{c}{Date Range}          &
    \multicolumn{1}{c}{Number of Observations} &
    \multicolumn{1}{c}{Resolution}          &
    \multicolumn{1}{c}{Observing Mode}          \\
}
\startdata
ANU~2.3\,m/WiFeS & 2012 May 13 & 1 & 3000 & Recon Spectral Type \\
ANU~2.3\,m/WiFeS & 2012 Aug 6--Oct 24 & 4 & 7000 & Recon RVs \\
du~Pont~2.5\,m/Echelle & 2012 Oct 25--26 & 2 & 30000 & Recon RVs/Spectral Type \\
Magellan~Clay~6.5\,m/PFS & 2013 Nov 7--16 & 7 & 100000 & High Precision RVs \\
MPG~2.2\,m/FEROS & 2013 Mar 24--May 11 & 8 & 48000 & High Precision RVs \\
ESO~3.6\,m/HARPS & 2013 Dec 7--10 & 31 & 115000 & High Precision RVs \\
\enddata 
\ifthenelse{\boolean{emulateapj}}{
    \end{deluxetable*}
}{
    \end{deluxetable}
}

\ifthenelse{\boolean{emulateapj}}{
    \begin{deluxetable*}{lrrrrrr}
}{
    \begin{deluxetable}{lrrrrrr}
}
\tablewidth{0pc}
\tablecaption{
    Relative radial velocities and bisector span measurements of
    \hatcur{}.
    \label{tab:rvs}
}
\tablehead{
    \colhead{BJD} & 
    \colhead{RV\tablenotemark{a}} & 
    \colhead{\ensuremath{\sigma_{\rm RV}}\tablenotemark{b}} & 
    \colhead{BS} & 
    \colhead{\ensuremath{\sigma_{\rm BS}}} & 
        \colhead{Phase} &
        \colhead{Instrument}\\
    \colhead{\hbox{(2\,456\,000$+$)}} & 
    \colhead{(\ms)} & 
    \colhead{(\ms)} &
    \colhead{(\ms)} &
    \colhead{} &
        \colhead{} &
        \colhead{}
}
\startdata
$ 377.58801 $ & $    30.48 $ & $    32.00 $ & $  115.0 $ & $   14.0 $ & $   0.961 $ & FEROS \\
$ 378.57524 $ & $  -148.52 $ & $    33.00 $ & $  112.0 $ & $   14.0 $ & $   0.258 $ & FEROS \\
$ 401.48495 $ & $    -4.52 $ & $    33.00 $ & $  182.0 $ & $   14.0 $ & $   0.147 $ & FEROS \\
$ 406.48479 $ & $   -14.52 $ & $    35.00 $ & $  437.0 $ & $   16.0 $ & $   0.651 $ & FEROS \\
$ 424.46313 $ & $  -176.52 $ & $    69.00 $ & $ -187.0 $ & $   29.0 $ & $   0.057 $ & FEROS \\
$ 603.77801 $\tablenotemark{c} & $    21.86 $ & $     7.09 $ & \nodata      & \nodata      & $   0.982 $ & PFS \\
$ 604.76307 $ & $   -44.62 $ & $     7.36 $ & $  692.4 $ & $  205.7 $ & $   0.278 $ & PFS \\
$ 604.82373 $\tablenotemark{d} & \nodata      & \nodata      & $  -69.7 $ & $  122.4 $ & $   0.297 $ & PFS \\
$ 605.81892 $ & $    49.82 $ & $     5.79 $ & $  -15.7 $ & $  315.3 $ & $   0.596 $ & PFS \\
$ 608.85541 $ & $   -37.49 $ & $     5.80 $ & $ -296.4 $ & $  243.2 $ & $   0.509 $ & PFS \\
$ 609.79629 $ & $    47.03 $ & $     7.77 $ & $    0.0 $ & $  107.1 $ & $   0.792 $ & PFS \\
$ 611.84038 $ & $   -51.46 $ & $     8.74 $ & $  388.1 $ & $  808.0 $ & $   0.407 $ & PFS \\
$ 612.85205 $ & $    81.60 $ & $     8.84 $ & $ -481.3 $ & $  735.3 $ & $   0.711 $ & PFS \\
$ 633.62669 $ & $   -44.47 $ & $    58.38 $ & \nodata      & \nodata      & $   0.958 $ & HARPS \\
$ 635.80096 $ & $    18.01 $ & $    59.69 $ & \nodata      & \nodata      & $   0.612 $ & HARPS \\
$ 636.79588 $ & $    64.54 $ & $    37.05 $ & \nodata      & \nodata      & $   0.912 $ & HARPS \\

    [-1.5ex]
\enddata
\tablenotetext{a}{
        The zero-point of these velocities is arbitrary. An overall
        offset $\gamma_{\rm rel}$ fitted separately to the PFS, HARPS
        and FEROS velocities in \refsecl{analysis} has been
        subtracted.
}
\tablenotetext{b}{
        Internal errors excluding the component of
        astrophysical/instrumental jitter considered in
        \refsecl{analysis}.
}
\tablenotetext{c}{
        The CCF peak height was too low in the orders where we
        computed the BS to be able to extract a BS measurement for
        this observation.
}
\tablenotetext{d}{
        This PFS observation was taken without the iodine cell to
        be used as a template. The RV is not measured for this
        observations, but BS value is measured.
}
\ifthenelse{\boolean{emulateapj}}{
    \end{deluxetable*}
}{
    \end{deluxetable}
}

\section{Analysis}
\label{sec:analysis}

\subsection{Excluding blend scenarios}

In order to rule out the possibility that \hatcur{} is not a planetary
system, but is instead a blend between an eclipsing binary star and
another source, we carried out a blend analysis similar to that done
in \citet{hartman:2012:hat39hat41}, with a difference being that in
this case we use the Dartmouth \citep{dotter:2008} stellar evolution
models to calculate the properties of simulated blended systems. We
find that although there exist blend models involving a eclipsing
binary blended with a brighter foreground M dwarf which match the
light curves and absolute photometry, in all such cases both the
foreground source and the primary in the background binary have an
apparent magnitude difference $|\Delta V| < 1$\,mag, so that
cross-correlation functions (CCFs) computed from the PFS, FEROS and
HARPS spectra would show obvious secondary peaks and RV variations
greater than $1$\,\kms. The simulated CCFs are grossly inconsistent
with the observed CCFs, so we conclude that this is not a blended
eclipsing binary system, and is instead a transiting planet system.

As is often the case we cannot exclude the possibility that \hatcur{}
is an unresolved binary system with one component having a transiting
planet. High resolution imaging would provide constraints on any such
wide binary companions. For the analysis presented here we assume that
this is an isolated star. If future observations reveal that this is a
binary system, corrections to the planet mass and radius would
increase their values from those presented in this paper.

\subsection{Global Fit of Light Curves and RV Measurements}
\label{sec:globmod}

In order to determine the physical parameters of the \hatcur{} system
we carried out an analysis similar to that described in
\cite{bakos:2010:hat11,penev:2013:hats1}. All light curves (HATSouth
data and follow-up data) and RV measurements are simultaneously fitted
using a \citet{mandel:2002} transit light curve model and a Keplerian
RV orbit. 

The light curve model is extended using a model for instrumental
variations such that the total model can be expressed as:
\begin{eqnarray}
m_{k,i} & = & m_{k,0} + \Delta m(t_{i};T_{0},T_{N_{t}},\zrstar,\rprstar,b,a_{{\rm LD},k},b_{{\rm LD},k}) \nonumber \\
 & & + \sum_{j=1}^{N_{{\rm EPD},k}}c_{{\rm EPD},k,j}x_{k,j,i} + \sum_{j=1}^{N_{\rm TFA}}c_{{\rm TFA},j}y_{k,j,i} 
\end{eqnarray}
where $m_{k,i}$ is the measured magnitude to be modeled for
observation $i$ of light curve $k$; $m_{k,0}$ is the zero-point
magnitude for light curve $k$ (which is a free parameter in the
model); $\Delta m(t_{i};T_{0},T_{N_{t}},\zrstar,\rprstar,b,a_{{\rm
    LD},k},b_{{\rm LD},k})$ is the physical \citet{mandel:2002} model
evaluated at time $t_{i}$ and parameterized by initial and final
transit epochs $T_{0}$ and $T_{N_{t}}$ (the period is then given by
$(T_{N_{t}}-T_{0})/N_{t}$), reciprocal of the half duration of the
transit $\zrstar$, ratio of the planetary and stellar radii
$\rprstar$, normalized impact parameter $b$, and quadratic limb
darkening coefficients $a_{{\rm LD},k}$ and $b_{{\rm LD},k}$
appropriate for the filter of light curve $k$ (except for $a_{{\rm
    LD},k}$ and $b_{{\rm LD},k}$, which are fixed using the
tabulations of \citet{claret:2004}, these parameters are varied in the
fit); there are $N_{{\rm EPD},k}$ EPD parameters applied to light
curve $k$ with $c_{{\rm EPD},k,j}$ being the free coefficient fitted
for EPD parameter series $j$ applied to light curve $k$, and
$x_{k,j,i}$ being the value of EPD parameter series $j$ at observation
$i$ for light curve $k$; and there are $N_{\rm TFA}$ TFA templates
used to fit the light curve, with $c_{{\rm TFA},j}$ being the free
coefficient fitted for template $j$ and $y_{k,j,i}$ being the value of
template $j$ at observation $i$ for light curve $k$. For the HATSouth
light curve we do not include the EPD and TFA terms in the fit, and
instead model the light curve that was pre-processed through these
filtering routines {\em without} accounting for the transits. In this
case we also include an instrumental blending factor (varied in the
fit) which scales the depth of the \citet{mandel:2002} model applied
to the HATSouth light curve by a factor between 0 and 1 (assuming a
uniform prior between these limits) to account for both blending from
nearby stars as well as the artificial dilution of signals due to the
filtering.

For the RV model we allow an independent RV zero-point, and an
independent RV jitter for each of the three instruments used. The
jitter is a term added in quadrature to the formal RV uncertainties
for each instrument and is varied in the fit following
\cite{hartman:2012:hat39hat41}.

We use a Differential Evolution Markov-Chain Monte Carlo (DEMCMC)
procedure \citep{terbraak:2006,eastman:2013} to explore the fitness of
the model over parameter space and produce a chain of parameters drawn
from the posterior distribution. This chain is then used to estimate
the most likely value (taken as the median value over the chain)
together with the 68.3\% ($1\sigma$) confidence interval for each of
the physical parameters.

The fit is performed both allowing the eccentricity to vary, and
fixing it to zero. We find that without additional constraints, the
free-eccentricity model strongly prefers a non-zero eccentricity of
$e=\hatcurRVecceneccendartmouth{}$. This is entirely due to the PFS
velocities which closely follow such an eccentric orbit with a
near-zero jitter for the PFS RVs of
$\hatcurRVjitterAeccendartmouth{}$\,\ms. When the eccentricity is
fixed to zero, on the other hand, the PFS RVs are consistent with a
circular orbit, but in this case require a jitter of
$\hatcurRVjitterAcircdartmouth{}$\,\ms. Due to the faintness of
\hatcur{} in the optical band-pass, and consequent sky contamination,
such a high ``jitter'' is not unreasonable, and may simply reflect an
underestimation of the formal RV uncertainties. As we discuss below,
the stellar parameters inferred for the high eccentricity solution are
inconsistent with the spectroscopic parameters and broad-band
photometric colors of the star. When the photometric observations are
directly folded into our light curve and RV modelling, as we discuss
in Section~\ref{sec:empirical}, the preferred eccentricity is
consistent with zero ($e = \hatcurRVecceneccenempirical$).

\subsection{Determining the Physical Parameters of the Star and Planet}
\label{sec:stelparam}

To determine the mass and radius of the transiting planet from the
physical parameters measured above requires knowledge of the stellar
mass and radius. For a non-binary star such as \hatcur{} these
parameters are not easy to measure directly and instead must be
inferred by comparing other measurable parameters, such as the surface
temperature and bulk stellar density, with theoretical stellar
evolution models (requiring the metallicity, a color indicator, and a
luminosity indicator to identify a unique stellar model), or with
empirical relations calibrated using binary stars. We considered both
methods, discussed in turn below.

\subsubsection{Dartmouth Models}
\label{sec:dartmouth}

Because the star is a cool dwarf we make use of the Dartmouth stellar
evolution models \citep{dotter:2008} which appear to provide the best
match to M dwarf and late K dwarf stars
\citep[e.g.][]{feiden:2011,sandquist:2013}. We also use the effective
temperature and metallicity measured from the PFS I$_{2}$-free
template spectrum.

We use the results from our DEMCMC analysis of the light curve and
radial velocity data (Section~\ref{sec:globmod}) together with the
Dartmouth isochrones to determine the stellar parameters. For each
density measurement in the posterior parameter chain we associate
T$_{\rm eff}$ and \feh\ measurements drawn from Gaussian
distributions. We look up a matching stellar model from the Dartmouth
isochrones, interpolating between the tabulated models, and append the
set of stellar parameters associated with this model to the
corresponding link in the posterior parameter chain. Other planetary
parameters, such as the mass and radius, which depend on the stellar
parameters are then calculated for each link in the chain.

\begin{figure}[!ht]
\plotone{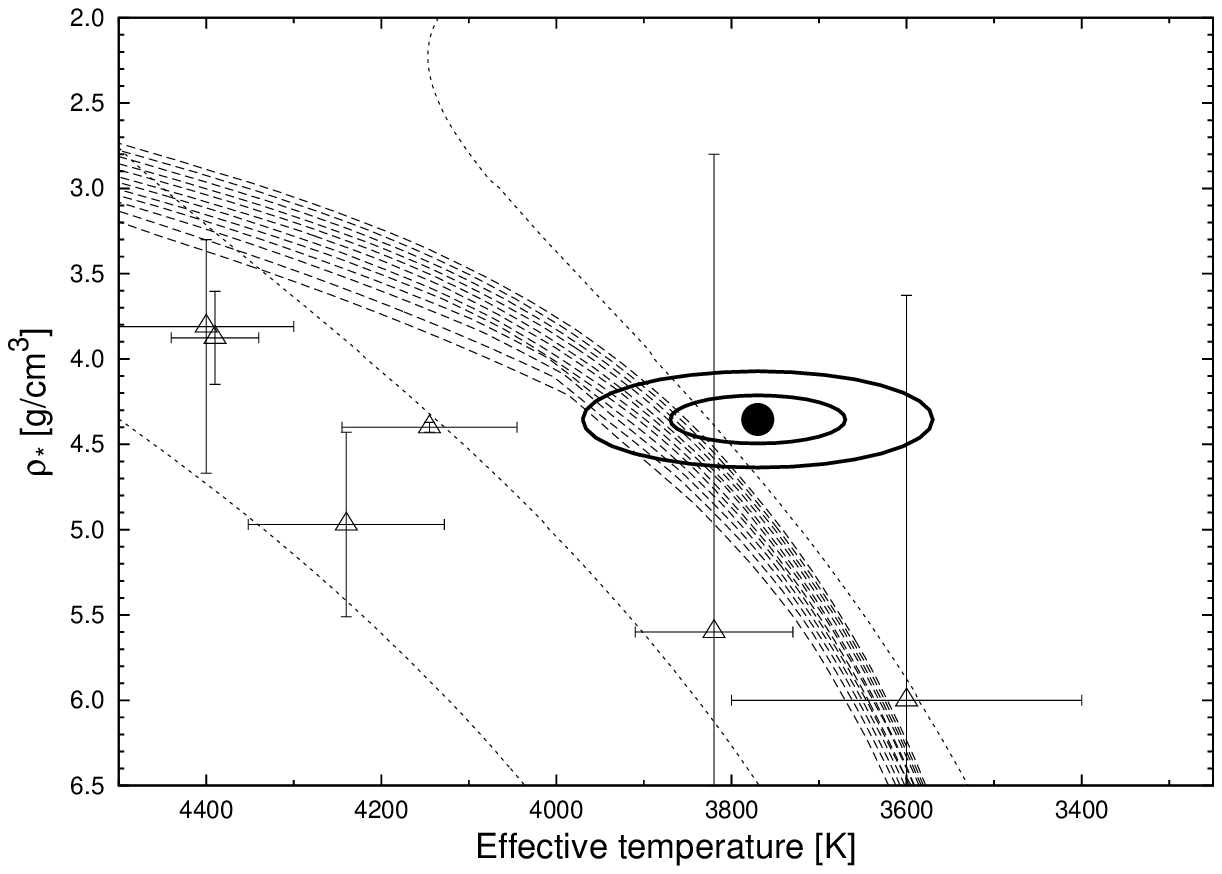}
\plotone{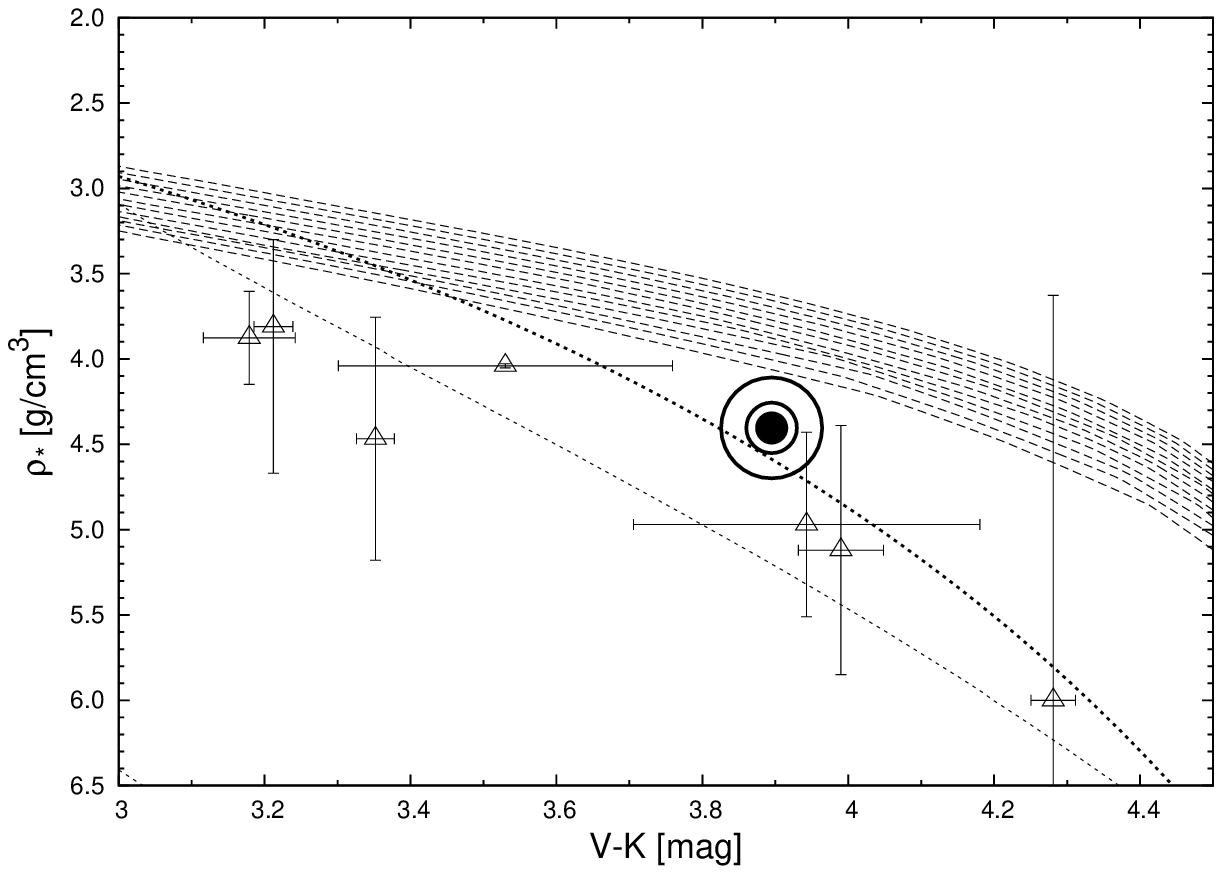}
\plotone{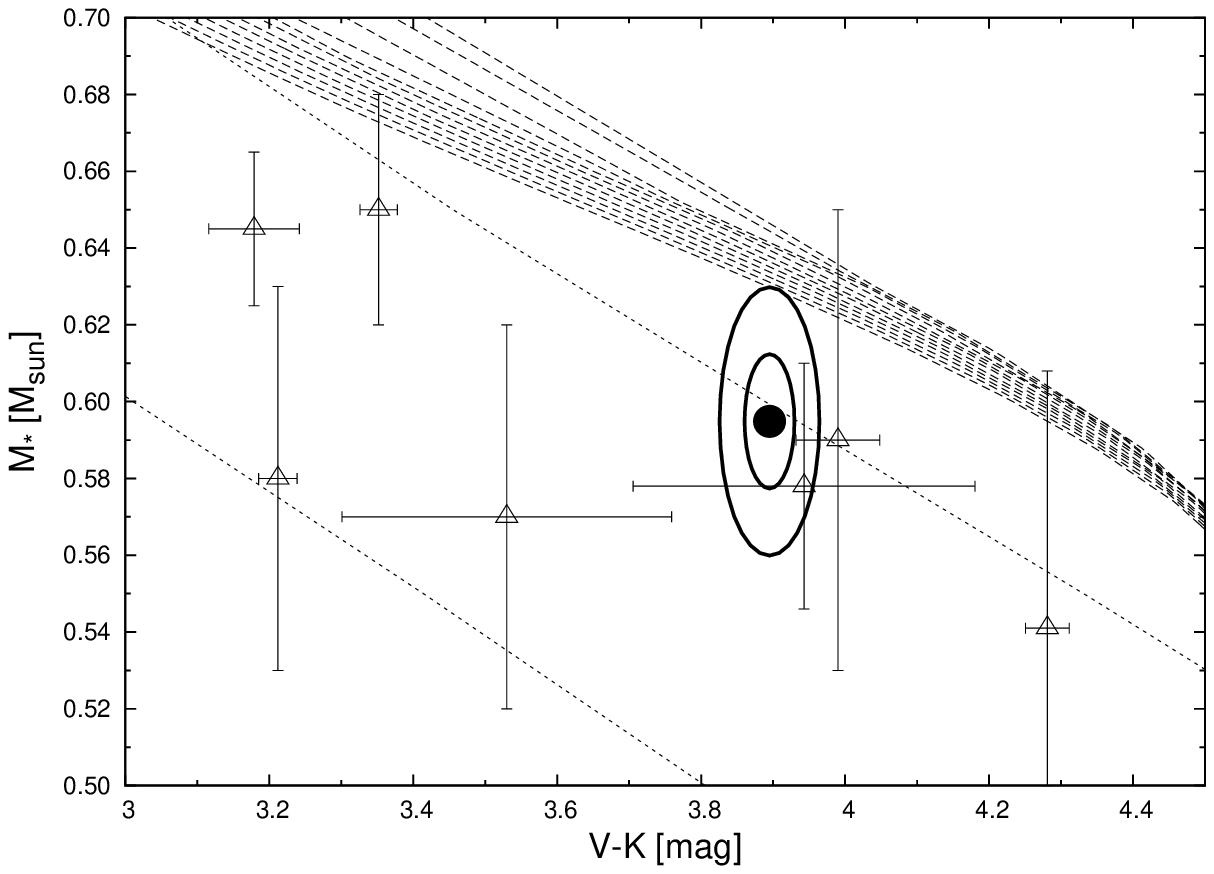}
\caption{
    Model isochrones (dashed lines) from \citet{dotter:2008} for the
    spectroscopically determined metallicity of \hatcur\ and ages of 1
    to 13\,Gyr in 1\,Gyr steps (showing $\rhostar$ vs $T_{\rm eff}$ at
    top, $\rhostar$ vs $V-K$ in the middle, and $\mstar$ vs $V-K$ at
    the bottom). The measured values of $T_{\rm eff}$, $V-K$, and
    $\rhostar$ for \hatcur{}, and the value of $\mstar$ inferred from
    comparison to the stellar models, are shown using the large filled
    circles together with their 1$\sigma$ and 2$\sigma$ confidence
    ellipsoids. The open triangles show other transiting planet host
    stars with measured $T_{\rm eff}$, $V-K$, $\rhostar$ and $\mstar$
    values similar to \hatcur. Near the K/M spectral type boundary
    ($V-K \sim 4$) the models predict somewhat lower densities for a
    given $V-K$ or $T_{\rm eff}$ than are seen among transit hosts
    (Table~\ref{tab:kmtepstars}). For comparison we also show the
    relations derived from our empirical model (dotted lines;
    Section~\ref{sec:empirical}). In the top panel we show the three
    dotted lines are the median relation, and the $1\sigma$ lower and
    upper bounds. In the bottom two panels the lower dotted line is
    the median relation, while the upper dotted line is the $1\sigma$
    upper bound. The $1\sigma$ lower bound from the empirical model
    lies just outside the range of these plots (below the bottom left
    corner in each case). The empirical models appear to be more
    consistent with the transiting planet hosts than the theoretical
    models, but also cover a broader range of parameter space than
    seen among the transit hosts.\\
\label{fig:iso}}
\end{figure}

Figure~\ref{fig:iso} compares the measured $T_{\rm eff}$ and
$\rhostar$ values for \hatcur\ to the interpolated [Fe/H]$=0.20$ Dartmouth
isochrones. We also compare the $V-K$ and $\rhostar$ values to these
same isochrones. For reference we show seven transiting-planet-hosting
stars with $\teffstar$, $V-K$, $\rhostar$ and $\mstar$ similar to that
of \hatcur{}. These are listed as well in
Table~\ref{tab:kmtepstars}. At fixed $T_{\rm eff}$ or $V-K$ the models
predict $\rhostar$ values that are slightly lower than those seen
amongst the transit hosts. This suggests that using $V-K$ as an input
in constraining the mass of the star through the Dartmouth models may
lead to a slightly overestimated stellar mass (lower densities yield
higher masses). For comparison we also show the relations derived from
the empirical relations discussed below. The empirical relations
appear to provide a better match to the stars shown in
Figure~\ref{fig:iso}.


\begin{figure*}[!ht]
\plottwo{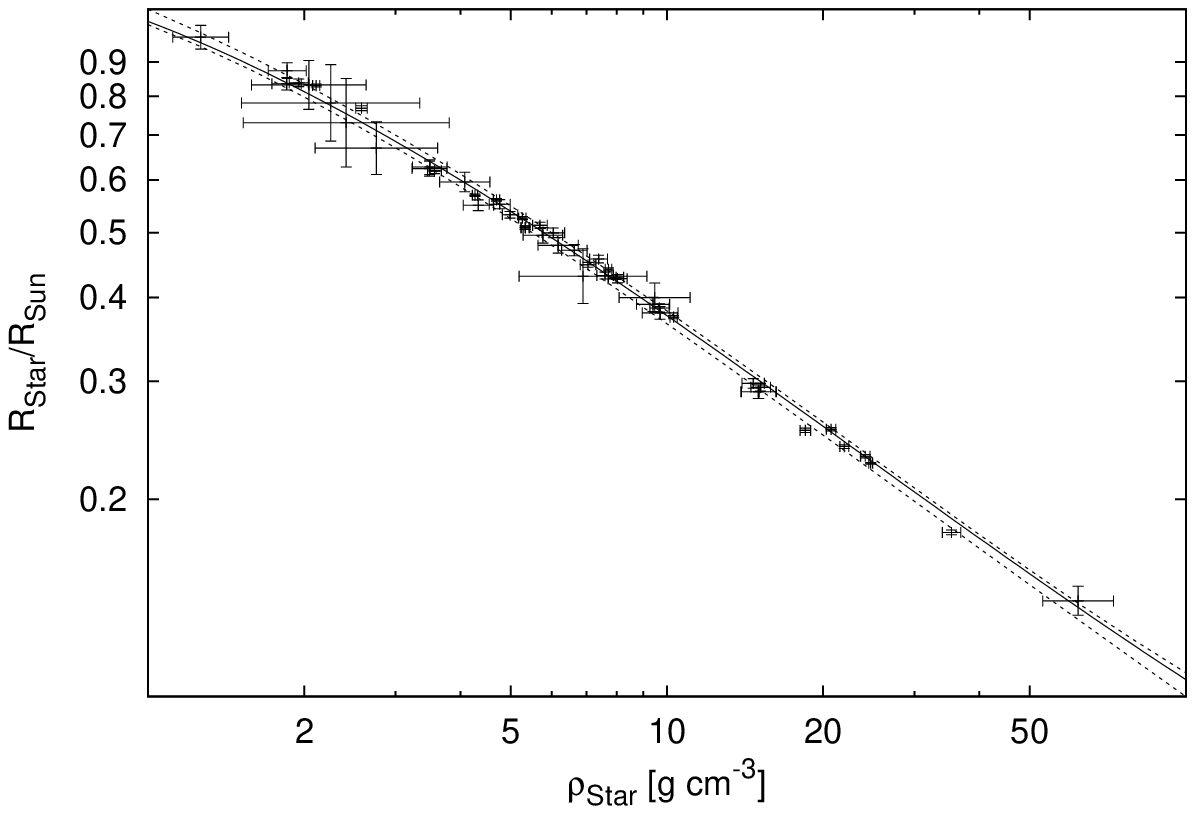}{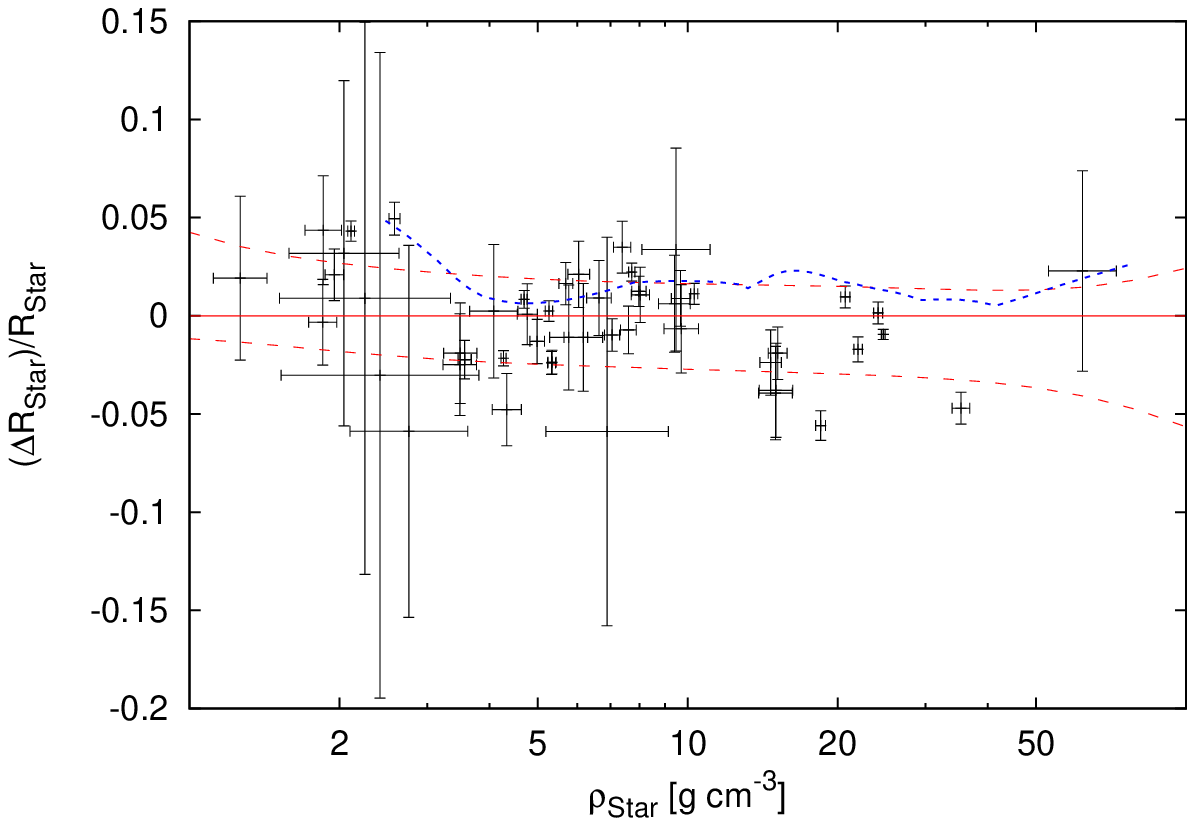}
\plottwo{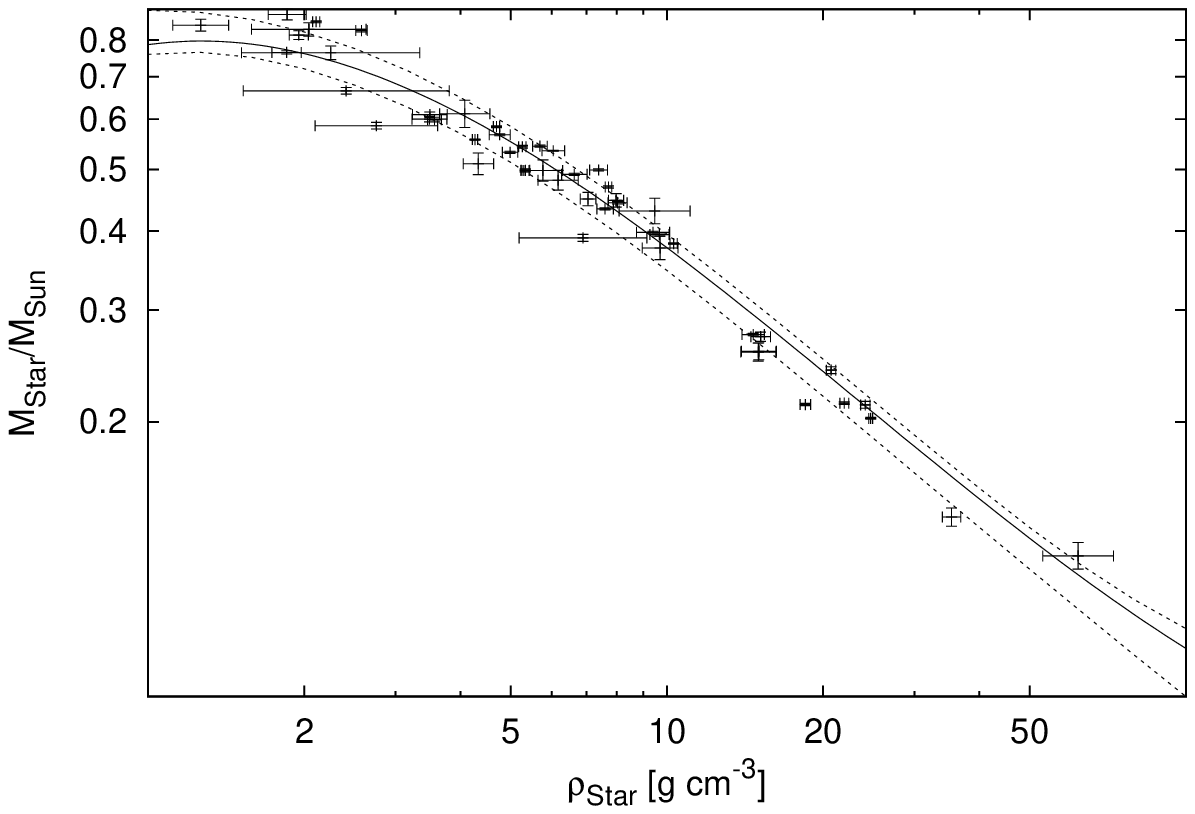}{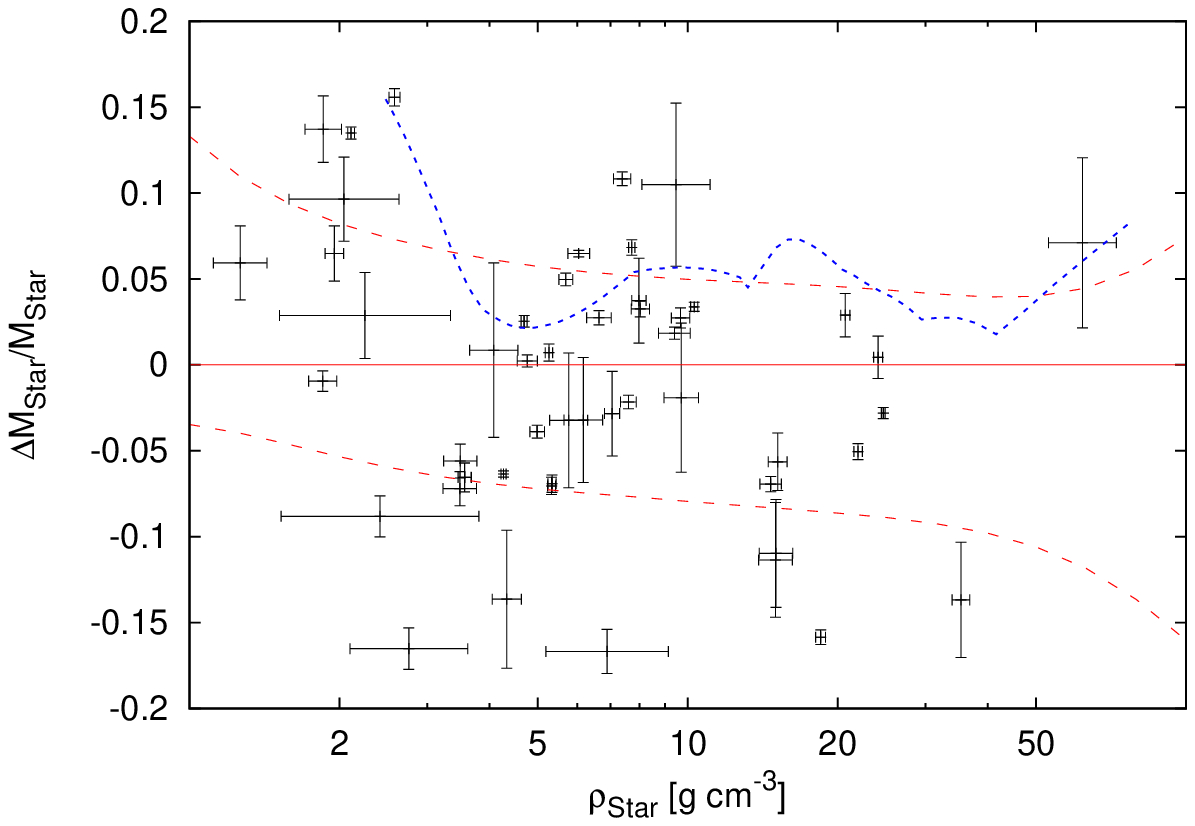}
\plottwo{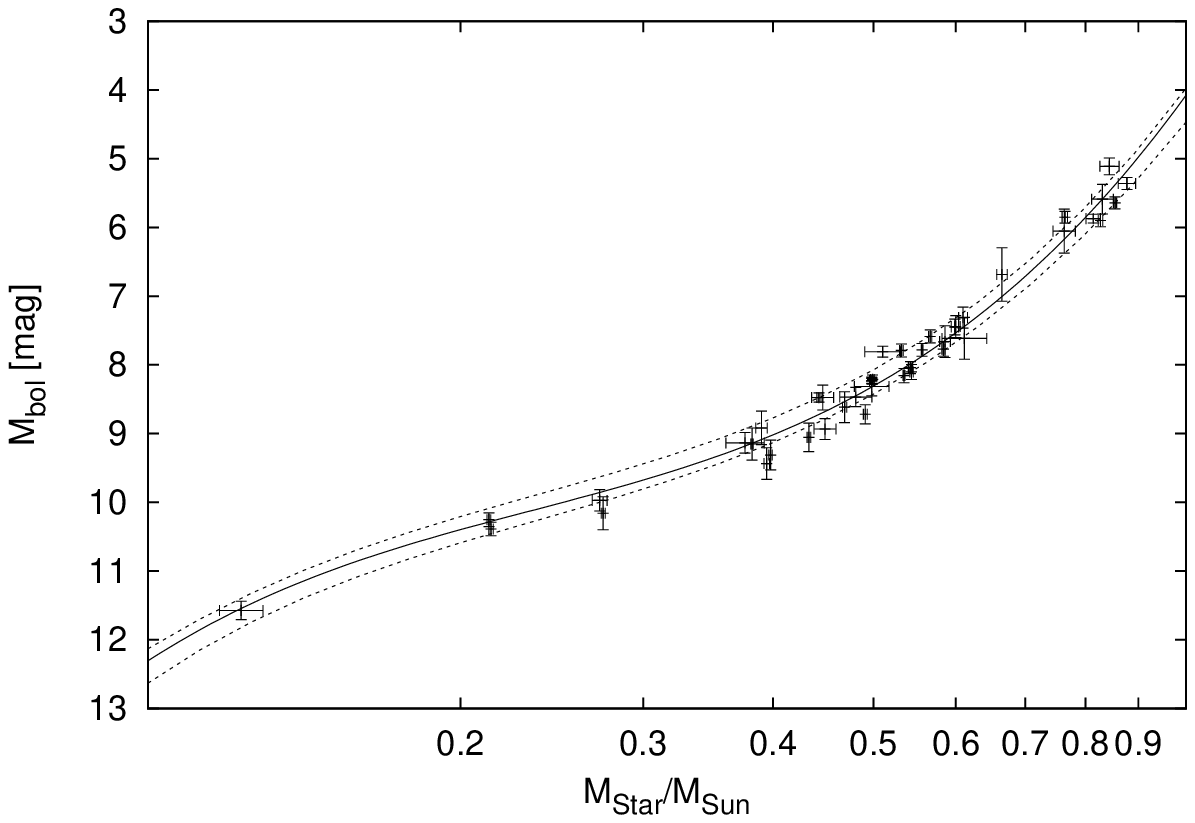}{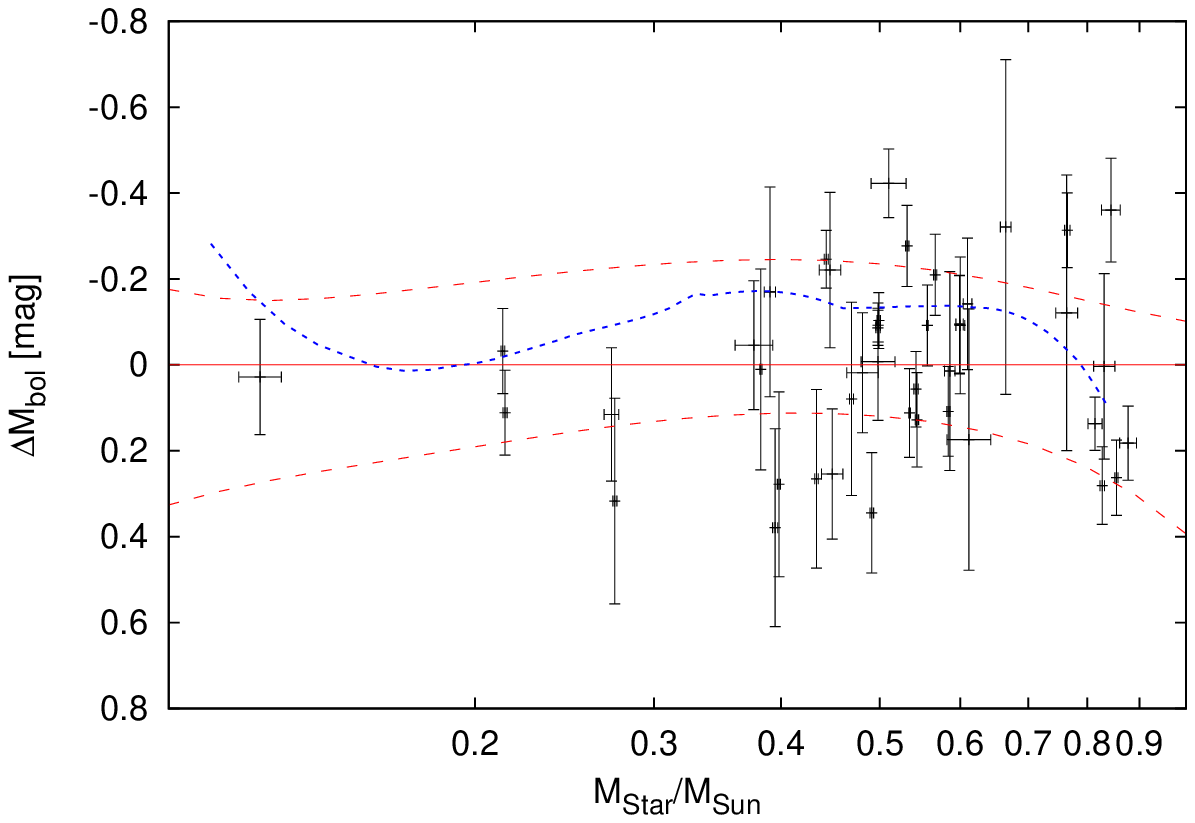}
\caption{
    Top: Empirical relation between stellar density and radius
    (equation~\ref{eqn:empiricalmodelfunctionslR}, with parameters
    given in Table~\ref{tab:EmpiricalModelParameters}). The solid line
    shows the best-fit relation with $dR = 0$, the dashed lines show
    the expected $1\sigma$ spread for this relation allowing $dR$ to
    be drawn from a Gaussian distribution with standard deviation
    $S_{R}$. The points show the eclipsing binary components from
    Table~\ref{tab:MEBs} used to fit this relation. Middle: Same as
    top, here we show the empirical relation between stellar density
    and mass. Bottom: Same as top, here we show the empirical relation
    between stellar mass and bolometric magnitude. On the right-hand
    panels we show the fractional residuals from the best-fit models,
    the short-dashed (blue) lines show the expected relations from the
    Dartmouth stellar models for an age of 4.5\,Gyr and solar
    metallicity, while the longer dashed (red) lines show the expected
    $1\sigma$ spread from the empirical model. The solar-metallicity
    4.5\,Gyr Dartmouth models predict systematically larger radii at
    fixed density, and brighter bolometric magnitudes at fixed mass,
    than observed in the eclipsing binary sample. The empirical
    models provide a good fit to these data, by design.\\
\label{fig:EmpiricalModelFit}}
\end{figure*}

As noted in Section~\ref{sec:globmod} when we allow the eccentricity
to vary in the fit the PFS data pull the model toward a high
eccentricity solution ($e = \hatcurRVecceneccendartmouth$) which also
yields a high stellar density of
$\hatcurISOrhoeccendartmouth$\,\gcmc. The combination of the high
stellar density, hot effective temperature ($\teffstar =
\hatcurSMEiteffcircdartmouth$\,K) and high metallicity ([Fe/H]$=
\hatcurSMEizfehcircdartmouth$) cannot be matched by the Dartmouth
isochrones. If we instead use $V-K$ as the temperature indicator and
draw metallicities from the \citet{haywood:2001} solar neighborhood
metallicity distribution we find that only very low metallicity models
($[Fe/H] = -0.63 \pm 0.07$) match the observations. Such a metallicity
is at odds with the spectroscopic measurement. Moreover, taking the
parameters from the eccentric model, and assuming $Q_{P} = 10^6$ the
circularization timescale \citep[e.g.][]{adams:2006} is only
$\hatcurPPtcirceccendartmouth$\,Myr. No planets with $t_{\rm circ} <
1$\,Gyr have been found with eccentricities greater than 0.1
\citep[e.g.~see Fig.~12 of][]{bakos:2012:hat34to37}, making it
unlikely that \hatcurb{} has such a high eccentricity.

\subsubsection{Empirical Relations}
\label{sec:empirical}

As an alternative method to determine the stellar parameters, and to
better understand the degree of systematic errors in these parameters,
we also develop a set of empirical relations between stellar density,
which is directly measured for a transiting planet system, and other
stellar parameters.  Such relations have been developed and employed
in transiting exoplanet studies previously
\citep{torres:2010,enoch:2010,southworth:2011}. The
\citet{torres:2010} and \citet{enoch:2010} relations only considered
stars with $M > 0.6$\,\msun, making them inapplicable in this
case. The \citet{southworth:2011} relations consider stars over the
range $0.2\,\msun < M < 3.0\,\msun$. They present two relations, one
for mass as a function of temperature, density and metallicity, the
other for radius as a function of temperature, density and
metallicity. Fitting these as two independent functions ignores the
fact that density, mass, and radius must satisfy the relation $M =
\frac{4}{3}\pi R^{3} \rho$. Moreover, one should not expect the
scaling of mass and radius with metallicity or temperature to be
independent of stellar mass over such a broad range in mass. And,
since metallicity is available for only very few M dwarf eclipsing
binaries, the fit performed by \citet{southworth:2011} effectively
imposes the metallicity scaling for A through G stars on the M
dwarfs. We therefore consider it worthwhile to revisit these relations
for K and M dwarf stars.

\citet{johnson:2011} and \citet{johnson:2012} have also developed
empirical relations for characterizing M dwarf planet hosts, applying
them to the characterization of LHS~6343~AB and Kepler-45,
respectively. Our approach is similar to theirs in that we make use of
relations between mass and absolute magnitudes based on data from
\citet{delfosse:2000}, and we also consider the empirical mass--radius
relation based on eclipsing binaries, however we differ in the sample
of eclipsing binaries that we consider, and we adopt a different
parameterization of the problem. Moreover, while \citet{johnson:2012}
use several empirical relations which were independently fit using
different data sets, our approach is to self-consistently determine
all of the relations through a joint analysis of the available
data. We discuss this comparison in more detail below.

\begin{figure*}[!ht]
\plottwo{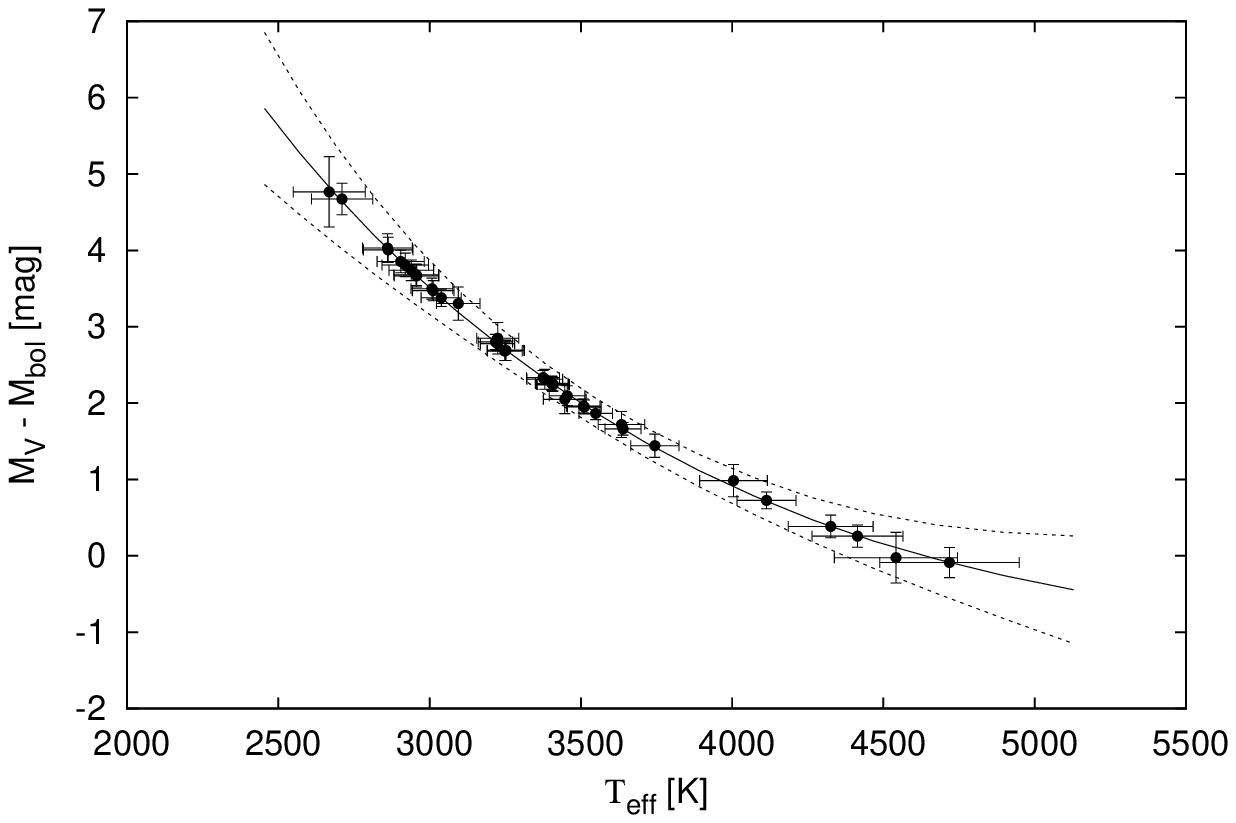}{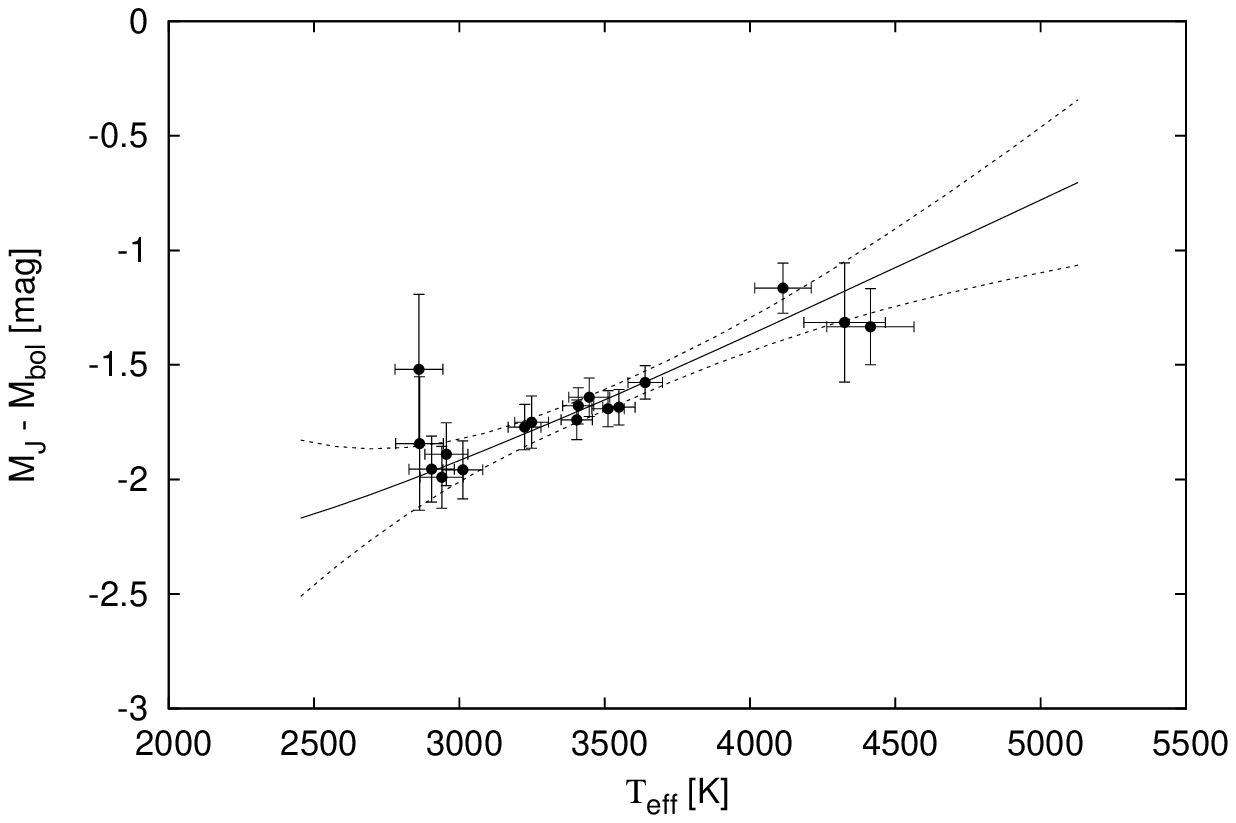}
\plottwo{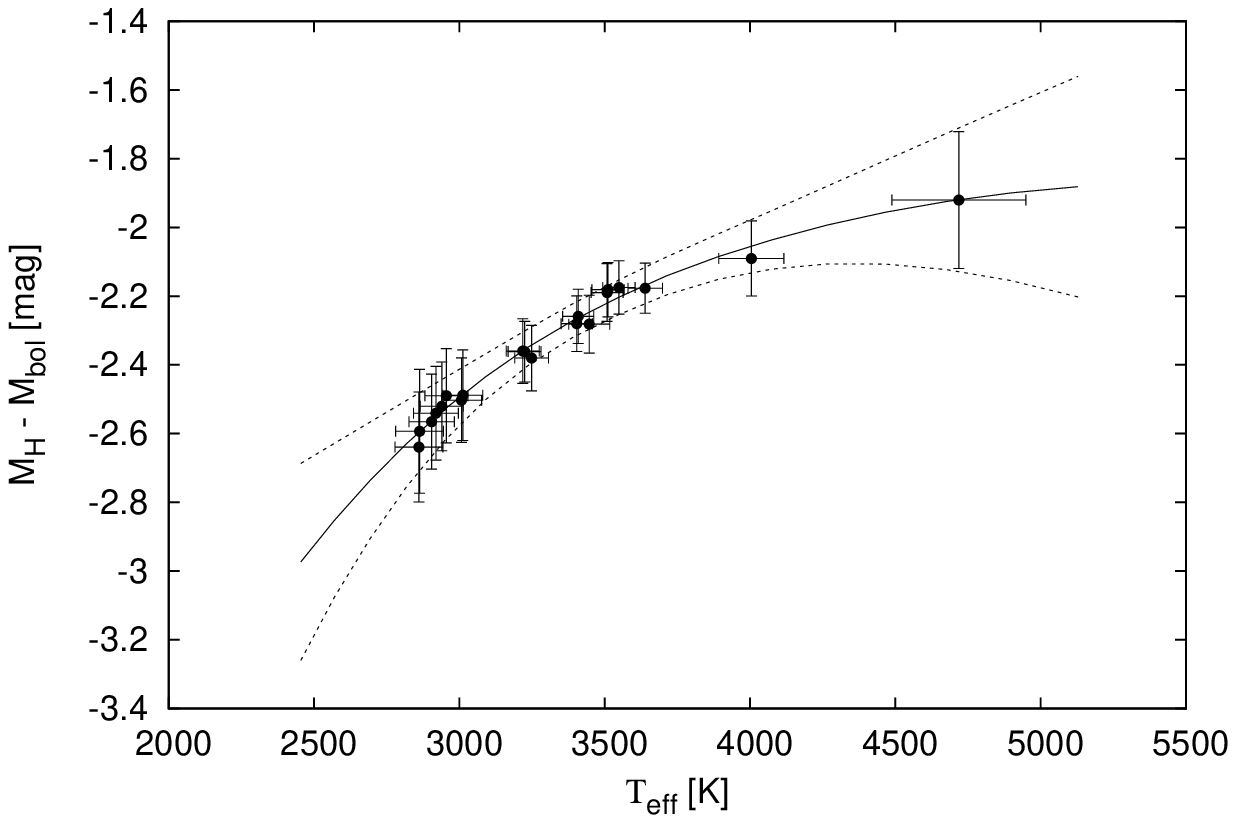}{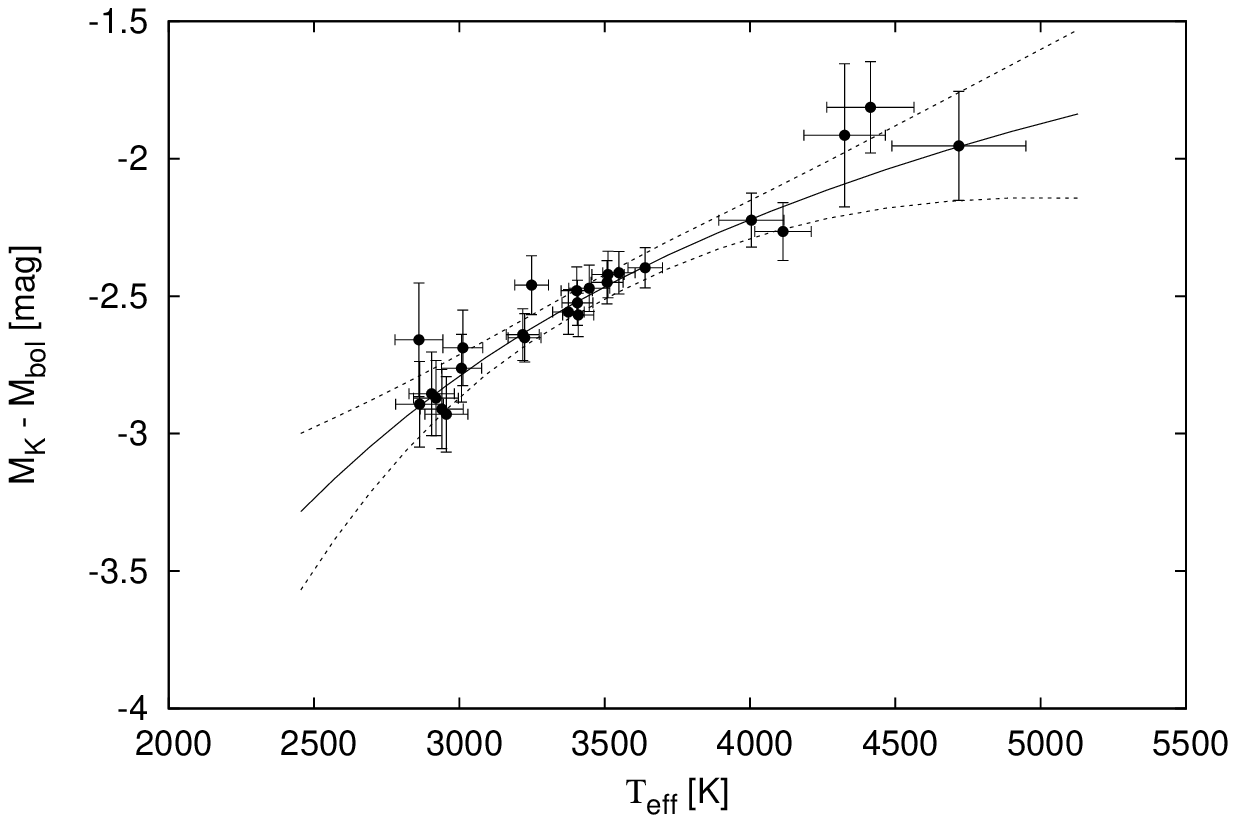}
\caption{
    Top: Empirical relation between effective temperature and the
    bolometric correction in $V$-band (upper left), $J$-band (upper
    right), $H$-band (lower left) and $K$-band (lower right). In each
    case the solid line is the median relation, while the dashed lines
    show the $1\sigma$ confidence region about the median
    relation. The points in this plot correspond to stars in
    Table~\ref{tab:photMbinaries} with T$_{\rm eff}$ and $M_{\rm bol}$
    for each star determined as part of the empirical model fitting
    procedure. Because T$_{\rm eff}$ and $M_{\rm bol}$ are not
    directly observed parameters for these stars, but are rather
    determined through the modelling, and largely constrained by the
    stellar mass, which is observed, the data appear to follow the
    model very closely. This should not be taken as validation of
    the model (see instead Figure~\ref{fig:EmpiricalModelFitMags}),
    rather the purpose of showing these plots is to demonstrate the form
    of the bolometric correction, its uncertainty, as well as the
    range of temperatures over which the model is constrained.\\
\label{fig:EmpiricalModelFitBC}}
\end{figure*}

\begin{figure*}[!ht]
\plottwo{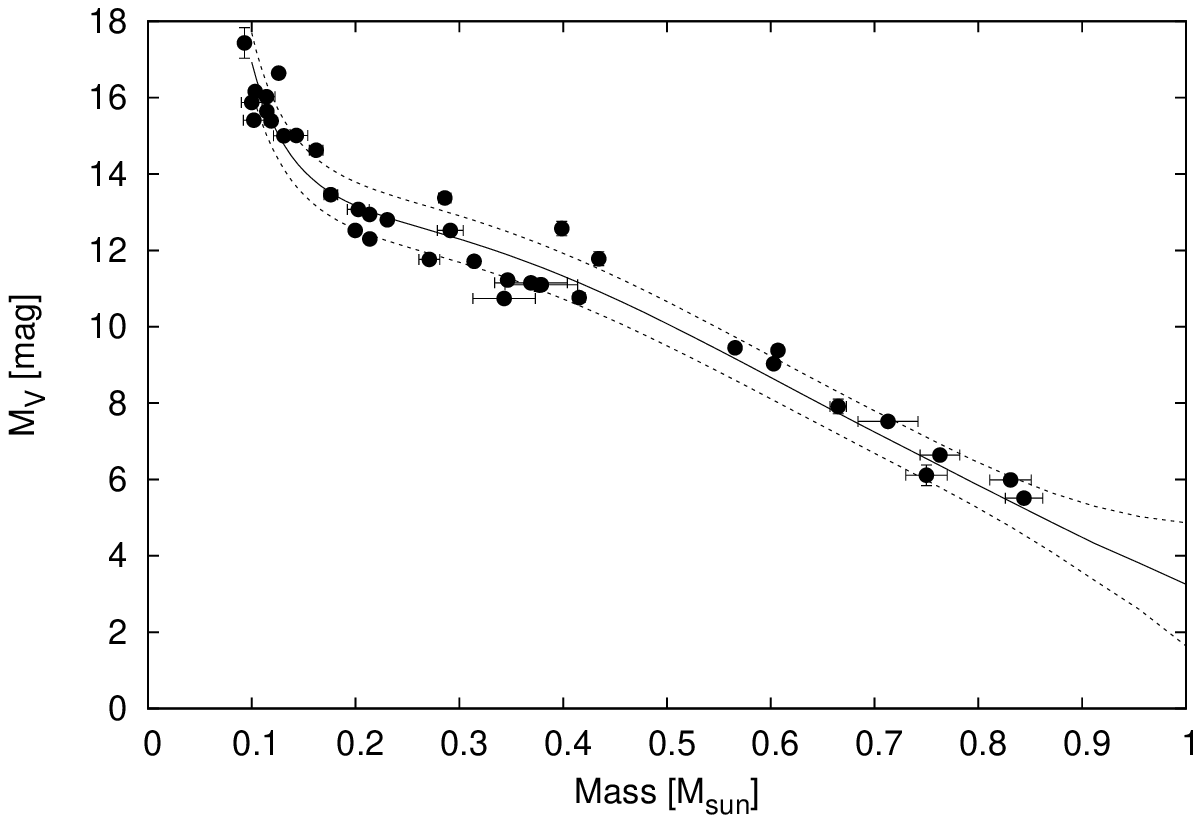}{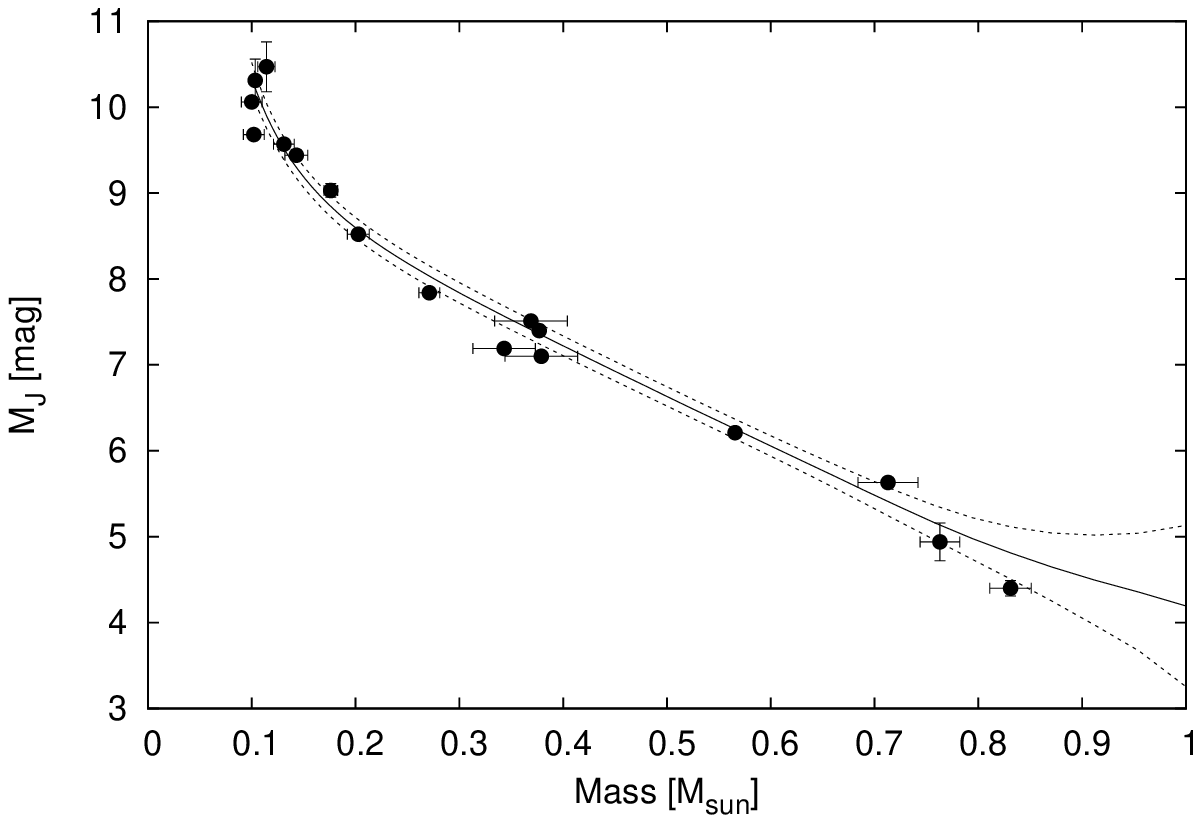}
\plottwo{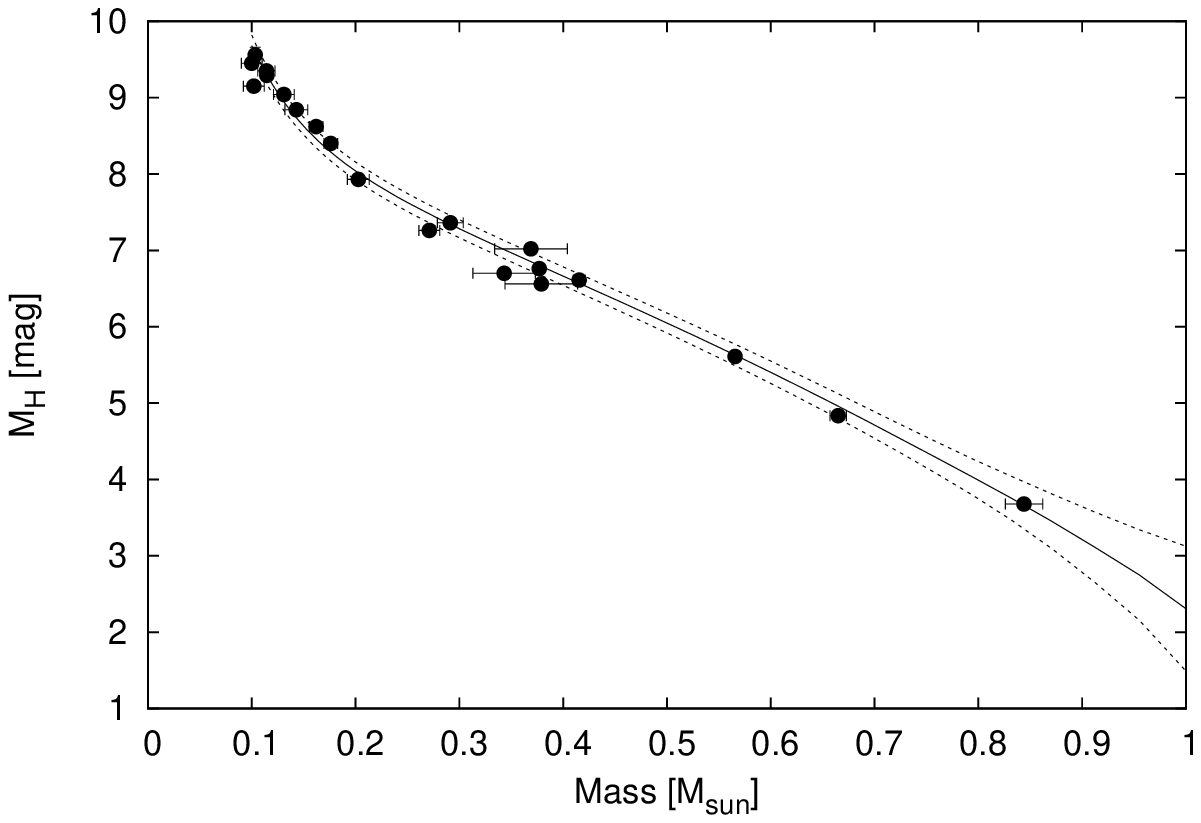}{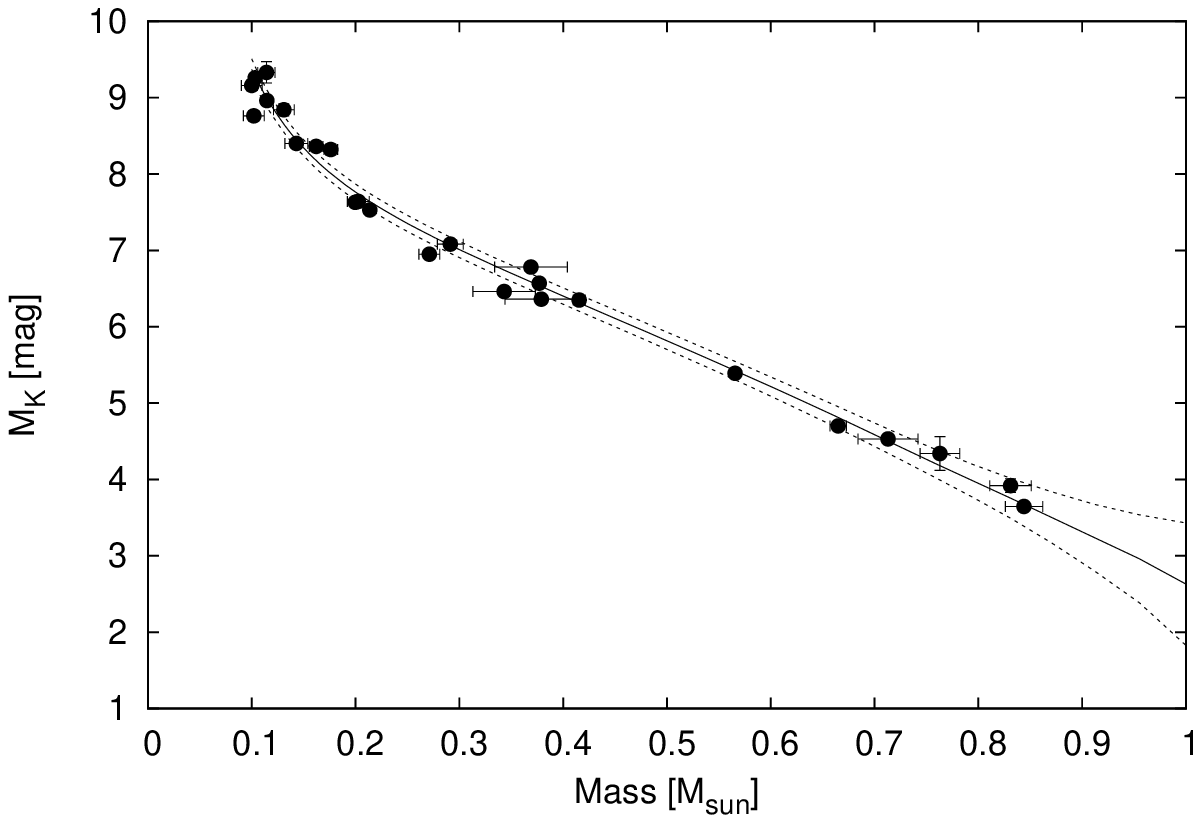}
\caption{
    Top: Same as Figure~\ref{fig:EmpiricalModelFitBC}, here we show
    the empirical relation between stellar mass and the absolute
    magnitudes in the $V$, $J$, $H$ and $K$ bands. The residuals from
    these fits are shown in
    Figure~\ref{fig:EmpiricalModelFitMagsResid}.\\
\label{fig:EmpiricalModelFitMags}}
\end{figure*}

We look for the following relations: (1) $\rhostar \rightarrow \rstar$
(which also defines a $\rhostar \rightarrow \mstar$ relation), (2)
$\mstar \rightarrow M_{\rm bol}$, (which together with relation 1 also
defines a relation between $\teffstar$ and the other parameters) (3)
$\teffstar \rightarrow M_{\rm bol} - M_{V}$, (4) $\teffstar
\rightarrow M_{\rm bol} - M_{J}$, (5) $\teffstar \rightarrow M_{\rm
  bol} - M_{H}$, (6) $\teffstar \rightarrow M_{\rm bol} - M_{K}$. The
motivation for choosing this particular formulation is that $\rhostar$
is typically a well-measured parameter for transiting planet systems,
and the relation between $\rhostar$ and $\rstar$ is tighter than for
other relations involving $\rhostar$. The other relations are then
based on physical dependencies (bolometric magnitude depends
primarily on stellar mass, with age and metallicity being secondary
factors, and bolometric corrections depend primarily on effective
temperature, with metallicity being a secondary factor). These
relations are parameterized as follows:
\begin{widetext}
\begin{equation}
\label{eqn:empiricalmodelfunctionslR}
lR = \left( \begin{array}{c}
a_{\rho,0} \\
a_{\rho,1} \\
a_{\rho,2} \\
a_{\rho,3} \end{array} \right)^{T} \left( \begin{array}{cccc}
   0.131674 & -0.605624 &  0.739127 & -0.263765 \\
  -0.494952 &  0.614479 &  0.437886 & -0.430922 \\
  -0.765425 & -0.291096 &  0.099548 &  0.565224 \\
   0.389626 &  0.413398 &  0.502032 &  0.652117 \end{array} \right)
\left( \begin{array}{c}
 1 \\
 l\rho \\
 l\rho^2 \\
 l\rho^3 \end{array} \right) + dlR
\end{equation}

\begin{equation}
\label{eqn:empiricalmodelfunctionsMbol}
M_{\rm bol} = \left( \begin{array}{c}
b_{m,0} \\
b_{m,1} \\
b_{m,2} \\
b_{m,3} \end{array} \right)^{T} \left( \begin{array}{cccc}
  -0.042114 & -0.349391 & -0.787638 & -0.505746 \\
   0.170316 &  0.723492 &  0.095153 & -0.662192 \\
  -0.309788 & -0.505183 &  0.591641 & -0.546610 \\
  -0.934479 &  0.315080 & -0.143296 &  0.083309 \end{array} \right)
\left( \begin{array}{c}
 1 \\
 lM \\
 lM^2 \\
 lM^3 \end{array} \right) + dM_{\rm bol}
\end{equation}

\begin{equation}
\label{eqn:empiricalmodelfunctionsBCV}
BC_{\lambda} = \left( \begin{array}{c}
c_{\lambda,0} \\
c_{\lambda,1} \\
c_{\lambda,2} \end{array} \right)^{T} \left( \begin{array}{ccc}
   0.871464 & -0.485768 &  0.067675 \\
   0.484681 &  0.831841 & -0.270418 \\
   0.075065 &  0.268461 &  0.960361 \end{array} \right)
\left( \begin{array}{c}
 1 \\
 lT \\
 lT^2 \end{array} \right)
\end{equation}
\end{widetext}
where $lR = \log_{10}(R)$, $l\rho = \log_{10}(\rho)$, $lM =
\log_{10}(M)$, $lT = \log_{10}(T_{\rm eff})$ are the logarithms of the
stellar radius in solar units, density in cgs units, mass in solar
units and temperature in Kelvin, respectively. The sets of
coefficients in the vectors to the left of the matrices (e.g.,
$a_{\rho,0}$, $a_{\rho,1}$, etc.; optimal values are given in
Table~\ref{tab:EmpiricalModelParameters}) are varied in the fit, and
the matrices are chosen to minimize correlations between these
parameters. The matrices are determined by initially fitting the
relations using simple polynomials (i.e.~we replace the matrices in
the relations above with the identity matrix), and then performing a
Principal Component Analysis on the resulting Markov Chains to
determine the transformation to a linearly uncorrelated set of
parameters. These matrices are held fixed in a subsequent fit where we
determine the optimized values of the parameters $a_{\rho,0}$,
$a_{\rho,1}$, etc.~(see below for more discussion of the fitting
procedure). The term $\lambda$ is either $V$, $J$, $H$ or $K$. To
allow for intrinsic scatter in the relations, the terms $dlR$ and
$dM_{\rm bol}$ represent real deviations in the radius or bolometric
magnitude from the model. Such deviations are expected due to
additional parameters, such as the metallicity or age, which are not
included in this model. Every star has its own value of $dlR$ and
$dM_{\rm bol}$, assumed to be drawn from Gaussian distributions with
standard deviations of $S_{R}$ and $S_{Mbol}$, respectively. The
following auxiliary relations are also used to relate the stellar
density to the mass, the bolometric magnitude and radius to the
effective temperature, and the bolometric magnitude and bolometric
correction to the absolute magnitude in a given filter
($M_{\lambda}$):
\begin{eqnarray}
lM & = & l\rho - 0.14968 + 3lR(l\rho) \\
lT & = & (42.227 - 5lR - M_{bol})/10 \\
M_{\lambda} & = & M_{\rm bol} + BC_{\lambda}.
\end{eqnarray}

\begin{figure*}[!ht]
\plottwo{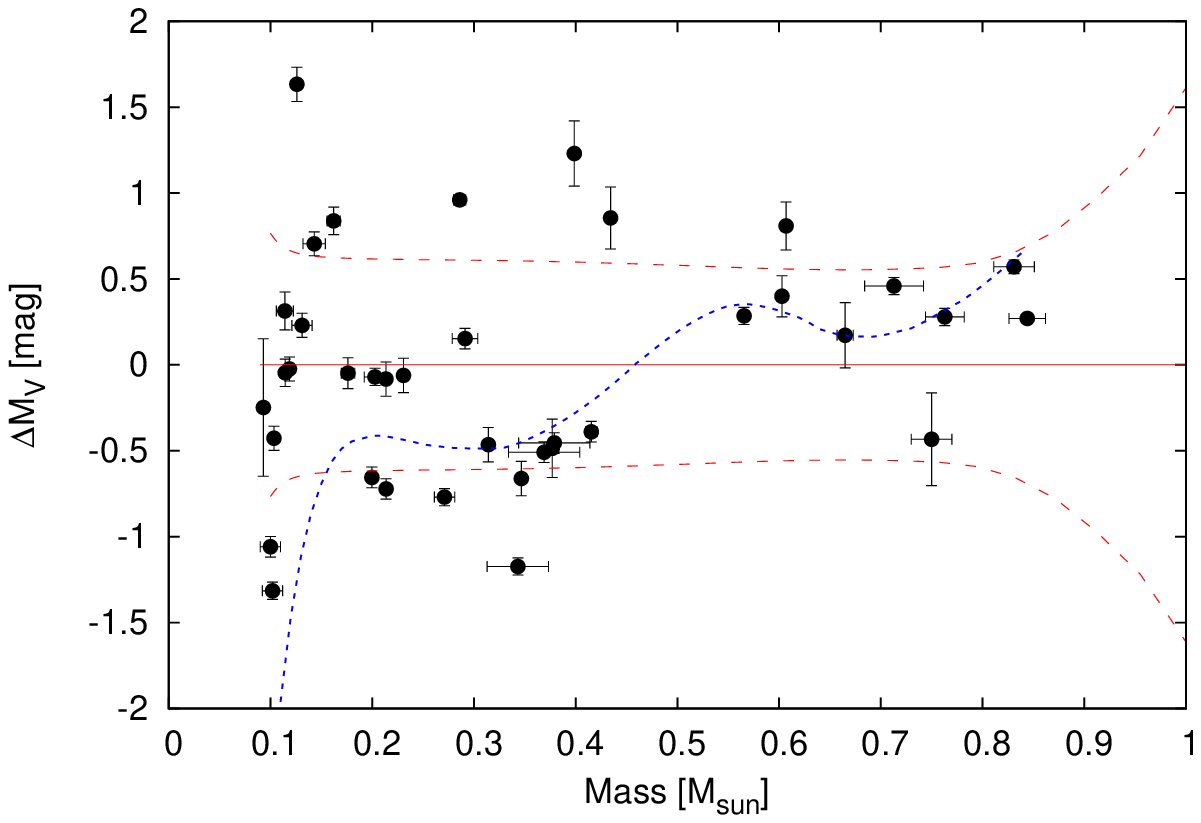}{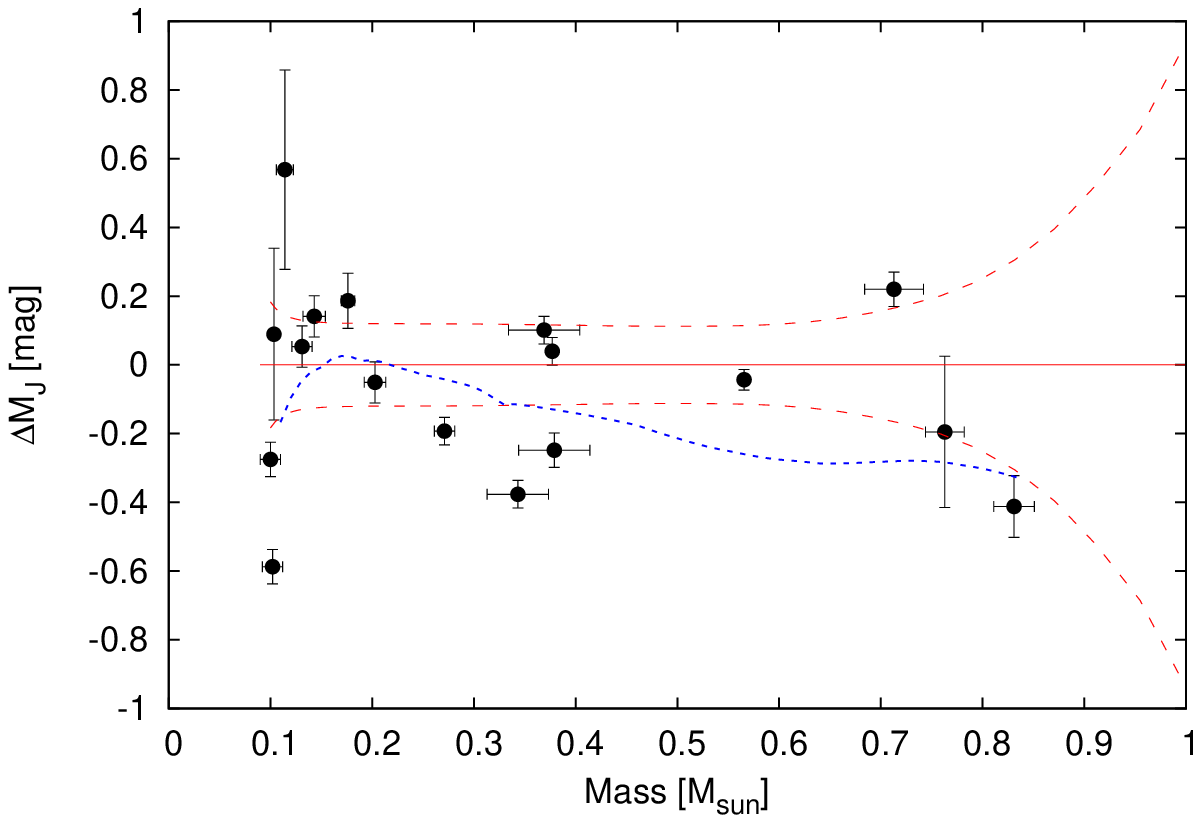}
\plottwo{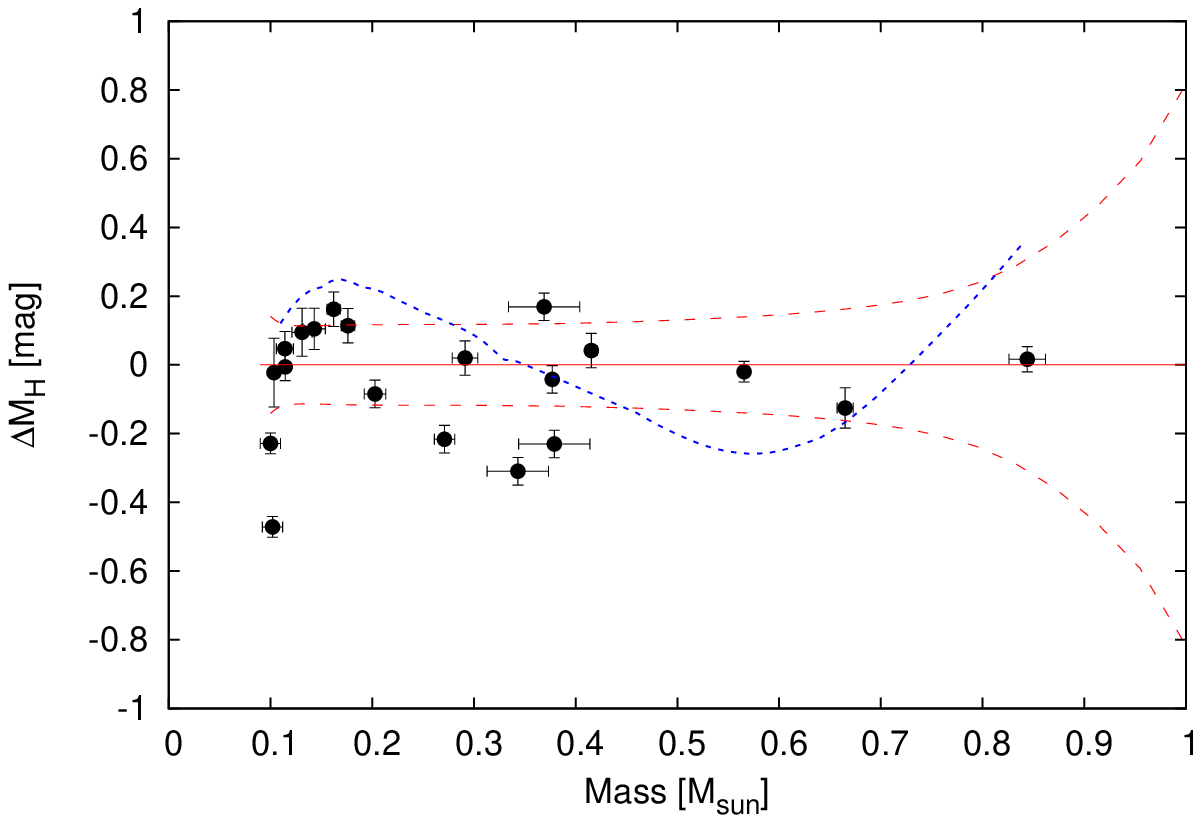}{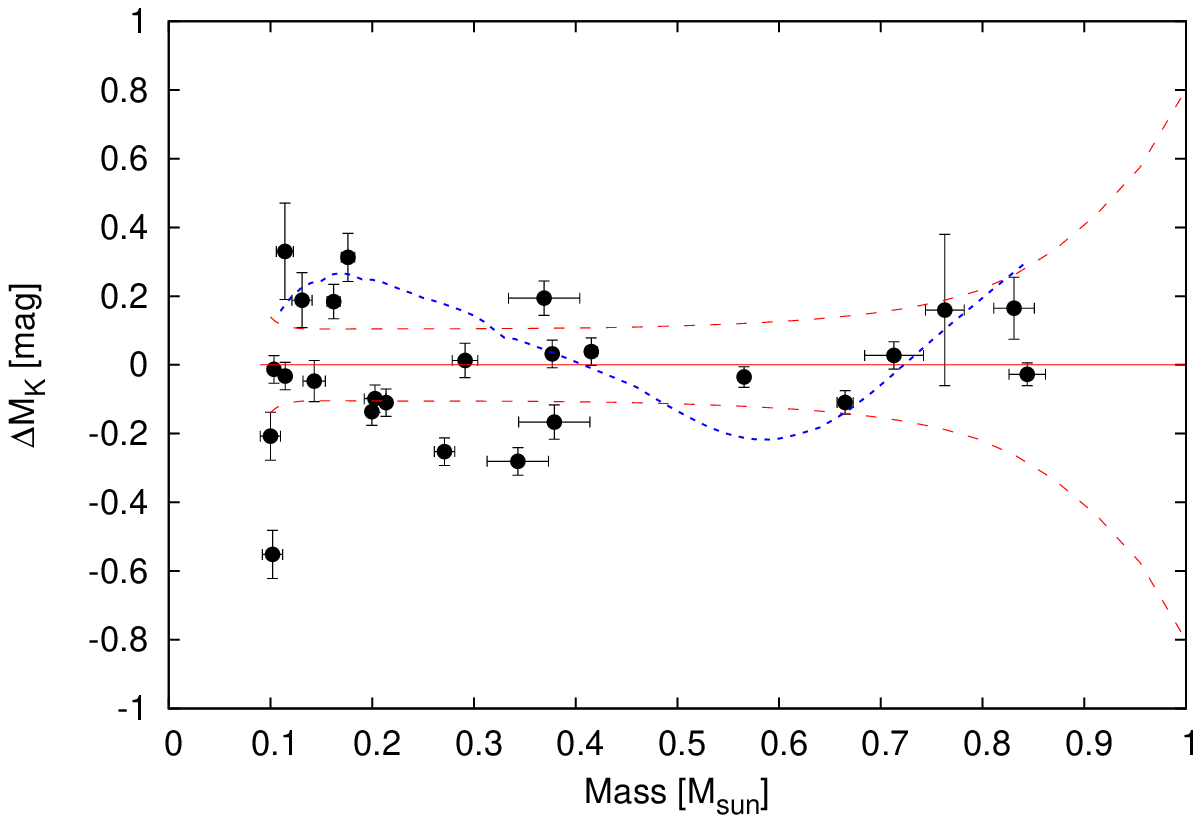}
\caption{
    Same as Figure~\ref{fig:EmpiricalModelFitMags}, here we show
    residuals from the median empirical relation. The short-dash
    (blue) lines show the relation from the 4.5\,Gyr, Solar
    metallicity Dartmouth models, while the long-dash (red) lines show
    the expected $1\sigma$ spread of the empirical model.\\ \\
\label{fig:EmpiricalModelFitMagsResid}}
\end{figure*}

\begin{figure*}[!ht]
\plottwo{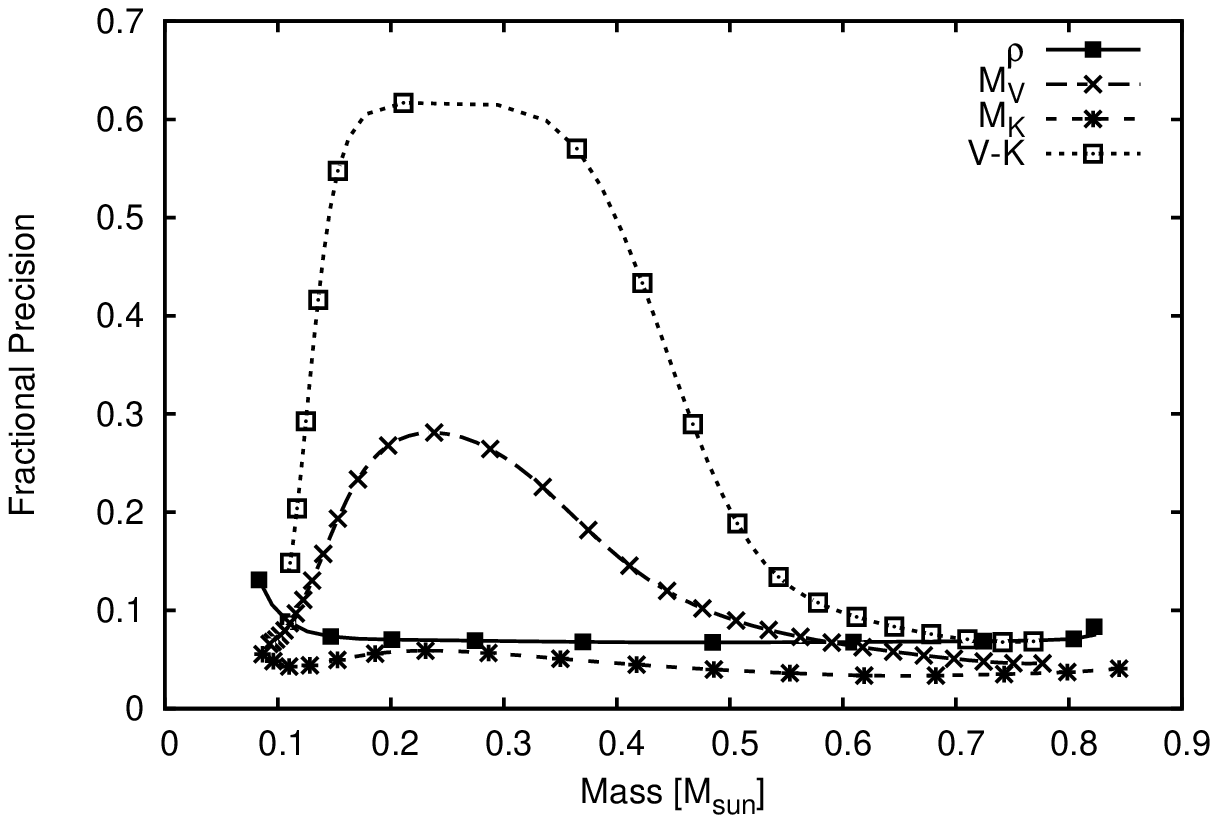}{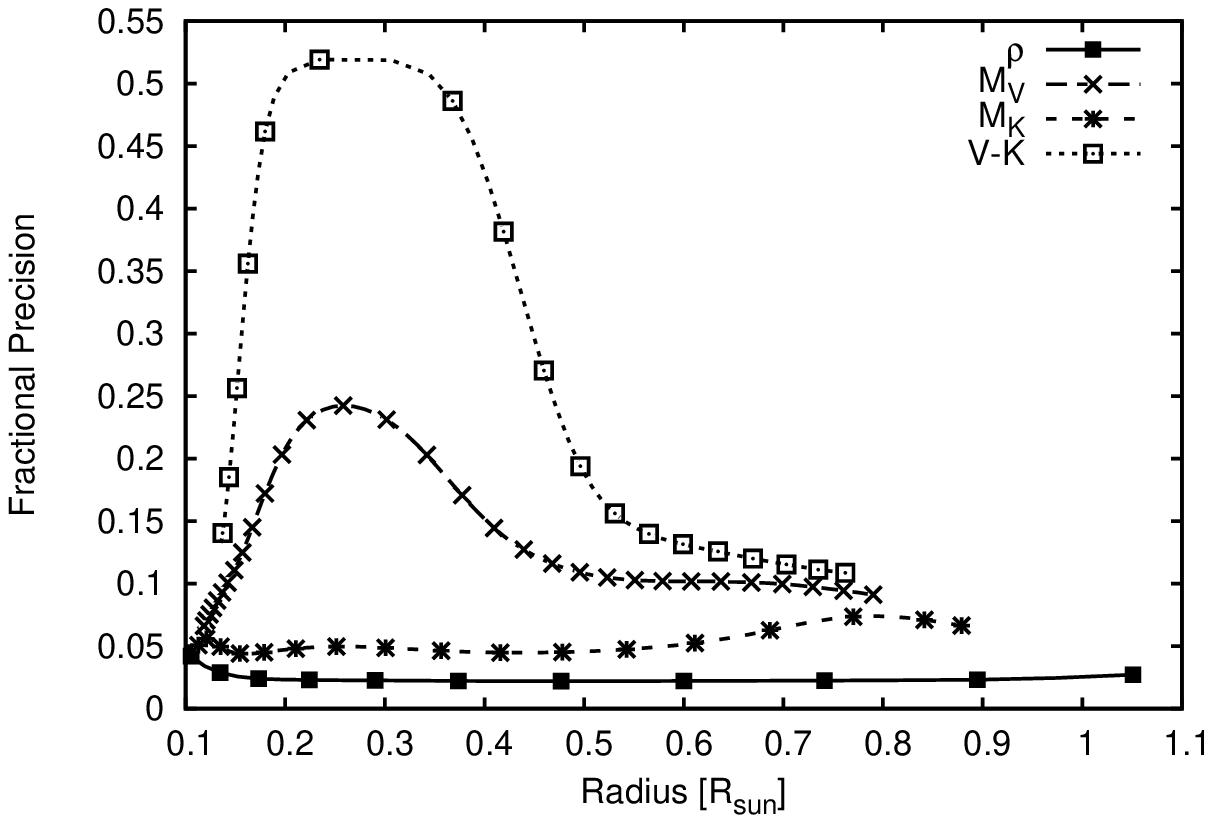}
\plottwo{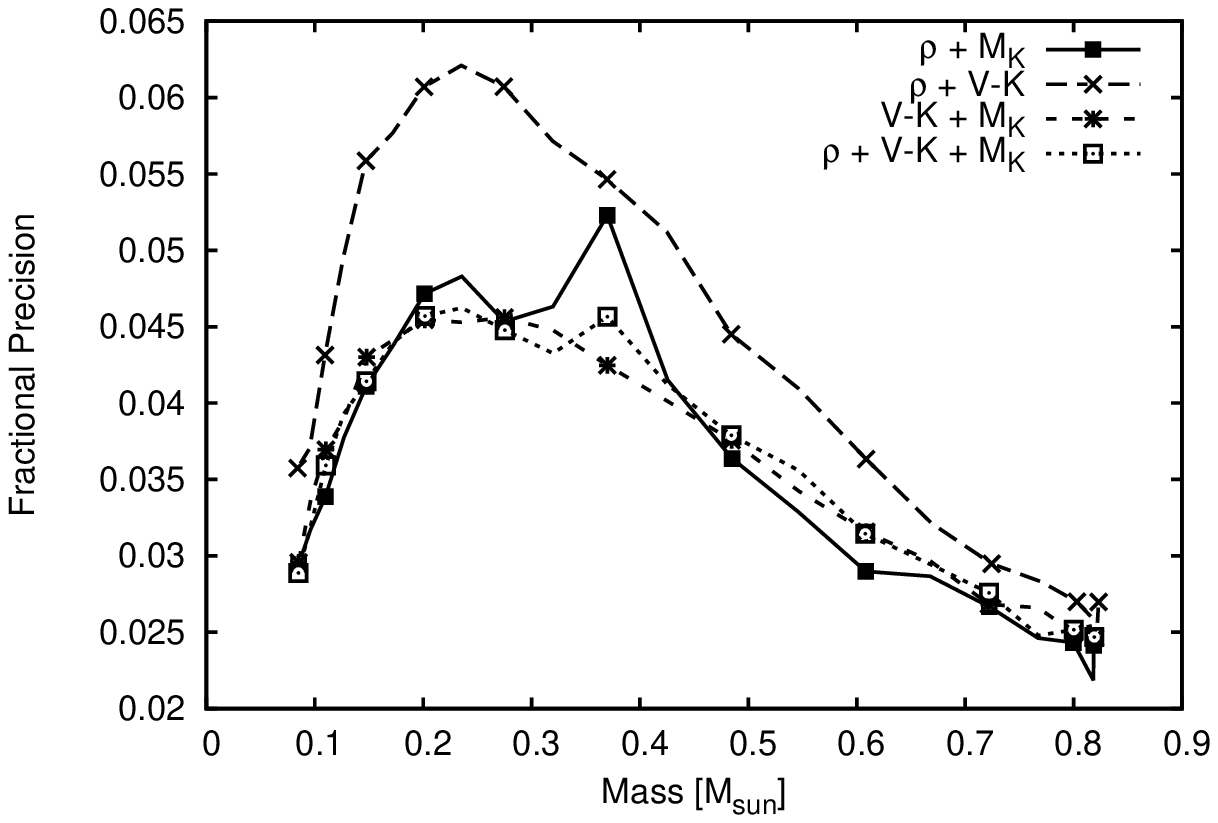}{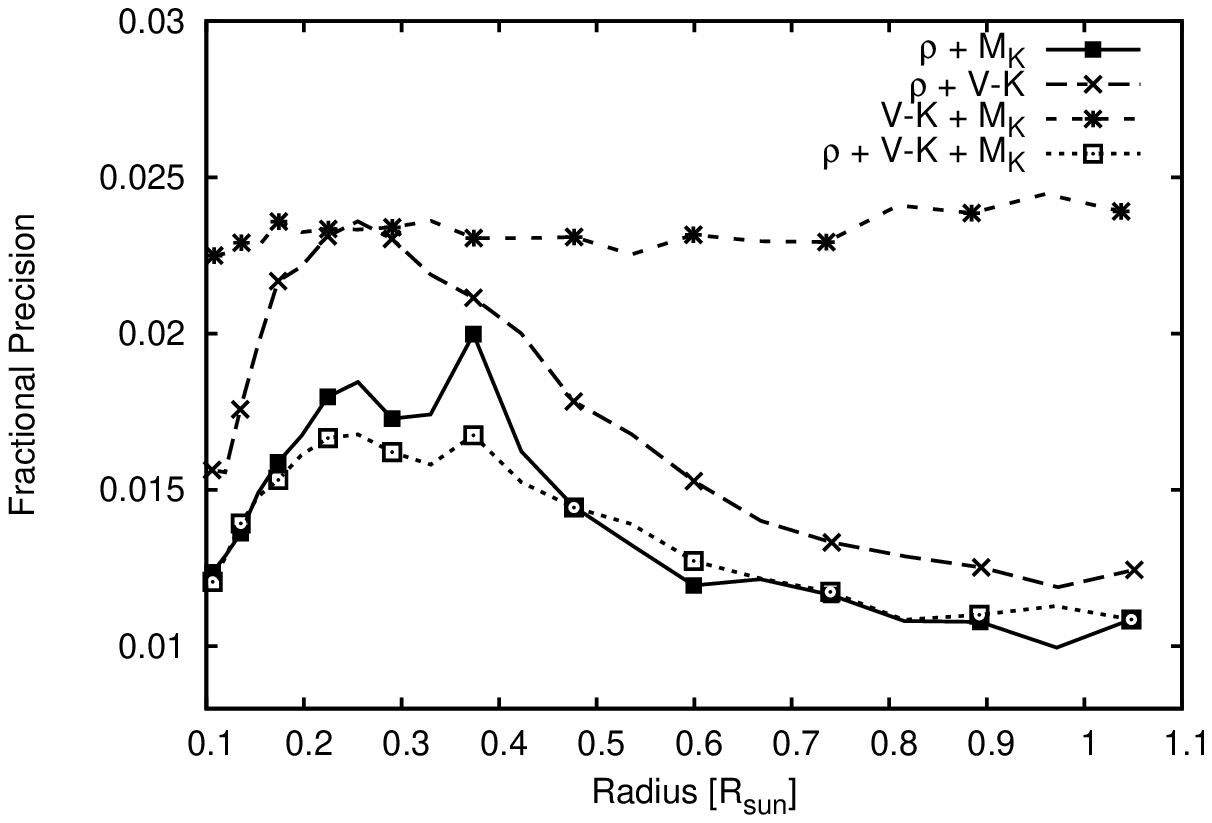}
\caption{
    Top: Comparison of the precision with which stellar mass (left)
    and stellar radius (right) may be measured using either
    $\rhostar$, $M_{V}$, $M_{K}$, or $V-K$ as an observed
    parameter. For each observable we step through a range of values
    determining the median stellar mass (radius) and its standard
    deviation using the empirical relations given in
    equations~\ref{eqn:empiricalmodelfunctionslR}
    through~\ref{eqn:empiricalmodelfunctionsBCV}.  We find that
    $M_{K}$ is the highest precision stellar mass indicator, allowing
    a precision that varies between $\sim 3$\% and $6$\%, whereas
    $\rhostar$ is the highest precision stellar radius indicator,
    allowing a precision of $\sim 2$\% over the range $0.2\,\rsun < R
    < 0.9\,\rsun$. Bottom: Comparison of the precision with which
    stellar mass (left) and stellar radius (right) may be measured
    using the combinations of parameters indicated in the
    figure. Combining the density with an absolute magnitude indicator
    allows for a mass precision between 2.5\% and 4.5\%, and a radius
    precision between $\sim 1$\% and $\sim 2$\%.\\
\label{fig:compareprecision}}
\end{figure*}

The stars that we use in the fit are listed in Tables~\ref{tab:MEBs}
and~\ref{tab:photMbinaries}.  The eclipsing binaries are compiled from
tables in \cite{zhou:2014:mebs}, \cite{torres:2010}, and
\cite{nefs:2013}, and include non-pre-main-sequence stars with $\mstar
< 0.85\,\msun$ and with $\mstar$ and $\rstar$ measured to better than
5\% accuracy. The upper mass limit corresponds roughly to the mass for
which stars older than the pre-main-sequence phase, and with $-0.5 <
[Fe/H] < 0.5$, will have $\log g > 4.4$ throughout the $\sim 10$\,Gyr
age of the Galactic disk. Stars below this mass show tight
correlations between the density and the mass, radius and
luminosity. Because individual broad-band photometric magnitudes are
not available for most of the eclipsing binary components, we use
resolved binaries for which masses and component magnitudes are
available to constrain relations (3) through (6) above. Most of these
stars are taken from the compilation by \cite{delfosse:2000}, with a
few objects from the list of \cite{torres:2010} to extend into the
late K dwarf regime.

To fit the above relations we use a likelihood function of the form:
\begin{widetext}
\begin{eqnarray}
\label{eqn:likelihoodEmpirical}
\ln L & = & -\frac{1}{2}\sum_{i=1}^{N_{\star}}\left( \frac{lR(l\rho_{i}) - lR_{{\rm obs},i}}{\sigma_{lR,i}}\right) ^{2} \\
 & & -\frac{1}{2}\sum_{i=1}^{N_{\star}}\left( \frac{lM(l\rho_{i}) - lM_{{\rm obs},i}}{\sigma_{lM,i}}\right) ^{2} \nonumber \\
 & & -\frac{1}{2}\sum_{i=1}^{N_{\star}}\left( \frac{lT(lR(l\rho_{i}),M_{\rm bol}(lM(l\rho_{i}))) - lT_{{\rm obs},i}}{\sigma_{lT,i}}\right) ^{2} \nonumber \\
 & & -\frac{1}{2}\sum_{\lambda}\sum_{i=1}^{N_{\star}}\left( \frac{M_{\rm bol}(lM(l\rho_{i}))+BC_{\lambda}(lT(lR(l\rho_{i}),M_{\rm bol}(lM(l\rho_{i})))) - M_{\lambda,obs,i}}{\sigma_{\lambda,i}} \right) ^2 \nonumber \\
 & & -\frac{1}{2}\sum_{i=1}^{N_{\star}}\left( \ln(S_{R}^2) + \left( \frac{dlR_{i}}{S_{R}} \right) ^2 \right) \nonumber \\
 & & -\frac{1}{2}\sum_{i=1}^{N_{\star}}\left( \ln(S_{Mbol}^2) + \left( \frac{dM_{{\rm bol},i}}{S_{Mbol}} \right) ^2 \right) \nonumber
\end{eqnarray}
\end{widetext}
where the sum on $\lambda$ is over the four filters $V$, $J$, $H$ and
$K$, and for each star we exclude terms from the sums if the relevant
measurement ($lR_{\rm obs}$, $lM_{\rm obs}$, $lT_{\rm obs}$,
$M_{V,obs}$, $M_{J,obs}$, $M_{H,obs}$, or $M_{K,obs}$) is not
available. Here $l\rho_{i}$, $dlR_{i}$, and $dM_{{\rm bol},i}$ are
free parameters for each star $i$, and the other free parameters are
$S_{R}$, $S_{Mbol}$, and the coefficients in
equations~\ref{eqn:empiricalmodelfunctionslR}--\ref{eqn:empiricalmodelfunctionsBCV}. We
use the DEMCMC procedure to explore this likelihood
function. Table~\ref{tab:EmpiricalModelParameters} provides the values
for the best-fit model together with their approximate $1\sigma$
(uncorrelated) uncertainties.\footnote{A C code implementing these relations to determine the properties of K and M dwarf stars from various observed quantities is available at \url{http://www.astro.princeton.edu/\~jhartman/kmdwarfparam.html}.}

\ifthenelse{\boolean{emulateapj}}{
    \begin{deluxetable*}{lrrrr}
}{
    \begin{deluxetable}{lrrrr}
}
\tablewidth{0pc}
\tabletypesize{\scriptsize}
\tablecaption{
    K and M Dwarf Binary Components Used to Fit Empirical Relations \tablenotemark{a}
    \label{tab:MEBs}
}
\tablehead{
    \multicolumn{1}{c}{Star}          &
    \multicolumn{1}{c}{Mass} &
    \multicolumn{1}{c}{Radius} &
    \multicolumn{1}{c}{T$_{\rm eff}$} & 
    \multicolumn{1}{c}{Reference} \\
    \multicolumn{1}{c}{} &
    \multicolumn{1}{c}{(\msun)} &
    \multicolumn{1}{c}{(\rsun)} & 
    \multicolumn{1}{c}{(K)} & 
    \multicolumn{1}{c}{}
}
\startdata
1~RXS~J154727.5+450803~A & $0.2576\pm0.0085$ & $0.2895\pm0.0068$ & $\ldots$ & 1 \\
1~RXS~J154727.5+450803~B & $0.2585\pm0.0080$ & $0.2895\pm0.0068$ & $\ldots$ & 1 \\
ASA~J011328--3821.1~A & $0.612\pm0.03$ & $0.596\pm0.02$ & $3750\pm250$ & 2 \\
CG~Cyg~B & $0.814\pm0.013$ & $0.838\pm0.011$ & $4720\pm60$ & 3,4 \\
CM~Dra~A & $0.2130\pm0.0009$ & $0.2534\pm0.0019$ & $3130\pm70$ & 5 \\
CM~Dra~B & $0.2141\pm0.0010$ & $0.2396\pm0.0015$ & $3120\pm70$ & 5 \\
CU~Cnc~A & $0.4333\pm0.0017$ & $0.4317\pm0.0052$ & $3160\pm150$ & 6 \\
CU~Cnc~B & $0.3980\pm0.0014$ & $0.3908\pm0.0094$ & $3130\pm150$ & 6 \\
GJ~3236~A & $0.376\pm0.016$ & $0.3795\pm0.0084$ & $3310\pm110$ & 7 \\
GJ~551 \tablenotemark{b} & $0.123\pm0.006$ & $0.141\pm0.007$ & $3098\pm56$ & 8,9,10 \\
GU~Boo~A & $0.610\pm0.006$ & $0.627\pm0.016$ & $3920\pm130$ & 11 \\
GU~Boo~B & $0.600\pm0.006$ & $0.624\pm0.016$ & $3810\pm130$ & 11 \\
HATS550-016 & $0.110^{+0.005}_{-0.006}$ & $0.147^{+0.003}_{-0.004}$ & $\ldots$ & 12 \\
HATS551-021 & $0.132^{+0.014}_{-0.005}$ & $0.154^{+0.006}_{-0.008}$ & $\ldots$ & 12 \\
HD~195987~A & $0.844\pm0.018$ & $0.98\pm0.04$ & $5200\pm100$ & 13 \\
J1219-39~B & $0.091\pm0.002$ & $0.1174^{+0.0071}_{-0.0050}$ & $\ldots$ & 14 \\
Kepler-16~B & $0.20255^{+0.00066}_{-0.00065}$ & $0.22623^{+0.00059}_{-0.00055}$ & $\ldots$ & 15 \\
KIC~1571511~B & $0.14136^{+0.0051}_{-0.0042}$ & $0.17831^{+0.0013}_{-0.0016}$ & $\ldots$ & 16 \\
KOI-126~B & $0.2413\pm0.0030$ & $0.2543\pm0.0014$ & $\ldots$ & 17 \\
KOI-126~C & $0.2127\pm0.0026$ & $0.2318\pm0.0013$ & $\ldots$ & 17 \\
LSPM~J1112+7626~A & $0.3946\pm0.0023$ & $0.3860\pm0.0055$ & $3060\pm160$ & 18 \\
LSPM~J1112+7626~B & $0.2745\pm0.0012$ & $0.2978\pm0.0049$ & $2950\pm160$ & 18 \\
MG1-116309~A & $0.567\pm0.002$ & $0.552\pm0.0085$ & $3920\pm80$ & 19 \\
MG1-116309~B & $0.532\pm0.002$ & $0.532\pm0.006$ & $3810\pm80$ & 19 \\
MG1-1819499~A & $0.557\pm0.001$ & $0.569\pm0.0022$ & $3690\pm80$ & 19 \\
MG1-1819499~B & $0.535\pm0.001$ & $0.500\pm0.0085$ & $3610\pm80$ & 19 \\
MG1-2056316~A & $0.4690\pm0.0021$ & $0.441\pm0.002$ & $3460\pm180$ & 19 \\
MG1-2056316~B & $0.382\pm0.001$ & $0.374\pm0.002$ & $3320\pm180$ & 19 \\
MG1-506664~A & $0.584\pm0.002$ & $0.560\pm0.0025$ & $3730\pm90$ & 19 \\
MG1-506664~B & $0.544\pm0.002$ & $0.513\pm0.0055$ & $3610\pm90$ & 19 \\
MG1~646680A & $0.499\pm0.002$ & $0.457\pm0.006$ & $3730\pm50$ & 19 \\
MG1-646680~B & $0.443\pm0.002$ & $0.427\pm0.006$ & $3630\pm50$ & 19 \\
MG1-78457~B & $0.491\pm0.002$ & $0.471\pm0.009$ & $3270\pm100$ & 19 \\
NSVS~01031772~A & $0.5428\pm0.0027$ & $0.526\pm0.0028$ & $3615\pm72$ & 20 \\
NSVS~01031772~B & $0.4982\pm0.0025$ & $0.5087\pm0.0031$ & $3520\pm30$ & 20 \\
NSVS~6550671~A & $0.510\pm0.02$ & $0.550\pm0.01$ & $3730\pm60$ & 21 \\
T-Cyg1-01385~B & $0.43\pm0.02$ & $0.40\pm0.02$ & $\ldots$ & 22 \\
UV~Psc~B & $0.7644\pm0.0045$ & $0.835\pm0.018$ & $4750\pm80$ & 23 \\
V568~Lyr~B & $0.8273\pm0.0042$ & $0.7679\pm0.0064$ & $4900\pm100$ & 24 \\
WOCS~23009~B & $0.447\pm0.011$ & $0.4292\pm0.0033$ & $3620\pm150$ & 25 \\
WTS19b-2-01387~A & $0.498\pm0.019$ & $0.496\pm0.013$ & $3498\pm100$ & 26 \\
WTS19b-2-01387~B & $0.481\pm0.017$ & $0.479\pm0.013$ & $3436\pm100$ & 26 \\
YY~Gem~A & $0.599\pm0.005$ & $0.619\pm0.006$ & $3819\pm98$ & 27 \\
YY~Gem~B & $0.599\pm0.005$ & $0.619\pm0.006$ & $3819\pm98$ & 27
\enddata
\tablerefs{1: \cite{hartman:2011:kmdwarf}; 2: \cite{helminiak:2012}; 3: \cite{bedford:1987}; 4: \cite{popper:1994}; 5: \cite{morales:2009}; 6: \cite{ribas:2003}; 7: \cite{irwin:2009}; 8: \cite{segransan:2003}; 9: \cite{valenti:2005}; 10: \cite{demory:2009}; 11: \cite{lopezmorales:2005}; 12: \cite{zhou:2014:mebs}; 13: \cite{torres:2002:hd195987}; 14: \cite{triaud:2013:EBLM}; 15: \cite{doyle:2011}; 16: \cite{ofir:2012}; 17: \cite{carter:2011}; 18: \cite{irwin:2011}; 19: \cite{kraus:2011}; 20: \cite{lopezmorales:2006}; 21: \cite{dimitrov:2010}; 22: \cite{fernandez:2009}; 23: \cite{popper:1997}; 24: \cite{grundahl:2008}; 25: \cite{sandquist:2013}; 26: \cite{birkby:2012}; 27: \cite{torres:2002}}
\tablenotetext{a}{Data compiled primarily from tables given in \cite{torres:2010}, \cite{nefs:2013} and \cite{zhou:2014:mebs}, we provide the original references for each source in the table.}
\tablenotetext{b}{This is a single star, with an interferometric radius measurement, and mass estimated assuming a mass-luminosity relation.\\}
\ifthenelse{\boolean{emulateapj}}{
    \end{deluxetable*}
}{
    \end{deluxetable}
}

\ifthenelse{\boolean{emulateapj}}{
    \begin{deluxetable*}{lrrrrr}
}{
    \begin{deluxetable}{lrrrrr}
}
\tablewidth{0pc}
\tabletypesize{\scriptsize}
\tablecaption{
    K and M Dwarf Stars with Absolute Magnitudes and Measured Masses Used in Fitting the Bolometric Corrections in the Empirical Stellar Model \tablenotemark{a}
    \label{tab:photMbinaries}
}
\tablehead{
    \multicolumn{1}{c}{Star}          &
    \multicolumn{1}{c}{Mass} &
    \multicolumn{1}{c}{$M_{V}$} &
    \multicolumn{1}{c}{$M_{J}$} & 
    \multicolumn{1}{c}{$M_{H}$} &
    \multicolumn{1}{c}{$M_{K}$} \\
    \multicolumn{1}{c}{} &
    \multicolumn{1}{c}{(\msun)} &
    \multicolumn{1}{c}{(mag)} & 
    \multicolumn{1}{c}{(mag)} & 
    \multicolumn{1}{c}{(mag)} &
    \multicolumn{1}{c}{(mag)}
}
\startdata
Gl~866~C & $0.0930\pm0.0008$ & $17.43\pm0.40$ & $\ldots$ & $\ldots$ & $\ldots$ \\
Gl~65~B & $0.100\pm0.010$ & $15.87\pm0.06$ & $10.06\pm0.05$ & $9.45\pm0.03$ & $9.16\pm0.07$ \\
Gl~65~A & $0.102\pm0.010$ & $15.41\pm0.05$ & $9.68\pm0.05$ & $9.15\pm0.03$ & $8.76\pm0.07$ \\
Gl~234~B & $0.1034\pm0.0035$ & $16.16\pm0.07$ & $10.31\pm0.25$ & $9.56\pm0.10$ & $9.26\pm0.04$ \\
Gl~623~B & $0.1142\pm0.0083$ & $16.02\pm0.11$ & $10.47\pm0.29$ & $9.35\pm0.05$ & $9.33\pm0.14$ \\
Gl~866~B & $0.1145\pm0.0012$ & $15.64\pm0.08$ & $\ldots$ & $9.29\pm0.04$ & $8.96\pm0.04$ \\
Gl~866~A & $0.1187\pm0.0011$ & $15.39\pm0.07$ & $\ldots$ & $\ldots$ & $\ldots$ \\
Gl~791.2~B & $0.126\pm0.003$ & $16.64\pm0.10$ & $\ldots$ & $\ldots$ & $\ldots$ \\
Gl~473~B & $0.131\pm0.010$ & $15.00\pm0.07$ & $9.57\pm0.06$ & $9.04\pm0.07$ & $8.84\pm0.08$ \\
Gl~473~A & $0.143\pm0.011$ & $15.01\pm0.07$ & $9.44\pm0.06$ & $8.84\pm0.06$ & $8.40\pm0.06$ \\
Gl~831~B & $0.1621\pm0.0065$ & $14.62\pm0.08$ & $\ldots$ & $8.62\pm0.05$ & $8.36\pm0.05$ \\
Gl~860~B & $0.1762\pm0.0066$ & $13.46\pm0.09$ & $9.03\pm0.08$ & $8.40\pm0.05$ & $8.32\pm0.07$ \\
Gl~747~B & $0.1997\pm0.0008$ & $12.52\pm0.06$ & $\ldots$ & $\ldots$ & $7.63\pm0.04$ \\
Gl~234~A & $0.2027\pm0.0106$ & $13.07\pm0.05$ & $8.52\pm0.06$ & $7.93\pm0.04$ & $7.64\pm0.04$ \\
CMDra~B & $0.2136\pm0.0010$ & $12.94\pm0.10$ & $\ldots$ & $\ldots$ & $\ldots$ \\
Gl~747~A & $0.2137\pm0.0009$ & $12.30\pm0.06$ & $\ldots$ & $\ldots$ & $7.53\pm0.04$ \\
CMDra~A & $0.2307\pm0.0010$ & $12.80\pm0.10$ & $\ldots$ & $\ldots$ & $\ldots$ \\
Gl~860~A & $0.2711\pm0.0100$ & $11.76\pm0.05$ & $7.84\pm0.04$ & $7.26\pm0.04$ & $6.95\pm0.04$ \\
Gl~791.2~A & $0.286\pm0.006$ & $13.37\pm0.03$ & $\ldots$ & $\ldots$ & $\ldots$ \\
Gl~831~A & $0.2913\pm0.0125$ & $12.52\pm0.06$ & $\ldots$ & $7.36\pm0.05$ & $7.08\pm0.05$ \\
Gl~644~Bb & $0.3143\pm0.0040$ & $11.71\pm0.10$ & $\ldots$ & $\ldots$ & $\ldots$ \\
Gl~623~A & $0.3432\pm0.0301$ & $10.74\pm0.05$ & $7.19\pm0.04$ & $6.70\pm0.04$ & $6.46\pm0.04$ \\
Gl~644~Ba & $0.3466\pm0.0047$ & $11.22\pm0.10$ & $\ldots$ & $\ldots$ & $\ldots$ \\
Gl~661~B & $0.369\pm0.035$ & $11.15\pm0.06$ & $7.51\pm0.04$ & $7.02\pm0.04$ & $6.78\pm0.05$ \\
Gl~570~C & $0.3770\pm0.0018$ & $11.09\pm0.17$ & $7.40\pm0.04$ & $6.76\pm0.04$ & $6.57\pm0.04$ \\
Gl~661~A & $0.379\pm0.035$ & $11.10\pm0.06$ & $7.10\pm0.05$ & $6.56\pm0.04$ & $6.36\pm0.05$ \\
GJ~2069~Ab & $0.3987\pm0.0007$ & $12.57\pm0.19$ & $\ldots$ & $\ldots$ & $\ldots$ \\
Gl~644~A & $0.4155\pm0.0057$ & $10.76\pm0.06$ & $\ldots$ & $6.61\pm0.05$ & $6.35\pm0.04$ \\
GJ~2069~Aa & $0.4344\pm0.0008$ & $11.78\pm0.18$ & $\ldots$ & $\ldots$ & $\ldots$ \\
Gl~570~B & $0.5656\pm0.0029$ & $9.45\pm0.05$ & $6.21\pm0.03$ & $5.61\pm0.03$ & $5.39\pm0.03$ \\
YYGem~A & $0.6028\pm0.0014$ & $9.03\pm0.12$ & $\ldots$ & $\ldots$ & $\ldots$ \\
YYGem~B & $0.6069\pm0.0014$ & $9.38\pm0.14$ & $\ldots$ & $\ldots$ & $\ldots$ \\
HD~195987~B & $0.6650\pm0.0079$ & $7.91\pm0.19$ & $\ldots$ & $4.835\pm0.059$ & $4.702\pm0.034$ \\
Gl~702~B & $0.713\pm0.029$ & $7.52\pm0.05$ & $5.63\pm0.05$ & $\ldots$ & $4.53\pm0.04$ \\
$\chi$~Dra~B & $0.750\pm0.02$ & $6.11\pm0.27$ & $\ldots$ & $\ldots$ & $\ldots$ \\
GJ~765.2~B & $0.763\pm0.019$ & $6.64\pm0.05$ & $4.94\pm0.22$ & $\ldots$ & $4.34\pm0.22$ \\
GJ~765.2~A & $0.831\pm0.020$ & $5.99\pm0.04$ & $4.40\pm0.09$ & $\ldots$ & $3.92\pm0.09$ \\
HD~195987~A & $0.844\pm0.018$ & $5.511\pm0.028$ & $\ldots$ & $3.679\pm0.037$ & $3.646\pm0.033$ \\
\enddata
\tablenotetext{a}{
  Data taken from \citet{delfosse:2000} and \citet{torres:2010}.\\
}
\ifthenelse{\boolean{emulateapj}}{
    \end{deluxetable*}
}{
    \end{deluxetable}
}

\ifthenelse{\boolean{emulateapj}}{
    \begin{deluxetable}{lrr}
}{
    \begin{deluxetable}{lrr}
}
\tablewidth{0pc}
\tabletypesize{\scriptsize}
\tablecaption{
    Fitted Parameters of Empirical Model for K and M Dwarf Properties
    \label{tab:EmpiricalModelParameters}
}
\tablehead{
    \multicolumn{1}{c}{Parameter}          &
    \multicolumn{1}{c}{Best Fit Value \tablenotemark{a}}     &
    \multicolumn{1}{c}{Median and $1\sigma$ Uncertainty \tablenotemark{b}}
}
\startdata
$a_{\rho,0}$ & $0.029$ & $0.054^{+0.048}_{-0.049}$ \\
$a_{\rho,1}$ & $-0.2924$ & $-0.2989^{+0.0083}_{-0.0084}$ \\
$a_{\rho,2}$ & $0.0817$ & $0.0815^{+0.0020}_{-0.0020}$ \\
$a_{\rho,3}$ & $-0.18187$ & $-0.18280^{+0.00075}_{-0.00074}$ \\
$b_{m,0}$   & $35.3$ & $32.6^{+3.1}_{-3.3}$ \\
$b_{m,1}$   & $-7.54$ & $-7.09^{+0.44}_{-0.44}$ \\
$b_{m,2}$   & $1.17$ & $1.24^{+0.10}_{-0.11}$ \\
$b_{m,3}$   & $-7.717$ & $-7.690^{+0.034}_{-0.034}$ \\
$c_{V,0}$   & $1070$ & $570^{+380}_{-370}$ \\
$c_{V,1}$   & $15.3$ & $16.5^{+2.2}_{-2.9}$ \\
$c_{V,2}$   & $0.099$ & $0.111^{+0.013}_{-0.013}$ \\
$c_{J,0}$   & $130$ & $100^{+150}_{-190}$ \\
$c_{J,1}$   & $-5.0$ & $-5.4^{+1.2}_{-1.1}$ \\
$c_{J,2}$   & $-0.1154$ & $-0.1105^{+0.0043}_{-0.0046}$ \\
$c_{H,0}$   & $-70$ & $-130^{+150}_{-140}$ \\
$c_{H,1}$   & $-3.96$ & $-3.46^{+0.96}_{-1.13}$ \\
$c_{H,2}$   & $-0.1619$ & $-0.1615^{+0.0044}_{-0.0041}$ \\
$c_{K,0}$   & $50$ & $-80^{+140}_{-140}$ \\
$c_{K,1}$   & $-5.08$ & $-4.83^{+0.92}_{-0.99}$ \\
$c_{K,2}$   & $-0.1786$ & $-0.1757^{+0.0039}_{-0.0038}$ \\
$S_{R}$     & $0.0069$ & $0.0091^{+0.0010}_{-0.0012}$ \\
$S_{Mbol}$  & $0.138$ & $0.169^{+0.022}_{-0.025}$ \\
\enddata
\tablenotetext{a}{
  A self-consistent set of values corresponding to the maximum likelihood model.
}
\tablenotetext{b}{
  The median and $1\sigma$ confidence interval for each parameter as
  determined from an MCMC analysis.\\
}
\ifthenelse{\boolean{emulateapj}}{
    \end{deluxetable}
}{
    \end{deluxetable}
}

Figure~\ref{fig:EmpiricalModelFit} shows the fit between $l\rho$, $lM$
and $lR$, and between $lM$ and $M_{\rm bol}$, together with the
residuals from this model, Figure~\ref{fig:EmpiricalModelFitBC} shows
the relations between effective temperature and the $V$, $J$, $H$, and
$K$ bolometric corrections, Figure~\ref{fig:EmpiricalModelFitMags}
shows the relation between mass and the absolute magnitudes in each of
these filters, and Figure~\ref{fig:EmpiricalModelFitMagsResid} shows
the residuals from the fits in
Figure~\ref{fig:EmpiricalModelFitMags}. Our model yields intrinsic
scatters of $\sim 2$\% in radius and $\sim 7$\% in mass, and $4$\% to
$10$\% in effective temperature (between $0.2$\,mag and $0.5$\,mag in
bolometric magnitudes) given a value for
$\rhostar$. Figure~\ref{fig:compareprecision} compares the precision
in the inferred mass and radius implied by this modeling when using a
single parameter to constrain the stellar properties (either
$\rhostar$, $M_{V}$, $M_{K}$ or $V-K$), and when using various
combinations of observables. Assuming no observational uncertainties,
$M_{K}$ is the highest precision stellar mass indicator, while
$\rhostar$ is the highest precision stellar radius indicator. By
combining $\rhostar$ with a photometric indicator, such as $M_{K}$, it
is possible to improve the precision in mass and radius by up to a
factor of two compared to the precision allowed by using only
$\rhostar$.

Note that due to the somewhat higher scatter in the stellar parameters
at the high mass end, and the small number of stars constraining the
fit at this end, we find that the stellar mass as a function of
density reaches a local maximum of $\sim 0.8$\,\msun\ at $\rho \sim
1.5$\,\gcmc. We therefore do not suggest using these models for stars
with $M > 0.8$\,\msun, or spectral types earlier than K5.

It is worth comparing our modeling to that done by
\citet{johnson:2011} and \citet{johnson:2012}. One significant
difference is the treatment of metallicity. \citet{johnson:2011}
include a correlation between metallicity and $\Delta M_{K}$
\citep{johnson:2009}, while \citet{johnson:2012} include an additional
correlation between metallicity and $\Delta J-K$ in their
modelling. We choose not to include metallicity explicitly in our
model due to the lack of a large sample of M dwarfs with
well-measured masses, radii and metallicities. Lacking such a dataset,
it is not possible to determine the effect of metallicity on all of
the relations considered. Instead the metallicity is assumed to
contribute to the intrinsic uncertainty in the $\rho \rightarrow R$
and $M \rightarrow M_{\rm bol}$ relations, measured using the $S_{R}$
and $S_{Mbol}$ parameters. While, like us, \citet{johnson:2011} use
eclipsing binaries in calibrating the mass--radius relation, their
sample is the \citet{ribas:2006} catalog which contained only 14 stars
with $M < 0.8$\,\msun, of which only 10 had masses and radii measured
to better than 5\% precision. \citet{johnson:2012}, on the other hand,
choose to use stars with interferometrically measured radii from
\citet{boyajian:2012} due to the possibility that the eclipsing
binaries are systematically inflated compared to single stars. While
these stars have directly measured radii, their masses must be
estimated using an assumed mass--luminosity relation. The few long
period, low-mass eclipsing binaries which have been studied have
parameters that are consistent with the shorter period binaries
\citep{irwin:2011,doyle:2011}, moreover \citet{boyajian:2012} find no
systematic difference between the mass--radius relation determined
from their single stars (for an assumed mass--luminosity relation),
and the mass--radius relation determined from short period eclipsing
binaries, so we choose to use the eclipsing binaries for which both
masses and radii are measured directly.

Having established our empirical stellar model, we use it to determine
the properties of \hatcur{} by incorporating it directly into our
light curve and RV curve fit. To do this we add the following terms to
our likelihood function:
\begin{eqnarray}
 & & -\frac{1}{2}\sum_{\lambda}\left( \frac{M_{\lambda}(l\rho,dlR,dM_{\rm bol}) + \Delta d + A_{\lambda} - m_{\lambda}}{\sigma_{\lambda}} \right) ^2 \nonumber \\
 & & -\frac{1}{2}\left( \left( \frac{dlR}{S_{R}}\right)^{2} + \left( \frac{dM_{\rm bol}}{S_{M}} \right)^2 \right)
\end{eqnarray}
Where $M_{\lambda}$ is the predicted absolute magnitude in filter
$\lambda$ given $dlR$, $dM_{\rm bol}$, and $l\rho$, which is determined from the other free
parameters in the model ($b^{2}$, $\zrstar$, $P$,
$\sqrt{e}\sin\omega$, $\sqrt{e}\cos\omega$, and $K$); $\Delta d$ is the distance modulus and $A_{\lambda}$ is the extinction; $m_{\lambda}$ is the observed magnitude; and
$\sigma_{\lambda}$ is the magnitude uncertainty. The sum is over the
filters $\lambda = V$, $J$, $H$, and $K$. The new parameters
introduced in the fit are thus $dlR$, $dM_{\rm bol}$, $\Delta d$ and
$A_{V}$. We restrict $A_{V}$ to be in the range $[0, 0.175]$ where the
upper limit is the total line-of-sight extinction from the
\citet{schlafly:2011} reddenning maps, and we assume the
\citet{cardelli:1989} extinction law to relate $A_{V}$ to the
extinction in the other band-passes. Given $l\rho$, $dlR$, and
$dM_{\rm bol}$ at each link in the resulting Markov Chain, we then
calculate all relevant stellar parameters, and dependent planetary
parameters.

Using the empirical model we find $\mstar =
\hatcurISOmcircempirical$\,\msun\ and $\rstar =
\hatcurISOrcircempirical$\,\rsun, assuming a circular orbit, and
$\mstar = \hatcurISOmeccenempirical$\,\msun\ and $\rstar =
\hatcurISOreccenempirical$\,\rsun, allowing for an eccentric
orbit. The mass and radius inferred assuming a circular orbit are
slightly lower than, though consistent with, the parameters inferred
using the Dartmouth isochrones ($\mstar =
\hatcurISOmcircdartmouth$\,\msun, and $\rstar =
\hatcurISOrcircdartmouth$\,\rsun). The precision allowed by both the
Dartmouth and empirical models are similar. When we allow the
eccentricity to vary with our empirical model the constraint from the
observed $V-K$ color of the star pulls the model to
a lower eccentricity solution, yielding parameters that are consistent
with those from the fixed-circular model. For the final parameters we
suggest adopting those from the empirical stellar model, assuming a
circular orbit. The adopted stellar parameters are listed in
Table~\ref{tab:stellar} while the planetary parameters are listed in
Table~\ref{tab:planetparam}.

\ifthenelse{\boolean{emulateapj}}{
    \begin{deluxetable*}{lrrrrrrl}
}{
    \begin{deluxetable}{lrrrrrrl}
}
\tablewidth{0pc}
\tabletypesize{\scriptsize}
\tablecaption{
    Transiting-planet hosting stars with similar properties to \hatcur{}.
    \label{tab:kmtepstars}
}
\tablehead{
    \multicolumn{1}{c}{Star}          &
    \multicolumn{1}{c}{Mass}             &
    \multicolumn{1}{c}{Radius}      &
    \multicolumn{1}{c}{$V-K$}         &
    \multicolumn{1}{c}{$\rhostar$}               &
    \multicolumn{1}{c}{Planet Mass}  &
    \multicolumn{1}{c}{Planet Period}  &
    \multicolumn{1}{c}{Reference}  \\
    \multicolumn{1}{c}{}          &
    \multicolumn{1}{c}{\msun}             &
    \multicolumn{1}{c}{\rsun}      &
    \multicolumn{1}{c}{mag}         &
    \multicolumn{1}{c}{\gcmc}               &
    \multicolumn{1}{c}{\mjup}  &
    \multicolumn{1}{c}{day}  &
    \multicolumn{1}{c}{}
}
\startdata
~~~~GJ~3470 & $0.541 \pm 0.067$ & $0.503 \pm 0.063$ & $4.281 \pm 0.030$ & $6.0 \pm 2.4$ & $0.0441 \pm 0.0053$ & $3.34$ & 1 \\
~~~~WASP-80 & $0.596 \pm 0.035$ & $0.593 \pm 0.012$ & $3.53 \pm 0.23$ & $4.040 \pm 0.012$ & $0.562 \pm 0.027$ & $3.07$ & 2,8 \\
~~~~KIC~10905746 & $0.578 \pm 0.032$ & $0.548 \pm 0.026$ & $3.94 \pm 0.24$ & $4.97 \pm 0.54$ & $\ldots$ & $9.88$ & 3 \\
~~~~WASP-43 & $0.580 \pm 0.050$ & $0.598^{+0.034}_{-0.042}$ & $3.212 \pm 0.027$ & $3.81^{+0.86}_{-0.51}$ & $1.78 \pm 0.10$ & $0.81$ & 4 \\
~~~~Kepler-45 & $0.570 \pm 0.059$ & $0.539 \pm 0.039$ & $3.990 \pm 0.058$ & $5.12 \pm 0.73$ & $0.500 \pm 0.061$ & $2.46$ & 5,9 \\
~~~~HAT-P-54 & $0.645 \pm 0.020$ & $0.617 \pm 0.013$ & $3.179 \pm 0.063$ & $3.88 \pm 0.27$ & $0.760 \pm 0.032$ & $3.80$ & 6 \\
~~~~Kepler-26 & $0.650 \pm 0.030$ & $0.590 \pm 0.030$ & $3.352 \pm 0.026$ & $4.47 \pm 0.71$ & $\ldots$ &  \tablenotemark{a} $12.28$ & 7 \\
[-1.5ex]
\enddata 
\tablerefs{1: \cite{bonfils:2012}; 2: \cite{triaud:2013}; 3: \cite{fischer:2012}; 4: \cite{hellier:2011}; 5: \cite{johnson:2012}; 6: \cite{bakos:2014:hat54}; 7: \cite{steffen:2012}; 8: \cite{mancini:2014}; 9: \cite{southworth:2012}}
\tablenotetext{a}{The period listed is for Kepler-26b, the innermost of the four planets transiting this star.\\}
\ifthenelse{\boolean{emulateapj}}{
    \end{deluxetable*}
}{
    \end{deluxetable}
}

\begin{deluxetable*}{lrrr}
\tablewidth{0pc}
\tabletypesize{\scriptsize}
\tablecaption{
    Stellar parameters for \hatcur{}
    \label{tab:stellar}
}
\tablehead{
    \multicolumn{1}{c}{~~~~~~~~Parameter~~~~~~~~}   &
    \multicolumn{1}{c}{Value Isochrones \tablenotemark{a}} &
    \multicolumn{1}{c}{\bf Value Empirical \tablenotemark{b}} &
    \multicolumn{1}{c}{Value Empirical \tablenotemark{b}} \\
    &
    \multicolumn{1}{c}{Circ.} &
    \multicolumn{1}{c}{\bf Circ.} &
    \multicolumn{1}{c}{Eccen.}
}
\startdata
\noalign{\vskip -3pt}
\sidehead{Identifying Information}
~~~~R.A. (h:m:s; 2MASS)                      &  \hatcurCCra{} & $\ldots$ & $\ldots$ \\
~~~~Dec. (d:m:s; 2MASS)                      &  \hatcurCCdec{} & $\ldots$ & $\ldots$ \\
~~~~2MASS ID                          &  \hatcurCCtwomass{} & $\ldots$ & $\ldots$ \\
\sidehead{Spectroscopic properties \tablenotemark{c}}
~~~~$\teffstar$ (K)\dotfill         &  \hatcurSMEiteffcircdartmouth & $\ldots$ & $\ldots$ \\
~~~~$\feh$\dotfill                  &  \hatcurSMEizfehcircdartmouth & $\ldots$ & $\ldots$ \\
\sidehead{Photometric properties}
~~~~$B$ (mag; APASS)\dotfill               &  \hatcurCCapassmB & \ldots & \ldots                \\
~~~~$V$ (mag; APASS)\dotfill               &  \hatcurCCapassmV & \ldots & \ldots                \\
~~~~$J$ (mag; 2MASS)\dotfill               &  \hatcurCCtwomassJmag & \ldots & \ldots           \\
~~~~$H$ (mag; 2MASS)\dotfill               &  \hatcurCCtwomassHmag & \ldots & \ldots          \\
~~~~$K_s$ (mag; 2MASS)\dotfill             &  \hatcurCCtwomassKmag & \ldots & \ldots          \\
\sidehead{Derived properties}
~~~~$\mstar$ ($\msun$)\dotfill      &  \hatcurISOmlongcircdartmouth & \hatcurISOmlongcircempirical & \hatcurISOmlongeccenempirical \\
~~~~$\rstar$ ($\rsun$)\dotfill      &  \hatcurISOrlongcircdartmouth & \hatcurISOrlongcircempirical & \hatcurISOrlongeccenempirical \\
~~~~$\teffstar$ (K) \tablenotemark{d}\dotfill      &  \hatcurISOteffcircdartmouth & \hatcurISOteffcircempirical & \hatcurISOteffeccenempirical \\
~~~~$\rhostar$ (cgs)\dotfill       &  \hatcurISOrholongcircdartmouth & \hatcurISOrholongcircempirical & \hatcurISOrholongeccenempirical \\
~~~~$\loggstar$ (cgs)\dotfill       &  \hatcurISOloggcircdartmouth & \hatcurISOloggcircempirical & \hatcurISOloggeccenempirical \\
~~~~$\lstar$ ($\lsun$)\dotfill      &  \hatcurISOlumcircdartmouth & \hatcurISOlumcircempirical & \hatcurISOlumeccenempirical \\
~~~~$M_V$ (mag)\dotfill             &  \hatcurISOmvcircdartmouth & \hatcurISOmvcircempirical & \hatcurISOmveccenempirical \\
~~~~$M_K$ (mag,\hatcurjhkfilset)\dotfill &  \hatcurISOMKcircdartmouth & \hatcurISOMKcircempirical & \hatcurISOMKeccenempirical \\
~~~~Age (Gyr)\dotfill               &  \hatcurISOagecircdartmouth & $\ldots$ & $\ldots$ \\
~~~~Distance (pc)\dotfill           &  \hatcurXdistcircdartmouth & \hatcurXdistcircempirical & \hatcurXdisteccenempirical \\
[-1.5ex]
\enddata
\tablenotetext{a}{
    Parameters based on combining the bulk stellar density determined
    from our fit to the light curves and RV data for \hatcur{}, the
    effective temperature and metallicity from the Magellan/PFS
    spectrum, together with the Dartmouth \citep{dotter:2008} stellar
    evolution models. We perform the fit two ways: allowing the
    eccentricity to vary (Eccen.), and keeping it fixed to zero
    (Circ.). The estimated value for $\rhostar$ differs significantly
    between these two fits. For the eccentric orbit fit the stellar
    density cannot be reproduced by the Dartmouth models given the
    measured effective temperature and metallicity. We therefore only
    list the parameters based on the Dartmouth models for the circular
    orbit fit.
}
\tablenotetext{b}{
    Parameters based on our empirical relations for K and M dwarf
    stellar properties, which are included directly in our global
    modelling of the light curves and RV data for \hatcur{}. Again we
    perform the fit two ways: allowing the eccentricity to vary, and
    keeping it fixed to zero. In this case including the constraint
    from the observed $V-K$ color of the star directly in the RV and
    light-curve modelling pulls the free-eccentricity model to a lower
    eccentricity solution, yielding stellar parameters that are
    consistent with those from the fixed-circular model.
}
\tablenotetext{c}{
   These parameters are determined from the I$_{2}$-free Magellan/PFS
   spectrum of \hatcur{} using the method of \cite{neves:2014}.
}
\tablenotetext{d}{
    The effective temperature listed here is derived from the
    Dartmouth stellar models, or from our empirical stellar relations,
    and is not measured directly from the spectrum.\\
}
\end{deluxetable*}

\begin{deluxetable*}{lrrr}
\tabletypesize{\scriptsize}
\tablecaption{Orbital and planetary parameters\label{tab:planetparam}}
\tablehead{
    \multicolumn{1}{c}{~~~~~~~~Parameter~~~~~~~~}   &
    \multicolumn{1}{c}{Value Isochrones \tablenotemark{a}} &
    \multicolumn{1}{c}{\bf Value Empirical \tablenotemark{b}} &
    \multicolumn{1}{c}{Value Empirical \tablenotemark{b}} \\
    \multicolumn{1}{c}{} &
    \multicolumn{1}{c}{Circ.} &
    \multicolumn{1}{c}{\bf Circ.} &
    \multicolumn{1}{c}{Eccen.}
}
\startdata
\noalign{\vskip -3pt}
\sidehead{\Lc{} parameters}
~~~$P$ (days)             \dotfill    & $\hatcurLCPcircdartmouth$ & $\hatcurLCPcircempirical$ & $\hatcurLCPeccenempirical$ \\
~~~$T_c$ (${\rm BJD}$)    
      \tablenotemark{c}   \dotfill    & $\hatcurLCTcircdartmouth$ & $\hatcurLCTcircempirical$ & $\hatcurLCTeccenempirical$ \\
~~~$T_{14}$ (days)
      \tablenotemark{c}   \dotfill    & $\hatcurLCdurcircdartmouth$ & $\hatcurLCdurcircempirical$ & $\hatcurLCdureccenempirical$ \\
~~~$T_{12} = T_{34}$ (days)
      \tablenotemark{c}   \dotfill    & $\hatcurLCingdurcircdartmouth$ & $\hatcurLCingdurcircempirical$ & $\hatcurLCingdureccenempirical$ \\
~~~$\arstar$              \dotfill    & $\hatcurPParcircdartmouth$ & $\hatcurPParcircempirical$ & $\hatcurPPareccenempirical$ \\
~~~$\zrstar$\tablenotemark{d}              \dotfill    & $\hatcurLCzetacircdartmouth$ & $\hatcurLCzetacircempirical$ & $\hatcurLCzetaeccenempirical$ \\
~~~$\rpl/\rstar$          \dotfill    & $\hatcurLCrprstarcircdartmouth$ & $\hatcurLCrprstarcircempirical$ & $\hatcurLCrprstareccenempirical$ \\
~~~$b \equiv a \cos i/\rstar$
                          \dotfill    & $\hatcurLCimpcircdartmouth$ & $\hatcurLCimpcircempirical$ & $\hatcurLCimpeccenempirical$ \\
~~~$i$ (deg)              \dotfill    & $\hatcurPPicircdartmouth$ & $\hatcurPPicircempirical$ & $\hatcurPPieccenempirical$ \\

\sidehead{Limb-darkening coefficients \tablenotemark{e}}
~~~$a_g$ (linear term)   \dotfill    & $\hatcurLBigcircdartmouth$ & $\cdots$ & $\cdots$ \\
~~~$b_g$ (quadratic term) \dotfill    & $\hatcurLBiigcircdartmouth$ & $\cdots$ & $\cdots$ \\
~~~$a_r$ (linear term)   \dotfill    & $\hatcurLBircircdartmouth$ & $\cdots$ & $\cdots$ \\
~~~$b_r$ (quadratic term) \dotfill    & $\hatcurLBiircircdartmouth$ & $\cdots$ & $\cdots$ \\
~~~$a_i$                 \dotfill    & $\hatcurLBiicircdartmouth$ & $\cdots$ & $\cdots$ \\
~~~$b_i$                  \dotfill    & $\hatcurLBiiicircdartmouth$ & $\cdots$ & $\cdots$ \\
~~~$a_z$                 \dotfill    & $\hatcurLBizcircdartmouth$ & $\cdots$ & $\cdots$ \\
~~~$b_z$                  \dotfill    & $\hatcurLBiizcircdartmouth$ & $\cdots$ & $\cdots$ \\
~~~$a_R$                 \dotfill    & $\hatcurLBiRcircdartmouth$ & $\cdots$ & $\cdots$ \\
~~~$b_R$                 \dotfill    & $\hatcurLBiiRcircdartmouth$ & $\cdots$ & $\cdots$ \\

\sidehead{RV parameters}
~~~$K$ (\ms)              \dotfill    & $\hatcurRVKcircdartmouth$ & $\hatcurRVKcircempirical$ & $\hatcurRVKeccenempirical$ \\
~~~$\sqrt{e}\cos\omega$ 
                          \dotfill    & $\ldots$ & $\ldots$ & $\hatcurRVrkeccenempirical$ \\
~~~$\sqrt{e}\sin\omega$
                          \dotfill    & $\ldots$ & $\ldots$ & $\hatcurRVrheccenempirical$ \\
~~~$e\cos\omega$ 
                          \dotfill    & $\ldots$ & $\ldots$ & $\hatcurRVkeccenempirical$ \\
~~~$e\sin\omega$
                          \dotfill    & $\ldots$ & $\ldots$ & $\hatcurRVheccenempirical$ \\

~~~$e$                    \dotfill    & $0$, fixed & $0$, fixed & $\hatcurRVecceneccenempirical$ \\
~~~$\omega$                    \dotfill    & $\ldots$ & $\ldots$ & $\hatcurRVomegaeccenempirical$ \\
~~~PFS RV jitter (\ms)\tablenotemark{f}        
                          \dotfill    & $\hatcurRVjitterAcircdartmouth$ & $\hatcurRVjitterAcircempirical$ & $\hatcurRVjitterAeccenempirical$ \\
~~~FEROS RV jitter (\ms)        
                          \dotfill    & $\hatcurRVjitterBcircdartmouth$ & $\hatcurRVjitterBcircempirical$ & $\hatcurRVjitterBeccenempirical$ \\
~~~HARPS RV jitter (\ms)        
                          \dotfill    & $\hatcurRVjitterCcircdartmouth$ & $\hatcurRVjitterCcircempirical$ & $\hatcurRVjitterCeccenempirical$ \\

\sidehead{Planetary parameters}
~~~$\mpl$ ($\mjup$)       \dotfill    & $\hatcurPPmlongcircdartmouth$ & $\hatcurPPmlongcircempirical$ & $\hatcurPPmlongeccenempirical$ \\
~~~$\rpl$ ($\rjup$)       \dotfill    & $\hatcurPPrlongcircdartmouth$ & $\hatcurPPrlongcircempirical$ & $\hatcurPPrlongeccenempirical$ \\
~~~$C(\mpl,\rpl)$
    \tablenotemark{g}     \dotfill    & $\hatcurPPmrcorrcircdartmouth$ & $\hatcurPPmrcorrcircempirical$ & $\hatcurPPmrcorreccenempirical$ \\
~~~$\rhopl$ (\gcmc)       \dotfill    & $\hatcurPPrhocircdartmouth$ & $\hatcurPPrhocircempirical$ & $\hatcurPPrhoeccenempirical$ \\
~~~$\log g_p$ (cgs)       \dotfill    & $\hatcurPPloggcircdartmouth$ & $\hatcurPPloggcircempirical$ & $\hatcurPPloggeccenempirical$ \\
~~~$a$ (AU)               \dotfill    & $\hatcurPParelcircdartmouth$ & $\hatcurPParelcircempirical$ & $\hatcurPPareleccenempirical$ \\
~~~$T_{\rm eq}$ (K)       \dotfill    & $\hatcurPPteffcircdartmouth$ & $\hatcurPPteffcircempirical$ & $\hatcurPPteffeccenempirical$ \\
~~~$\Theta$\tablenotemark{h}\dotfill  & $\hatcurPPthetacircdartmouth$ & $\hatcurPPthetacircempirical$ & $\hatcurPPthetaeccenempirical$ \\
~~~$\langle F \rangle$ ($10^{\hatcurPPfluxavgdim}$\ergscmsq) 
\tablenotemark{i}         \dotfill    & $\hatcurPPfluxavgcircdartmouth$ & $\hatcurPPfluxavgcircempirical$ & $\hatcurPPfluxavgeccenempirical$ \\
[-1.5ex]
\enddata
\tablenotetext{a}{
    Parameters based on combining the bulk stellar density determined
    from our fit to the light curves and RV data for \hatcur{}, the
    effective temperature and metallicity from the Magellan/PFS
    spectrum, together with the Dartmouth \citep{dotter:2008} stellar
    evolution models. We perform the fit two ways: allowing the
    eccentricity to vary (Eccen.), and keeping it fixed to zero
    (Circ.). The estimated value for $\rhostar$ differs significantly
    between these two fits. For the eccentric orbit fit the stellar
    density cannot be reproduced by the Dartmouth models given the
    measured effective temperature and metallicity. We therefore only
    list the parameters based on the Dartmouth models for the circular
    orbit fit.
}
\tablenotetext{b}{
    Parameters based on our empirical relations for K and M dwarf
    stellar properties, which are included directly in our global
    modelling of the light curves and RV data for \hatcur{}. Again we
    perform the fit two ways: allowing the eccentricity to vary, and
    keeping it fixed to zero. In this case including the constraint
    from the observed $V-K$ color of the star directly in the RV and
    light-curve modelling pulls the free-eccentricity model to a lower
    eccentricity solution, yielding parameters that are consistent
    with those from the fixed-circular model. We adopt the empirical
    models, with fixed circular orbit, for our final system
    parameters.
}
\tablenotetext{c}{
    \ensuremath{T_c}: Reference epoch of mid transit that minimizes the
    correlation with the orbital period. BJD is calculated from UTC.
    \ensuremath{T_{14}}: total transit duration, time between first to
    last contact;
    \ensuremath{T_{12}=T_{34}}: ingress/egress time, time between first
    and second, or third and fourth contact.
}
\tablenotetext{d}{
    Reciprocal of the half duration of the transit used as a jump
    parameter in our MCMC analysis in place of $\arstar$. It is
    related to $\arstar$ by the expression $\zrstar = \arstar
    (2\pi(1+e\sin \omega))/(P \sqrt{1 - b^{2}}\sqrt{1-e^{2}})$
    \citep{bakos:2010:hat11}.
}
\tablenotetext{e}{
        Values for a quadratic law given separately for the Sloan~$g$,
        $r$, $i$, and $z$ filters, as well as for the $R$
        filter. These values were adopted from the tabulations by
        \cite{claret:2004} assuming $T_{\rm eff} = 3700$\,K and
             [Fe/H]$=0$. We used the same limb darkening coefficients
             for all four of the models shown here.
}
\tablenotetext{f}{
    The jitter was added in quadrature to the RV uncertainties and
    varied in the fit following \citep{hartman:2012:hat39hat41}. We
    assumed an independent jitter for each instrument.
}
\tablenotetext{g}{
    Correlation coefficient between the planetary mass \mpl\ and radius
    \rpl.
}
\tablenotetext{h}{
    The Safronov number is given by $\Theta = \frac{1}{2}(V_{\rm
    esc}/V_{\rm orb})^2 = (a/\rpl)(\mpl / \mstar )$
    \citep[see][]{hansen:2007}.
}
\tablenotetext{i}{
    Incoming flux per unit surface area, averaged over the orbit.\\
}
\end{deluxetable*}

\section{Discussion}
\label{sec:discussion}

\begin{figure*}[!ht]
\plottwo{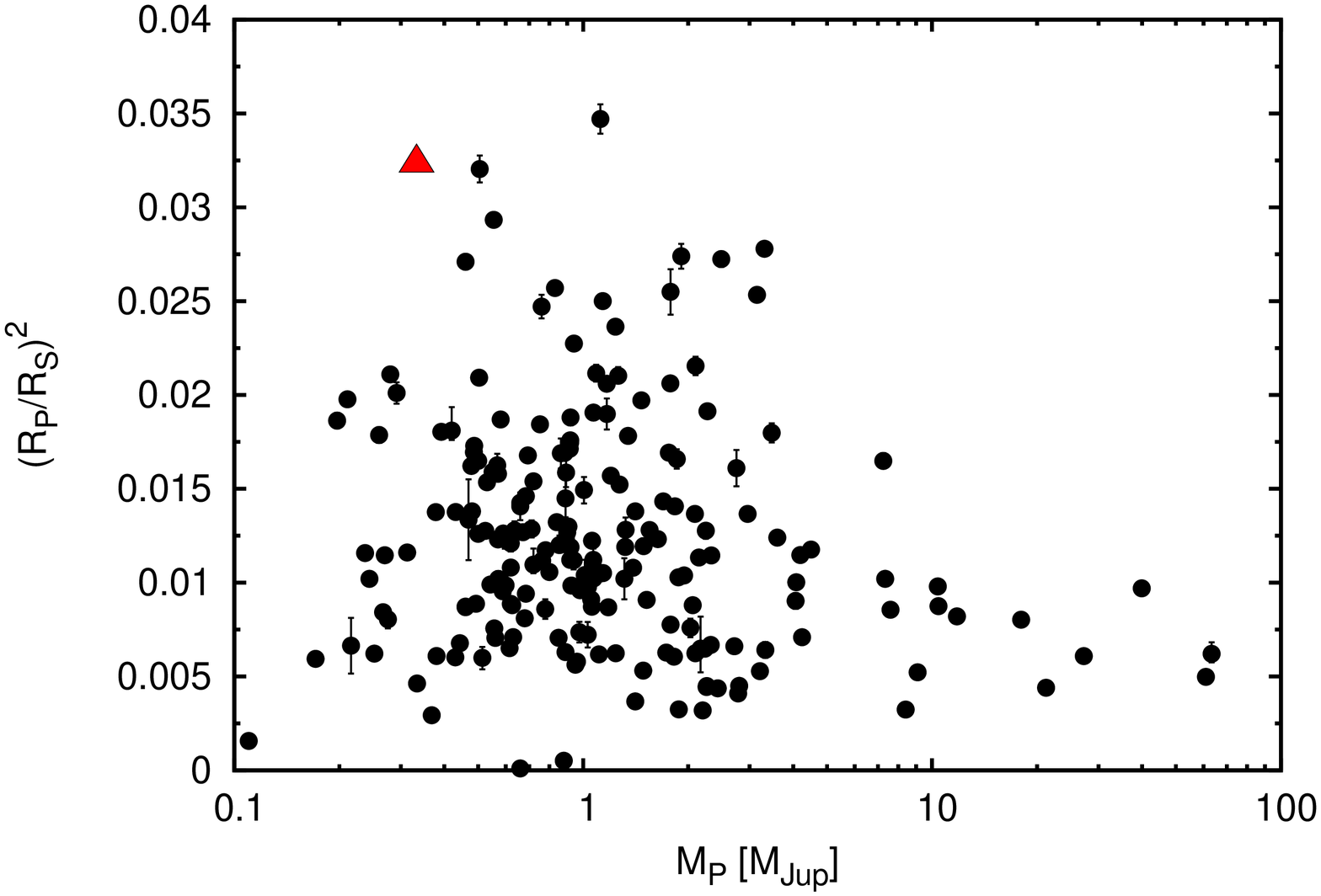}{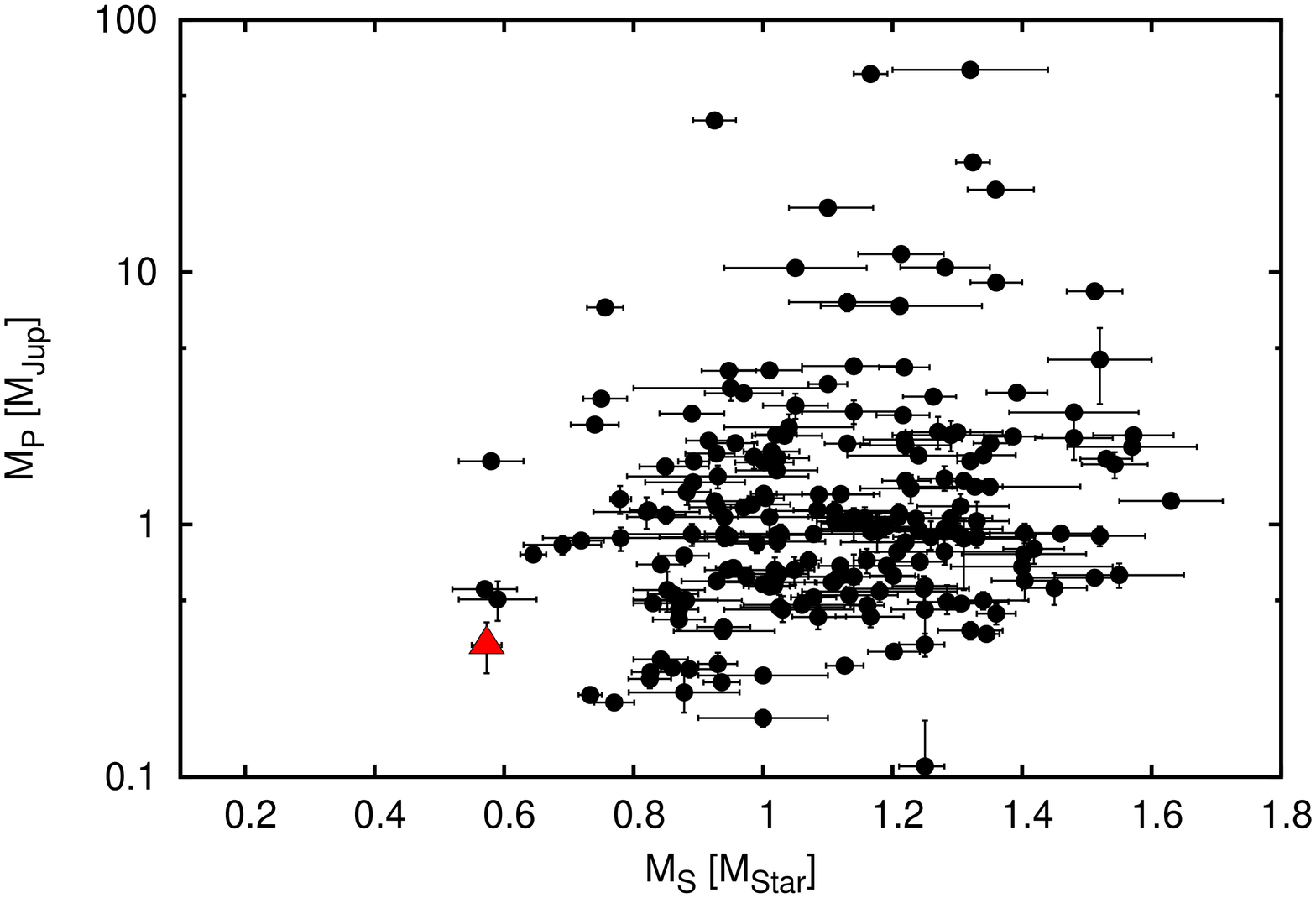}
\plottwo{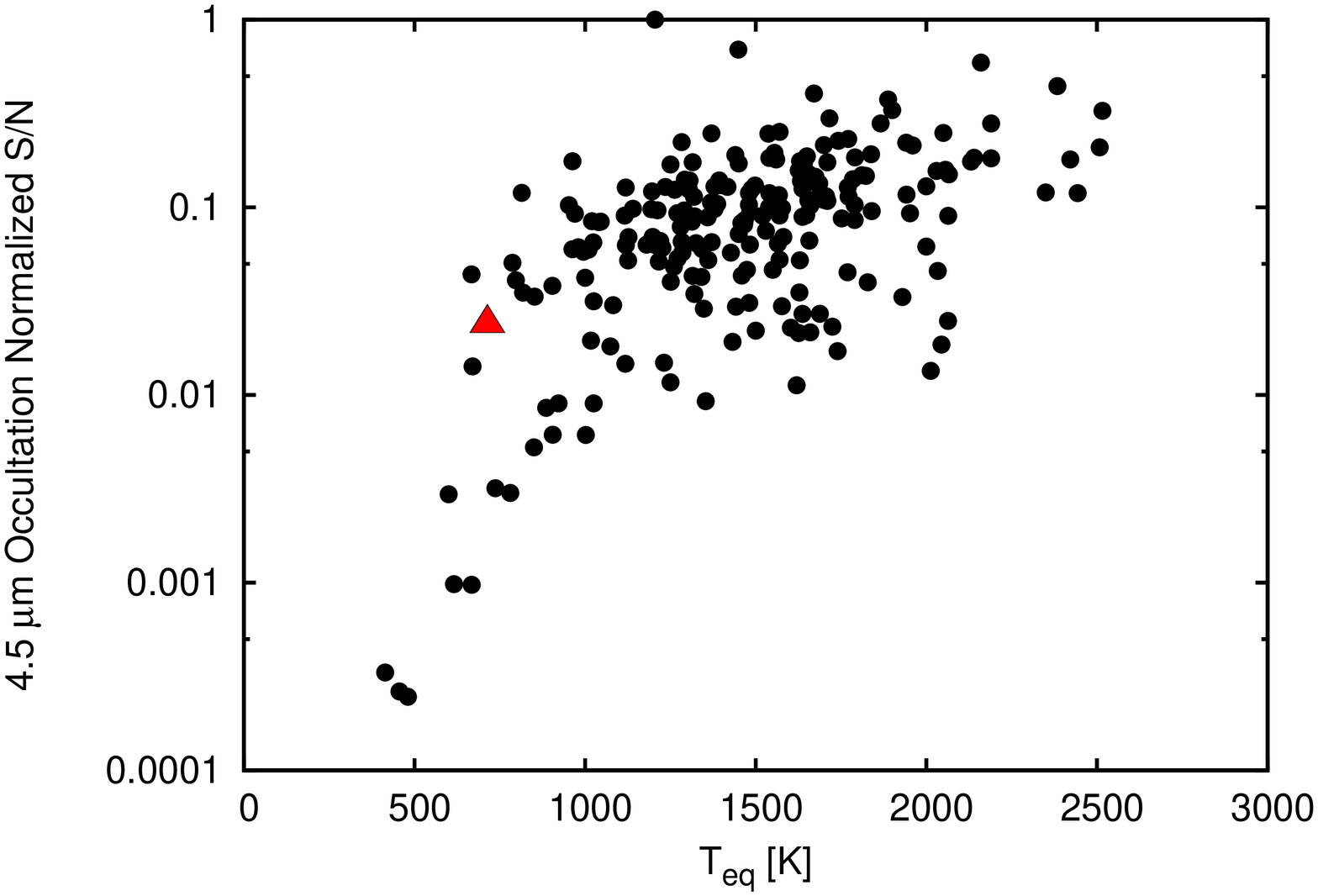}{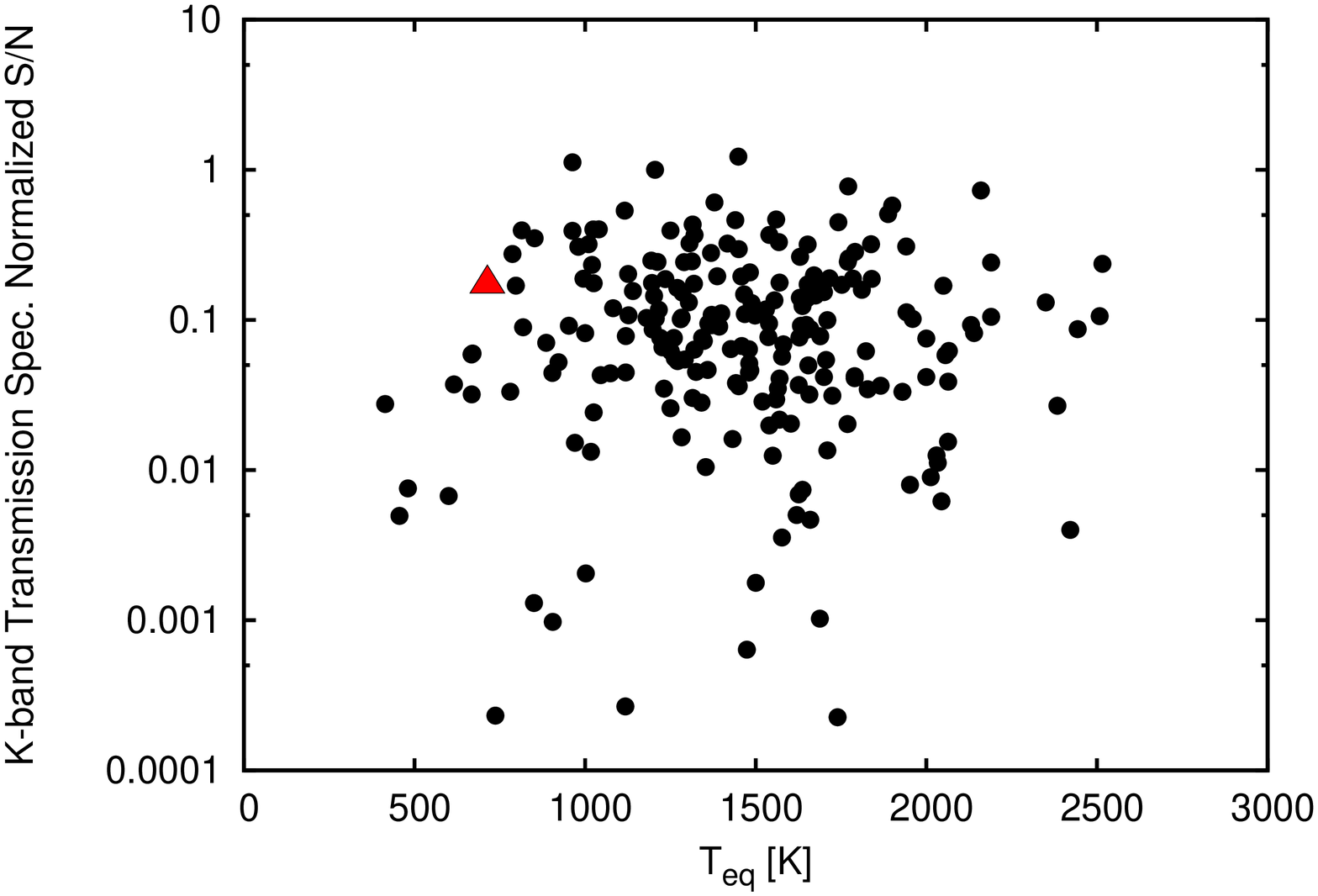}
\caption{
Comparison of the \hatcur{} system to other known transiting exoplanet
systems with $M_{P} > 0.1$\,\mjup. In each case \hatcur{} is indicated
by the filled red triangle, while other objects are indicated by
filled black circles. Top Left: Planet mass vs.~transit depth as
measured by $(\rprstar)^2$. Top Right: Host star mass vs.~planet
mass. Bottom Left: Estimated S/N of a single occultation event at
4.5\,$\mu$m normalized to that of HD~189733b. Bottom Right: Estimated
S/N of a transmission spectrum signal from one transit, assuming a
pure H$_{2}$ atmosphere, and observed at $K$-band, normalized to that
of HD~189733b. \hatcur{} has some of the deepest transits of known
transiting planet systems, it is also one of the lowest mass stars
known to host a transiting gas giant planet. In terms of atmospheric
characterization it is one of the highest S/N targets with $T_{\rm eq}
< 750$\,K for both occultation and transmission spectroscopy.\\
}
\label{fig:hats6properties}
\end{figure*}

In this paper we have presented the discovery of the \hatcur{}
transiting planet system by the HATSouth survey. We found that
\hatcurb{} has a mass of $\hatcurPPm{}$\,\mjup, radius of
$\hatcurPPr{}$\,\rjup\ and orbits a star with a mass of
$\hatcurISOm{}$\,\msun. \hatcur{} is one of only four stars with $M
\la 0.6$\,\msun\ known to host a short period gas giant planet (the
other three are WASP-80, WASP-43, and Kepler-45; see
Table~\ref{tab:kmtepstars}).  This is illustrated in
Figure~\ref{fig:hats6properties} where we plot planet mass against the
host star mass. We note that the published masses for these stars come
from a disparate set of methods. Applying the empirical relations
developed in this paper (Section~\ref{sec:empirical}) to these two
systems, using $\rho_{\star}$, $V$, $J$, $H$ and $K_{S}$ as
observables, we find masses of $0.614\pm0.031$\,\msun, and
$0.601\pm0.031$\,\msun, for WASP-43 and WASP-80, respectively,
slightly higher than the published values of
$0.580\pm0.050$\,\msun\ and $0.596\pm0.035$\,\msun. This same method
yields a mass of $0.534\pm0.034$\,\msun\ for Kepler-45, which is a bit
lower than the published value of $0.570\pm0.059$\,\msun.

\hatcurb{} also has some of the deepest transits known, with
$(\rprstar)^2 = 0.0323 \pm 0.0003$.  There are only two known planets
with deeper transits than \hatcurb{}: Kepler-45b ($(\rprstar)^2 =
0.0362\pm0.0054$; \citealp{southworth:2012}), and WTS-2b, a hot Jupiter
on a very short period orbit around a K2 dwarf star ($(\rprstar)^2 =
0.0347 \pm 0.0008$; \citealp{birkby:2014}).  One other system with
transit depths close to that of \hatcur{} is WASP-80b ($(\rprstar)^2 =
0.0292 \pm 0.0001$;
\citealp{mancini:2014}). Figure~\ref{fig:hats6properties} illustrates
this as well, plotting the transit depth vs.~planet mass.

With an early M dwarf host star, \hatcurb{} receives
substantially less stellar irradiation than planets with the same
orbital period around Sun-like stars. As a result the equilibrium
temperature, assuming zero albedo, is a relatively mild
$\hatcurPPteff{}$\,K. The atmospheric chemistry at this temperature is
expected to be notably different from other hotter planets, with a
greater abundance of CH$_{4}$ and associated hydro-carbon hazes
\citep{howe:2012,fortney:2013}. This makes \hatcurb{} a potentially
interesting target for studying the atmospheric properties of a warm
gas giant. In the bottom two panels of
Figure~\ref{fig:hats6properties} we plot the approximate $K$-band
transmission spectrum S/N, and $4.5$\,$\mu$m occultation spectrum S/N
vs.~T$_{\rm eq}$. In both cases we normalize the S/N to that of the
prototypical hot Jupiter HD~189733b. For the transmission spectrum S/N
we use the relation \citep[e.g.,][]{winn:2010}:
\begin{equation}
{\rm S/N} \propto \frac{2R_{P}HN_{H}}{R_{\star}^2}\sqrt{T_{14}F_{K}}
\label{eqn:sntrans}
\end{equation}
where $H = k_{B}T_{\rm eq}/\mu g_{P}$, $\mu$ is the mean molecular
weight (we assume $3.347\times 10^{-27}$\,kg for a pure H$_{2}$
atmosphere), $g_{P}$ is the surface gravity of the planet, $T_{14}$ is
the transit duration, $F_{K}$ is the $K$-band flux of the star in
Janskys, and $N_{H}$ is a constant of order unity (we assume 1)
depending on the particular spectroscopic feature under
consideration. The term $\sqrt{T_{14}F_{K}}$ is included because the
photometric noise is proportional to $1/\sqrt{F_{K}}$ and the number
of in transit measurements that can be made is proportional to
$T_{14}$ (meaning the noise goes as $1/\sqrt{T_{14}}$). We neglect the
constant of proportionality on the right-hand-side of
eq.~\ref{eqn:sntrans} needed to make the expression for S/N
unitless. For the occultation spectrum S/N we use the relation
\citep[e.g.,][]{winn:2010}:
\begin{equation}
{\rm S/N} \propto \left( \frac{R_{P}}{R_{\star}} \right)^2\frac{\exp(hc/k_{B}\lambda T_{\rm eff}) - 1}{\exp(hc/k_{B}\lambda T_{\rm eq}) - 1}\sqrt{T_{14}F_{4.5}}
\end{equation}
where $h$ is Planck's constant, $c$ is the speed of light, $k_{B}$ is
Boltmann's constant, $T_{\rm eff}$ is the stellar effective
temperature, $\lambda$ is the effective wavelength (we assume
4.5\,$\mu$m), and $F_{4.5}$ is the 4.5\,$\mu$m flux of the star in
Janskys.

We find that \hatcurb{} has the highest expected S/N transmission
spectrum among known gas giant planets with $T_{\rm eq} <
750$\,K. Regarding occultations, \hatcurb{} has the second highest
expected S/N at 4.5\,$\mu$m among gas giant planets with $T_{\rm
  eq} < 750$\,K after HD~80606b \citep[][the expected S/N for \hatcurb{} is
$\sim 50$\% that of HD~80606b]{laughlin:2009}. 

Perhaps the most relevant planet for comparison is WASP-80b, a hot
Jupiter with $T_{\rm eq} = 825 \pm 20$\,K \citep{mancini:2014}, with a
similar transit depth to \hatcurb{}, but orbiting a somewhat brighter
star. High-precision ground-based photometric transit observations of
this system over several optical and near infrared band-passes have
been presented by \citet{mancini:2014} and \citet{fukui:2014}, which
have provided suggestive evidence for the existence of a haze in the
atmosphere.  The expected transmission S/N for \hatcurb{} is
approximately $0.4$ times that of WASP-80b, so measuring the
transmission spectrum for \hatcurb{} will be challenging, but not out
of reach for ground-based facilities. \hatcurb{} is also a promising target for the
        {\em NASA James Webb Space Telescope} mission, currently
        scheduled for launch in 2018.

We checked whether the simultaneous optical multi-filter GROND
observations that we have already obtained for \hatcur{} can be used
to probe the transmission spectrum. We fit a \citet{mandel:2002}
transit model to each of the GROND $g$, $r$, $i$ and $z$ light curves,
fixing all transit parameters to the best-fit values for the system,
except for \rprstar, which we allow to vary independently for each
band-pass. We find that values for \rprstar\ values are consistent
with a flat transmission spectrum, to within the
uncertainties. Additional observations would be needed to detect
spectroscopic features in the atmosphere of \hatcurb.

In order to characterize the host star \hatcur{} we have also
presented a new set of empirical stellar models applicable to
main-sequence stars with $M < 0.80$\,\msun. We use eclipsing binaries
as well as resolved binaries with masses and measured absolute
magnitudes to determine empirical relations between $\rhostar$ and $R$
(and thus $\rhostar$ and $M$), between $M$ and $M_{\rm bol}$, and
between $T_{\rm eff}$ and the $V$, $J$, $H$ and $K$-band bolometric
corrections. We perform a global fit to the available data,
determining the intrinsic scatter in the relations in a
self-consistent manner. We find that from the density alone it is
possible to measure the mass of a $\sim 0.6$\,\msun\ star to $\sim
7$\% precision, and the radius to $\sim 2$\% precision. These
precisions may be improved by up to a factor of two, to $\sim 3$\% and
$\sim 1$\%, respectively, by incorporating additional photometric
data. The relations presented here are similar to those presented by
\citet{johnson:2011} and \citet{johnson:2012} except that we do not
attempt to incorporate metallicity directly into the model (it
contributes instead to the inherent scatter in the relations), and we
use a much larger sample of eclipsing binaries than previously
considered. Future improvements to our model may be made by
incorporating metallicity information directly into the model, but this
will require a large sample of stars with measured masses, radii,
absolute magnitudes, and metallicities.

Finally, we note that the discovery of \hatcurb{} is a demonstration
of the enhanced sensitivity of HATSouth to planets around low-mass K
and M dwarf stars relative to other wide-field ground based surveys,
such as HATNet or WASP. The reason for this enhanced sensitivity is
the larger aperture optics used by HATSouth, providing higher
photometric precision for fainter stars. HATSouth yields better than
2\% precision, per 240\,s integration, for stars down to $r \sim
14.5$. For comparison, if using the same integration time as HATSouth,
HATNet would yield 2\% precision for stars with $r \la 13$. Based on
the TRILEGAL Galactic models \citep{girardi:2005} we estimate that
there are $\sim 230000$ stars with $M < 0.6$\,\msun, $r < 14.5$ in the
Southern sky, and at least $10^{\circ}$ from the Galactic disk that
could be observed by HATSouth. This is nearly a factor of ten more
than the number of stars with $M < 0.6$\,\msun\ and $r < 13.0$ in the
Northern sky that could be observed by HATNet to the same photometric
precision. While in principle the discovery of \hatcurb{} may be used
to infer the occurrence rate of hot Jupiters around early M dwarfs
\citep[e.g.,][]{gaidos:2014}, in practice this requires a careful
calibration of our transit detection efficiency over the sample of M
dwarfs that we have observed. This is beyond the scope of the present
paper, but will be the subject of future work.

\acknowledgements 

Development of the HATSouth project was funded by NSF MRI grant
NSF/AST-0723074, operations have been supported by NASA grants NNX09AB29G and NNX12AH91H, and
follow-up observations receive partial support from grant
NSF/AST-1108686.
A.J.\ acknowledges support from FONDECYT project 1130857, BASAL CATA PFB-06, and project IC120009 ``Millennium Institute of Astrophysics (MAS)'' of the Millenium Science Initiative, Chilean Ministry of Economy. R.B.\ and N.E.\ are supported by CONICYT-PCHA/Doctorado Nacional. R.B.\ and N.E.\ acknowledge additional support from project IC120009 ``Millenium Institute of Astrophysics  (MAS)'' of the Millennium Science Initiative, Chilean Ministry of Economy.  V.S.\ acknowledges support form BASAL CATA PFB-06.  M.R.\ acknowledges support from FONDECYT postdoctoral fellowship 3120097.
This work is based on observations made with ESO Telescopes at the La
Silla Observatory.
This paper also uses observations obtained with facilities of the Las
Cumbres Observatory Global Telescope.
Work at the Australian National University is supported by ARC Laureate
Fellowship Grant FL0992131.
We acknowledge the use of the AAVSO Photometric All-Sky Survey (APASS),
funded by the Robert Martin Ayers Sciences Fund, and the SIMBAD
database, operated at CDS, Strasbourg, France.
Operations at the MPG~2.2\,m Telescope are jointly performed by the
Max Planck Gesellschaft and the European Southern Observatory.  The
imaging system GROND has been built by the high-energy group of MPE in
collaboration with the LSW Tautenburg and ESO\@.  We thank R\'egis
Lachaume for his technical assistance during the observations at the
MPG~2.2\,m Telescope. We thank Helmut Steinle and Jochen Greiner for
supporting the GROND observations presented in this manuscript.
We are grateful to P.Sackett for her help in the early phase of the
HATSouth project.




\end{document}